\documentclass[reprint,aps,prb,twocolumn,superscriptaddress]{revtex4-2}
\usepackage{}%\bibliographystyle{apsrev}
\usepackage{chngcntr}
\usepackage{amssymb}
\usepackage{upgreek}
\usepackage{graphicx}
\usepackage{mathrsfs}
\usepackage{textcomp}
\usepackage{mathcomp}
\usepackage[utf8]{inputenc}
\usepackage{color}
\usepackage{siunitx}
\usepackage{url}
\usepackage{multirow}
\usepackage{placeins}
\usepackage{gensymb}
\usepackage[utf8]{inputenc}
\usepackage[paperwidth=210mm,paperheight=297mm,centering,hmargin=2cm,vmargin=2cm]{geometry}
\usepackage{lipsum}
\usepackage{circuitikz}
\usepackage{fancyhdr}
\usepackage{mathtools}
\usepackage{bm}
\usepackage[normalem]{ulem}
\usepackage{upgreek}
\usepackage{comment}
\DeclareSIUnit\angstrom{\text {Å}}

\begin{document}

\title{Quantum critical point followed by Kondo-like behavior due to Cu substitution in itinerant, antiferromagnet ${\text{La}_{2}\text{(Cu}_{x}\text {Ni}_{1-x})_7}$}

\author{Atreyee Das}
\affiliation{Ames National Laboratory, U.S. DOE, Iowa State 
University, Ames, Iowa 50011, USA}
\affiliation{Department of Physics and Astronomy, Iowa State University, Ames, Iowa 50011, USA}

\author{Siham Mohamed}
\affiliation{Ames National Laboratory, U.S. DOE, Iowa State 
University, Ames, Iowa 50011, USA}
\affiliation{Department of Chemistry, Iowa State University, Ames, Iowa 50011, USA}

\author{Raquel A. Ribeiro}
\affiliation{Ames National Laboratory, U.S. DOE, Iowa State 
University, Ames, Iowa 50011, USA}
\affiliation{Department of Physics and Astronomy, Iowa State University, Ames, Iowa 50011, USA}

\author{Tyler J. Slade}
\affiliation{Ames National Laboratory, U.S. DOE, Iowa State 
University, Ames, Iowa 50011, USA}

\author{Juan Schmidt}
\affiliation{Ames National Laboratory, U.S. DOE, Iowa State 
University, Ames, Iowa 50011, USA}
\affiliation{Department of Physics and Astronomy, Iowa State University, Ames, Iowa 50011, USA}

\author{Chandan Setty}
\affiliation{Ames National Laboratory, U.S. DOE, Iowa State 
University, Ames, Iowa 50011, USA}
\affiliation{Department of Physics and Astronomy, Iowa State University, Ames, Iowa 50011, USA}

\author{Sergey L. Bud'ko}
\affiliation{Ames National Laboratory, U.S. DOE, Iowa State 
University, Ames, Iowa 50011, USA}
\affiliation{Department of Physics and Astronomy, Iowa State University, Ames, Iowa 50011, USA}

\author{Paul C. Canfield}
\affiliation{Ames National Laboratory, U.S. DOE, Iowa State 
University, Ames, Iowa 50011, USA}
\affiliation{Department of Physics and Astronomy, Iowa State University, Ames, Iowa 50011, USA}

\begin{abstract}
    $\text{La}_2 \text{Ni}_7$ is an itinerant magnetic system with a small ordered moment of $\sim$ 0.1 $\mu_{B}/\text{Ni}$ and a series of antiferromagnetic (AFM) transitions at $T_1$ = 61.0 K, $T_2$ = 56.5 K and $T_3$ = 42.2 K. $M(H)$, and $\rho(H)$ isotherms as well as constant field $M(T)$ and $\rho(T)$ measurements on single crystalline samples manifest a complex, anisotropic $H-T$ phase diagram with multiple phase lines. Here we present the growth and characterization of single crystals of the ${\text{La}_{2}\text{(Cu}_{x}\text {Ni}_{1-x})_7}$ series for 0 $\leq x \leq$ 0.181. We measured powder x-ray diffraction, and composition, as well as anisotropic temperature and field dependent resistivity, temperature and field dependent magnetization and temperature dependent heat capacity on these single crystals. Using the measured data we infer a transition temperature-composition $(T-x)$ phase diagram for this system to study the evolution of the AFM ordering upon Cu substitution. For ${0 \leq x \leq 0.097}$, the system remains magnetically ordered at base temperature with $x \leq$ 0.012, showing signs of multiple AFM ordering temperatures. For the higher substitution levels, ${0.125 \leq x \leq 0.181}$, there are no signatures of magnetic ordering, but anomalous features in resistance and heat capacity data are observed which are consistent with the Kondo effect in this system. The intermediate $x$ = 0.105 sample lies between the magnetic ordered and the Kondo regime and is in the vicinity of the AFM-quantum critical point (QCP). Thus, ${\text{La}_{2}\text{(Cu}_{x}\text {Ni}_{1-x})_7}$ is an example of a small moment system that can be tuned through a QCP. Given these data combined with the fact that the $\text{La}_2 \text{Ni}_7$ structure has kagome-like, Ni-sublattices running perpendicular to the crystallographic $c$ axis, and a predicted $3d$-electron flat band that contributes to the density of states near the Fermi energy, ${\text{La}_{2}\text{(Cu}_{x}\text {Ni}_{1-x})_7}$ becomes a promising system to host and study exotic physics. 
    
\end{abstract}

\maketitle

\section{Introduction}
\label{sec:Introduction}

Itinerant, metallic, magnetic systems with low transition temperatures are often explored with the anticipation that their ordering temperatures can be suppressed by application of pressure, changing the chemical composition (doping), or by magnetic field \cite{Canfield2016PreservedMagnetism}. Suppressing second-order phase transitions to zero temperature is associated with the emergence of several exotic physical phenomena; unconventional superconductivity and/or non Fermi liquid like temperature dependencies are often found in proximity of the quantum critical point (QCP) \cite{Canfield2016PreservedMagnetism, Brando2016MetallicFerromagnets, Pfleiderer2001Non-Fermi-liquidFerromagnets, Huy2007SuperconductivityUCoGe,  Dagotto1994, Paglione2010High-temperatureMaterials, Gegenwart2008QuantumMetals, Levy2007AcuteURhGe, Pfleiderer2004PartialMnSi, Ubaid-Kassis2010Quantum/math, Pfleiderer2009SuperconductingCompounds,  Cheng2015PressureMnP, QunatumcriticalityHVL}. Intensive studies have shown that although antiferromagnetic (AFM) transitions in many metallic systems can be continuously suppressed to zero temperature by the non-thermal tuning parameters mentioned above \cite{Canfield2016PreservedMagnetism, Shibauchi2014APnictides, Gegenwart2008QuantumMetals, Lohneysen1994Non-Fermi-liquidInstability, Friedemann2009DetachingInYbRh2Si2, Mun2013Magnetic-field-tunedYbPtBi},  the situation becomes quite different for ferromagnetic (FM) transitions. Theoretical as well as experimental studies suggest that stoichiometric systems with minimum disorder generally avoid a FM quantum critical point (QCP) at zero field \cite{Brando2016MetallicFerromagnets, Gati2021FormationLaCrGe3, Xiang2021AvoidedLa5Co2Ge3, Das2024Effect/math}. Although, hydrostatic pressure, and applied magnetic field are considered 'cleaner' choices for suppression of transition temperature, the reason being that they do not introduce any additional disorder, chemical substitution, despite inducing a degree of disorder, is another, often more versatile, tuning parameter to access QCP and, more generally, tune the ground state in these systems. 

\par

$\text{La}_2 \text{Ni}_7$, which belongs to the $\text{Ce}_2 \text{Ni}_7$ (space group P6/mmc, \#194) family, had been suspected to be a small moment antiferromagnet for decades, but prior studies on polycrystalline samples left the nature and even the number of its low temperature states unclear \cite{Virkar1969CrystalPhases, Buschow1970TheR2Ni7, Parker1983MagneticLa2Ni7, Gottwick1985TransportCompounds, Tazuke1993MagnetismCe,Tazuke1997MagneticSystem, Tazuke2004MetamagneticLa2Ni7, Fukase1999Successive/sub, Crivello2020RelationStructures}. Recently large, well-faceted, hexagonal single crystalline plates of $\text{La}_2 \text{Ni}_7$ were synthesized, and anisotropic transport, magnetization, and heat capacity measurements done on these crystals revealed a series of antiferromagnetic (AFM) transitions at $T_1$ = 61.0 K, $T_2$ = 56.5 K and $T_3$ = 42.2 K \cite{Ribeiro2022Small-moment/math}. With a saturated moment of $\sim$ 0.1 $\mu_{B}/\text{Ni}$ and an effective moment of $\sim$ 1.0 $\mu_{B}/\text{Ni}$, $\text{La}_2 \text{Ni}_7$, clearly is a small moment itinerant magnetic system. The anisotropic $H-T$ phase diagram constructed identified multiple magnetic regions. These findings served as the basis for further single crystal neutron diffraction \cite{Wilde2022Weak/math}, ARPES \cite{Lee2023Electronicsub7/sub} and NMR/NQR \cite{Ding2023} measurements on this system. %The small moment ordering in $\text{La}_2 \text{Ni}_7$ makes it a very promising candidate to study further by perturbation using non-thermal tuning parameters. 

\par

\begin{figure}[htbp]
    \centering
    \includegraphics[width=\linewidth]{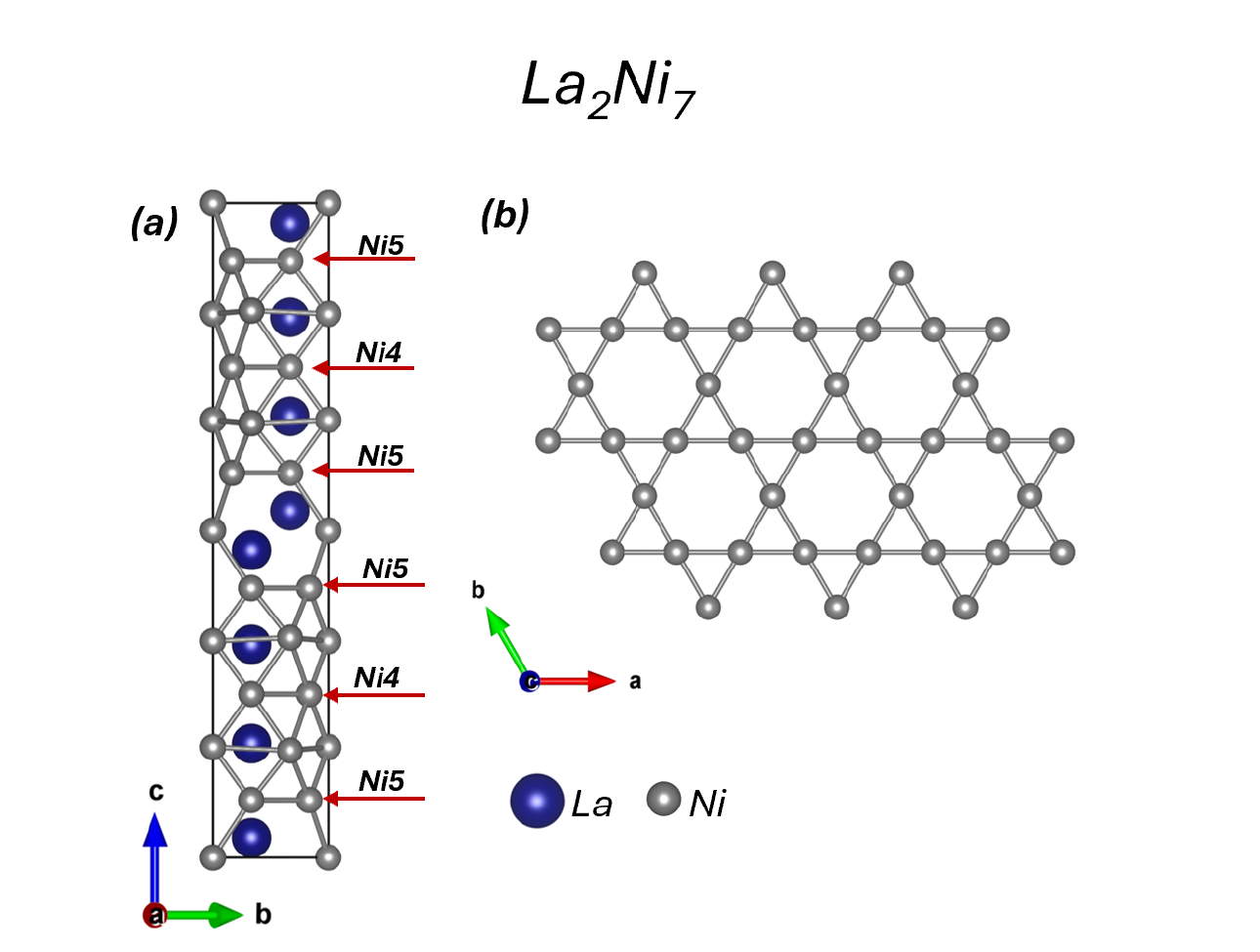}
    \caption{\footnotesize{(Color online) \textit{(a)} The crystal structure of $\text{La}_2 \text{Ni}_7$ as generated by VESTA \cite{Momma2011iVESTA3/iData}. The Ni4 and Ni5 planes are equivalent and have been marked by red arrows. \textit{(b)} The Ni4/Ni5 plane along the $ab-$ plane which form a kagome lattice.}}
    \label{fig:crystal}
\end{figure}

Whereas an initial study of $\text{La}_2 \text{Ni}_7$ under hydrostatic pressure found the three transitions to be only weakly changed up to 2 GPa \cite{Ribeiro2022Small-moment/math}, the small itinerant moment found in this system may suggest close proximity to a QCP and still make this material an attractive candidate for further investigations. In particular, Fig. \ref{fig:crystal} shows that the $Ni4$ and $Ni5$ atoms, occupying respectively the Wyckoff positions 12k and 6h, \cite{Buschow1970TheR2Ni7} in $\text{La}_2 \text{Ni}_7$ form Kagome-like networks. In such structures, destructive interference associated with frustrated electron hoping pathways is predicted to lead to flat bands, which if they are close enough to the Fermi level, may be related to magnetic (or other) instabilities. Indeed, density functional theory (DFT) calculations predicted a flat, $Ni-3d$ band near $E_F$ \cite{Lee2023Electronicsub7/sub}. We choose Cu substitution on the Ni site, which in a more local moment picture, replaces the moment-bearing Ni with non-magnetic Cu. In a more itinerant picture, Cu substitution changes the band filling, possibly moving the system away from some Stoner like criterion.

%Given that ARPES measurements did not detect this band \cite{Lee2023Electronicsub7/sub}, one possibility is that this flat band may be slightly above $E_F$, and may be accessible with electron doping. 

\par

In this paper, we report the synthesis and 
physical properties of electron doped ${\text{La}_{2}\text{(Cu}_{x}\text {Ni}_{1-x})_7}$ with ${0 \leq x \leq 0.181}$. Single crystals of ${\text{La}_{2}\text{(Cu}_{x}\text {Ni}_{1-x})_7}$ were grown and their temperature and field dependent magnetization, temperature dependent resistivity, and temperature dependent heat capacity were investigated. Our studies reveal that for ${0 \leq x \leq 0.097}$, the system remains magnetically ordered at base temperature, with the low dopings $x \leq$ 0.012 showing signs of multiple magnetic transitions. The magnetic transition temperatures decreases monotonically, and magnetic order is suppressed below $T$ = 1.8 K for $x >$ 0.097. The samples with higher substitution levels, ${0.125 \leq x \leq 0.181}$, manifest no signatures of magnetic ordering and instead have a low temperature upturn in $\rho(T)$ and $C_p/T$, which are features that can be associated with Kondo-like behavior. The intermediate substitution $x$ = 0.105 lies between the magnetic ordered and the Kondo regime and shows behavior characteristic of a non-Fermi liquid (nFL) in the resistivity data, and is in the vicinity of the QCP for this system.

\par

\section{Experimental Methods}
\label{sec:experimental methods}

\begin{figure*}[htbp]
    \centering
    \includegraphics[width=\textwidth]{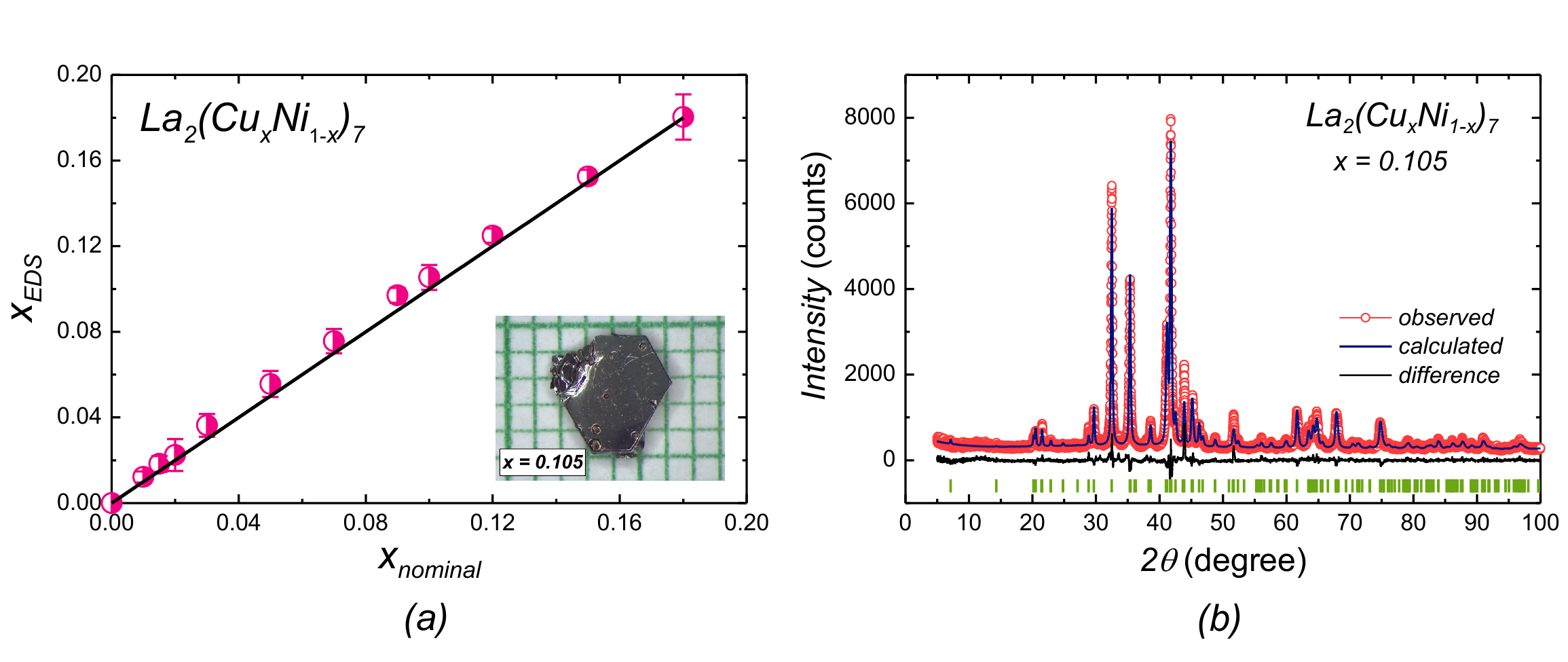}
    \caption{\footnotesize{\textit{(a)} Measured Cu substitution level, ($x_{EDS}$) values, for ${\text{La}_{2}\text{(Cu}_{x}\text {Ni}_{1-x})_7}$ crystals versus $x_{nominal}$ values used in the growth solution. The black line corresponds to the linear fit across the data points with the intercept fixed to (0,0), and a slope of essentially 1, clearly showing that $x_{EDS}$ $\sim$ with $x_{nominal}$. (Inset: typical crystals of ${\text{La}_{2}\text{(Cu}_{x}\text {Ni}_{1-x})_7}$ for $x_{EDS}$ = 0.105 on a mm grid.) \textit{(b)} Powder X-Ray diffraction data for ${\text{La}_{2}\text{(Cu}_{x}\text {Ni}_{1-x})_7}$ for $x_{EDS} = 0.105 $. The data for other $x$-values are qualitatively the same.}}
    \label{fig:EDS}
\end{figure*}

Single  crystals of ${\text{La}_{2}\text{(Cu}_{x}\text {Ni}_{1-x})_7}$; with ${x_{nominal}}$ = 0, 0.01, 0.015, 0.02, 0.03, 0.05, 0.07, 0.09, 0.10, 0.12, 0.15, and 0.18 were synthesized using a self flux, solution growth method \cite{Canfield2001High-temperatureQuasicrystals,Canfield2020NewPhysics,Canfield1992GrowthFluxes} in a manner similar to the growth of pure $\text{La}_{2}\text{Ni}_{7}$ \cite{Ribeiro2022Small-moment/math}. Small pieces of lanthanum (Materials Preparation Center - Ames National Laboratory $99.99\%$), nickel (Alfa Aesar $99.98\%$), and copper (Fine Metals Corporation $99.997\%$) with a starting composition of ${\text{La}_{33}\text{(Cu}_{x}\text {Ni}_{1-x})_{67}}$, were weighed, then welded into a tantalum, 3-cap-crucible \cite{Canfield2001High-temperatureQuasicrystals, Canfield2020NewPhysics}, and sealed in a fused silica ampoule under a partial argon atmosphere. The ampoule was heated in a box-furnace to $1180^{\circ}$C over 10 hours, held at the temperature for 20 hours to ensure a homogeneous melt, quickly cooled down to $1020^{\circ}$C followed by a very slow cool to $820^{\circ}$C over 200 hours. After dwelling at $820^{\circ}$C for a few hours, the excess solution was then decanted using a centrifuge \cite{Canfield1992GrowthFluxes, Canfield2020NewPhysics, Canfield2001High-temperatureQuasicrystals}. Well-faceted hexagonal plates of single crystalline ${\text{La}_{2}\text{(Cu}_{x}\text {Ni}_{1-x})_7}$ were obtained having typical dimensions of  ${\sim 4~\text{mm} \times \sim 4~\text{mm}}$ with an average thickness of ${\sim 1 ~\text{mm}}$ and of the same morphology as that of the undoped compound. (See the inset to Fig \ref{fig:EDS}(a), or inset to Fig 1 in Ref \cite{Ribeiro2022Small-moment/math} for pictures of representative crystals). Increasing the $x_{nominal}~ \geq$  0.20 in the ${\text{La}_{33}\text{(Cu}_{x}\text {Ni}_{1-x})_{67}}$ melt yielded single crystals of ${\text{La}\text{(Cu}_{x}\text {Ni}_{1-x})_5}$, with $x \sim 0.3$ and hence we stopped the attempts of substitution at ${x_{nominal}}$ = 0.18. \par

The Cu substitution levels $x_{EDS}$ of the ${\text{La}_{2}\text{(Cu}_{x}\text {Ni}_{1-x})_7}$ crystals were determined by Energy Dispersive Spectroscopy (EDS) quantitative chemical analysis. All SEM images were acquired with the ThermoFisher (FEI) Teneo Lovac FE-SEM, located at the Sensitive Instrument Facility, Ames National Laboratory. The data was analyzed using an Oxford Instruments Aztec System with a X-Max-80 detector, attached to the Teneo. An acceleration voltage of 15 kV, current of 1.6 nA, working distance of 10mm and take off angle of 35$^{\circ}$ were used for measurements. The composition of each crystal was measured at 7-8 different spots on the crystal's face, revealing good homogeneity of each crystal except for the highest doped sample ($x_{nominal}$ = 0.18), which had somewhat larger spot to spot variation, but still reasonable homogeneity. The standards used for reference are internal to the Oxford software.

The phase and crystal structure was confirmed using a Rigaku Miniflex-II powder diffractometer using Cu ${K_{{\alpha}}}$ radiation $(\lambda = 1.5406$~\r{A}). Single crystals were finely ground, and the powder was then mounted and measured on a single crystal Si, zero-background sample holder using a small amount of vacuum grease. Intensities were collected for 2$\theta$ ranging from 5$^{\circ}$ to 100$^{\circ}$ in steps of 0.01$^{\circ}$, counting each angle for 5 seconds. The patterns were refined and the lattice parameters were determined using GSAS II software. \cite{Toby2013Package}. 

\par

\begin{figure}[h]
    \centering
    \includegraphics[width=\linewidth]{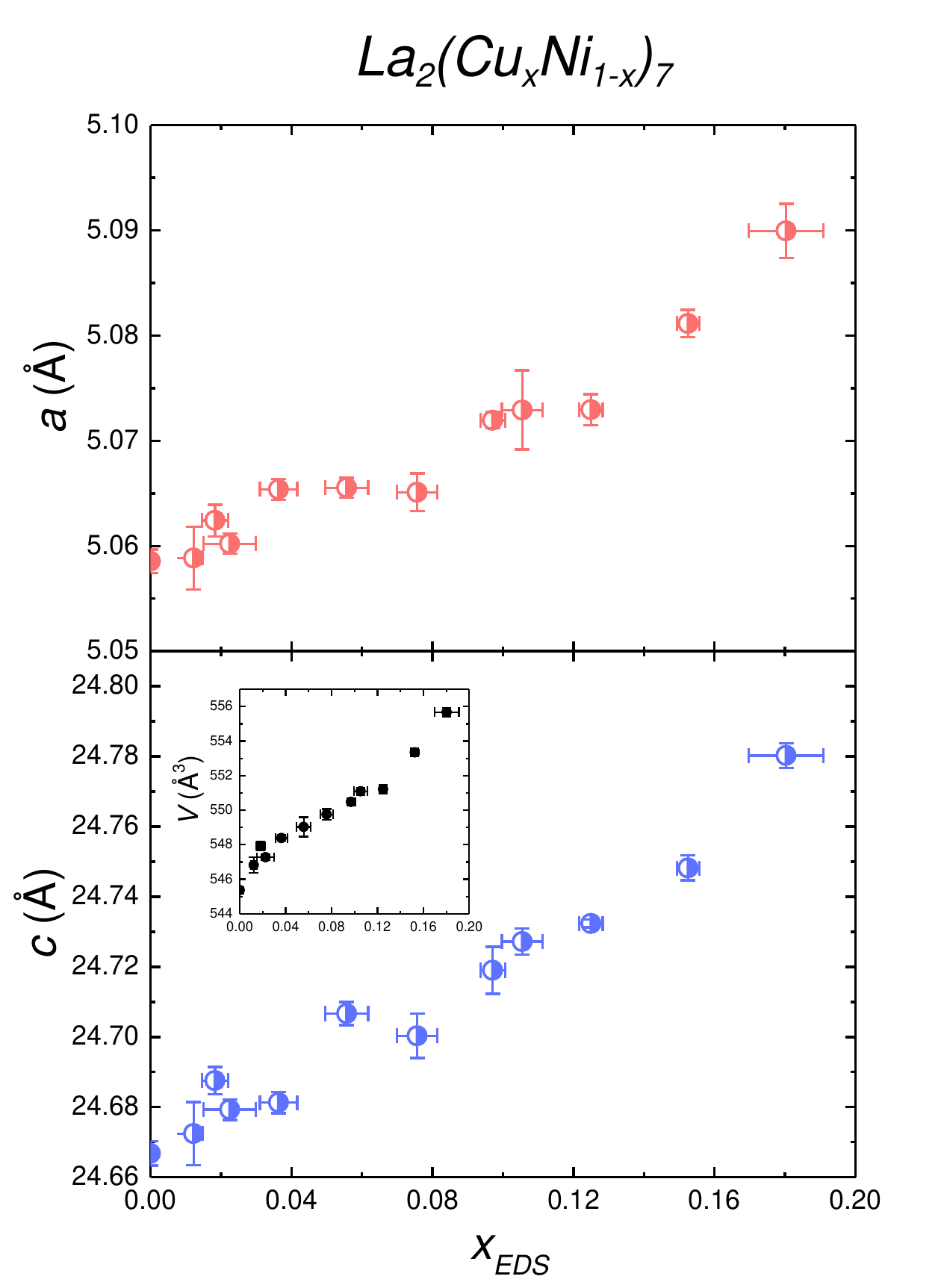}
    \caption{\footnotesize{Lattice parameters $a-$ on the top panel, and $c-$ on the bottom panel and the unit cell volume, $V$ (as an inset in the bottom panel) as a function of the Cu substitution level $(x_{EDS})$ in ${\text{La}_{2}\text{(Cu}_{x}\text {Ni}_{1-x})_7}$ single crystals obtained from the refinement of the powder x-ray diffraction data.}}
    \label{fig:lattice}
\end{figure}
  
Temperature and field dependent DC magnetization measurements ($M(T)$ and $M(H)$ respectively) were made in a Quantum Design Magnetic Property Measurement System (QD-MPMS classic) SQUID magnetometer with samples mounted on a diamagnetic Poly-Chloro-Tri-Fluoro-Ethylene (PCTFE) disk, which snugly fits inside a straw, using a small amount of superglue. Each sample was measured with the external field applied parallel ($H || c$) and perpendicular ($H \perp c$) to the hexagonal $c$ - axis as was done for the parent compound \cite{Ribeiro2022Small-moment/math}. Additionally, Curie-Weiss analysis was done on the magnetic susceptibility ($\chi = M(T)/H)$ data for $H || c$ and $H \perp c$ data as well as for a polycrystalline average of these data sets. The polycrystalline average susceptibility of the hexagonal samples was obtained by using $\chi_{poly}=\frac{1}{3}\chi_{||c}+\frac{2}{3}\chi_{\perp c}$ where $\chi_{||c}$ and $\chi_{\perp c}$ are the magnetic susceptibilities along $H ||c$ and $H \perp c$ respectively. The signal from the disk was measured beforehand and subtracted from the combined sample-disk magnetization data for both directions so as to obtain the value of the moment due to the sample for each $H || c$ and $H \perp c$. Both $M(H)$ and $M(T)$ measurements were done under zero-field cooled (ZFC) protocol where the sample was cooled down to 1.8 K before applying the external magnetic field. In some additional cases field-cooled (FC) protocol was also employed when the samples were cooled down in presence of the externally applied field. $M(H)$ measurements were done in magnetic fields from 0 kOe to 70 kOe whereas $M(T)$ measurements were done in the temperature range between 1.8 K and 300 K. 
\par

Temperature dependent ac resistivity measurements $(\rho(T))$ were done in the standard four-probe geometry in a Quantum Design Physical Property Measurement System (QD-PPMS) using a 3 mA excitation with a frequency of 17 Hz. $\rho(T)$ was measured with the current flowing in the basal plane (i.e. perpendicular to the c-axis). Hexagonal plate like samples of each $x$ of ${\text{La}_{2}\text{(Cu}_{x}\text {Ni}_{1-x})_7}$ were cut into bars with a wire saw and had dimensions of roughly 2 mm long, 0.8 mm wide, and 0.2 mm thick (along the c direcion). Electrical contacts were made using 25 $\mu m$ diameter platinum wires attached to the bar shaped samples using Epotek-H20E epoxy. $\rho(T)$ was measured in the range of $1.8~ \text{K} \leq T \leq 300~\text{K}$ under zero magnetic field and in some cases under a magnetic field ($H \perp I$, and with both $H\perp c$ and $H || c $ directions). 

Temperature dependent heat capacity measurements were performed in a QD-PPMS. Plate-like samples were mounted on the micro-calorimeter platform using a small amount of Apiezon N-grease and measured under zero magnetic field. The addenda (contribution from the grease and the sample platform) was measured separately and subtracted from the total data to obtain the contribution only due to the sample using the PPMS software.
\par

\begin{figure*}[htbp]
    \centering
    \includegraphics[width=1.0\textwidth]{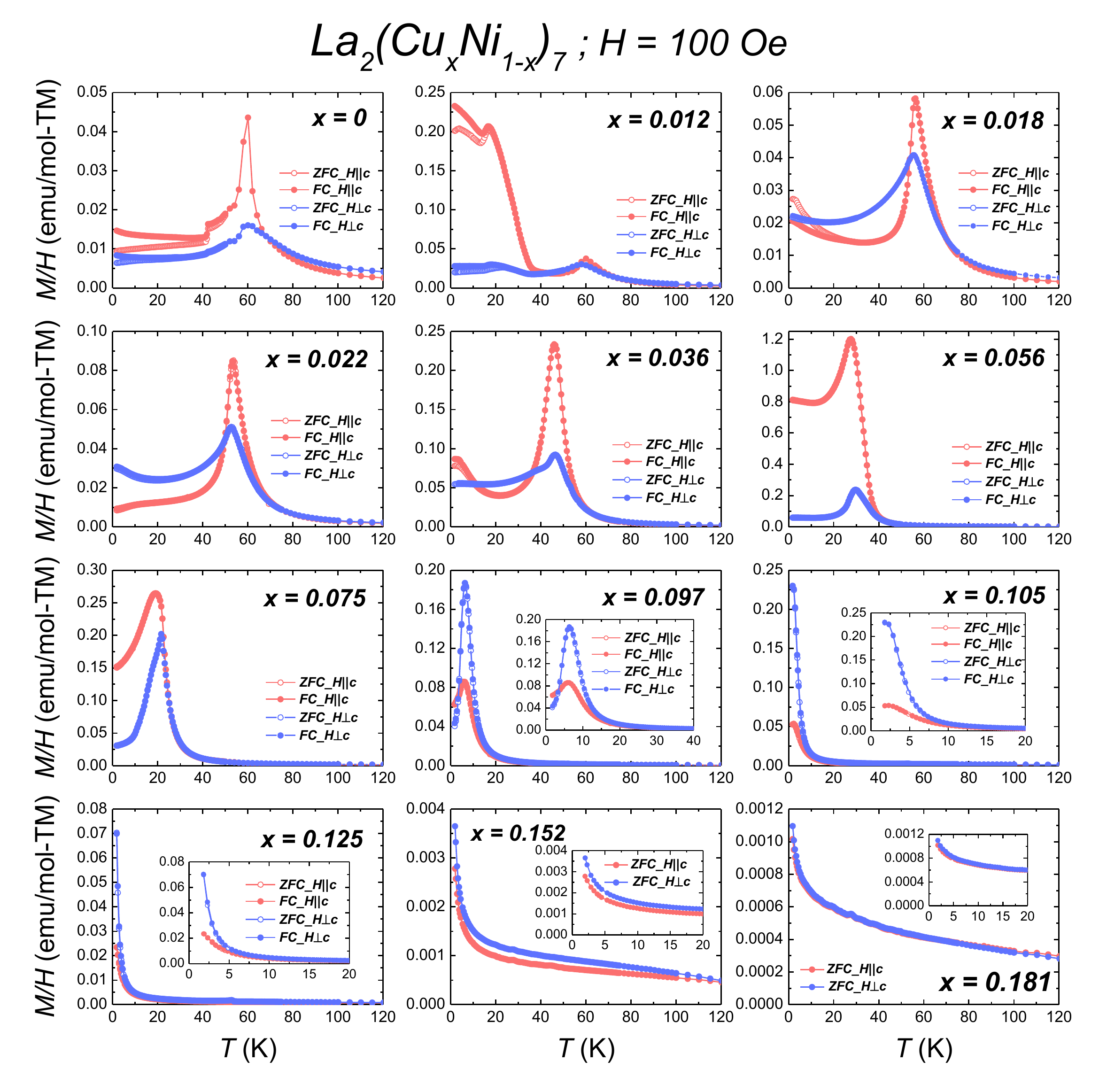}
    \caption{\footnotesize{(Color online) Anisotropic temperature dependent magnetization ($M(T)$) measured at $H$ = 100 Oe for ${\text{La}_{2}\text{(Cu}_{x}\text {Ni}_{1-x})_7}$ single crystals. $M(T)$ for each Cu substituted sample is plotted in separate panels for both $H||c$ and $H \perp c$. Although measurements were done up to $T = 300$ K, the data here is plotted only up to T$ = 120$ K to highlight the transition related features. The samples were measured for both ZFC-FC mode for 0 $\leq x \leq$ 0.125 and only in ZFC mode for $x$ = 0.152 and 0.181. Some of the panels (0.097 $\leq x \leq$ 0.181) have insets to show the low temperature behavior in greater detail. [NOTE: The $y$ axis scale varies from panel to panel.]}}
    \label{fig:M(T)_100 Oe}
\end{figure*}

\section{Results}
\label{sec:results}

\subsection*{Composition and lattice parameters}

The Cu substitution levels in ${\text{La}_{2}\text{(Cu}_{x}\text {Ni}_{1-x})_7}$ single crystals, determined by the EDS measurements $(x_{EDS})$ are plotted as a function of the nominal Cu levels $(x_{nominal})$, used for the growths are shown in Fig \ref{fig:EDS}(a). As $x_{nominal}$ is increased, $x_{EDS}$ increases in a monotonic and almost linear manner. The best fit line, going through the $(x_{EDS})$ data has a slope of 1.04(1), and can be considered as operationally indistinguishable from 1.0. From this point onwards, in this paper, $x$ will refer to the Cu substitution values determined by EDS measurements in ${\text{La}_{2}\text{(Cu}_{x}\text {Ni}_{1-x})_7}$. In the rare case when we need to refer to the nominal Cu concentration in the growth melt we will explicitly use the $x_{nominal}$ notation. 

\par

\begin{figure*}[htbp]
    \centering
    \includegraphics[width=0.95\textwidth]{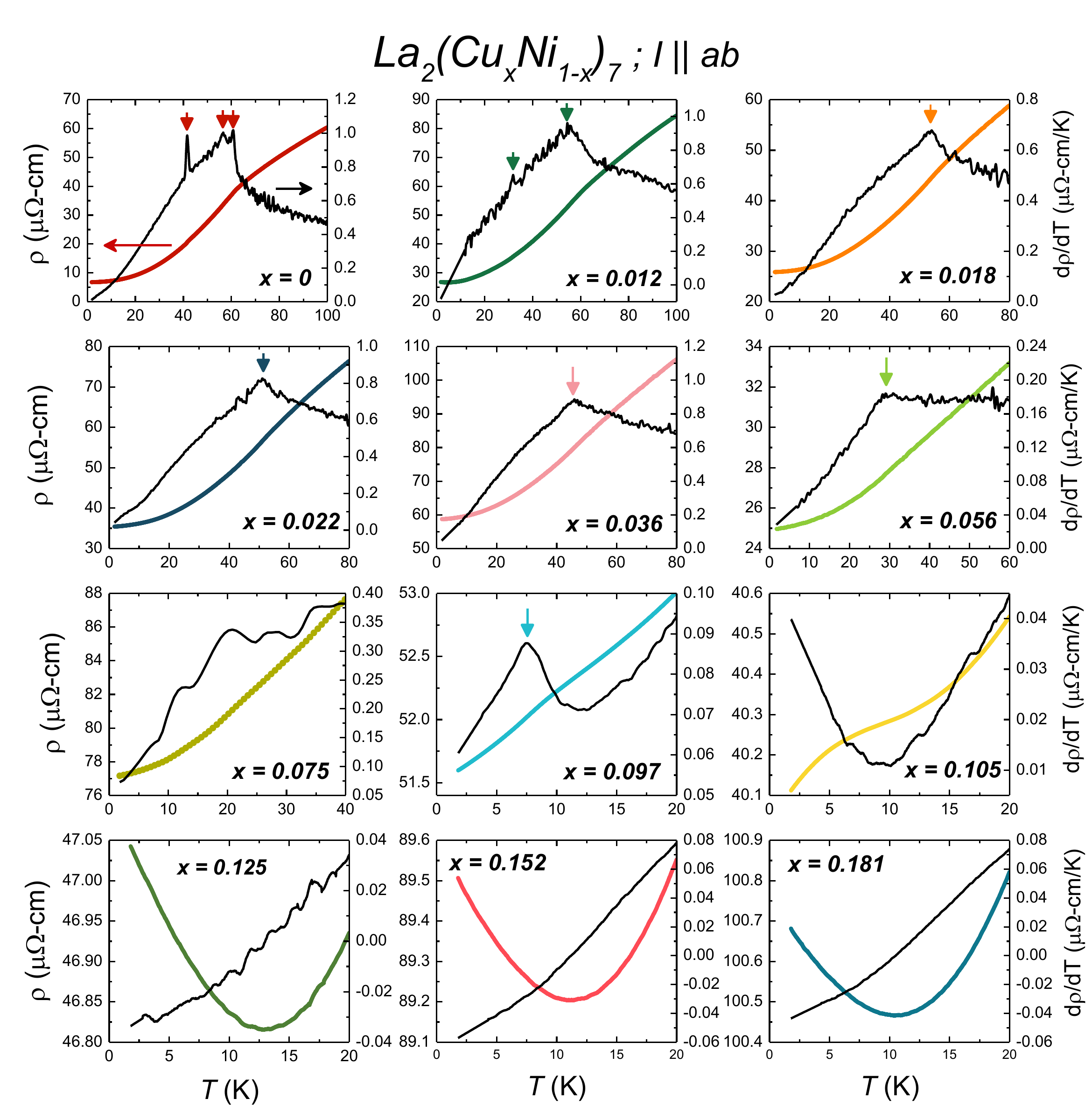}
    \caption{\footnotesize{Temperature dependent, zero field, in-plane resistivity on the left axis and its corresponding temperature-derivative on the right axis of ${\text{La}_{2}\text{(Cu}_{x}\text {Ni}_{1-x})_7}$ single crystals plotted in separate panels. The colored curves show the $\rho(T)$ data and the black curves the corresponding $\frac{d \rho}{dT}$. $\rho(T)$ for $0 \leq x \leq 0.097$ are plotted starting above their respective magnetic ordering temperatures down to 1.8 K whereas for the higher doped samples, $0.105 \leq x \leq 0.181$, which do not show magnetic ordering, resistivity is plotted from 20 K down to 1.8 K. The magnetic ordering temperatures determined from maxima in the derivative data are shown in each panel using arrows.}}
    \label{fig:R-T separate}
\end{figure*}

Figure \ref{fig:EDS}(b) shows a representative powder x-ray diffraction pattern for ${\text{La}_{2}\text{(Cu}_{x}\text {Ni}_{1-x})_7}$ for the $x = 0.105 $ sample. Similar x-ray patterns for crystals of each Cu substitution were refined and the lattice parameters $a=b$, and $c$ as well as the unit cell volume $V$ were obtained after refinement of the powder diffraction data. Figure \ref{fig:xrd_all} in the Appendix shows the powder x-ray pattern for all the Cu-substituted samples. The peaks are matched with the expected peaks of the ${\text{Ce}_{2}\text{Ni}_{7}}$ hexagonal structure with space group $P6_3/mmc$, (No. 194) structure type \cite{Buschow1970TheR2Ni7}. Fig \ref{fig:lattice} shows the change in the lattice parameters $a$, $c$, and the unit cell volume, $V$, of ${\text{La}_{2}\text{(Cu}_{x}\text {Ni}_{1-x})_7}$ versus $x$. There is an increase in $a$, $c$, and $V$ as we increase the Cu substitution level. Given that the difference in ionic radii between Ni and Cu is small and the substitution level is less than $x$ = 0.20, it is not surprising that the changes in the lattice parameters and volume are small, albeit clearly resolvable. The values of the lattice parameters for each of the $x$ values are shown in Table \ref{tab:lattice_La2Ni7} in the Appendix.

\subsection*{Zero and low-field data and $T-x$ phase diagram}

\subsubsection*{Low field $M(T)$}

Figure \ref{fig:M(T)_100 Oe} shows the anisotropic $M(T)$ data measured at 100 Oe for both $H || c$ and $H \perp c$ for each Cu substituted sample. For 0 $\leq x \leq$ 0.125, the data was measured in both ZFC-FC protocols, whereas for the two highest dopings $x$ = 0.152 and 0.181 the data was measured only in ZFC mode. For the parent compound ($x$ = 0) there are signs of three magnetic transitions \cite{Ribeiro2022Small-moment/math}. For the lowest doped sample ($x$ = 0.012), we also observe signs of multiple (primarily two) phases, but there are differences between the signatures of the phases seen for the parent compound and this Cu-substituted sample. This difference becomes more evident when $M(T)$ measurements are done at a field of $H$ = 1 kOe. Figure \ref{fig:MT_0 and 1} in the Appendix highlights this difference. Increasing the Cu substitution level suppresses the multiple transitions, so that when $x$ reaches 0.018, only one transition is detected. This transition is further suppressed with Cu substitution, but is observable up to $x$ = 0.097. We also observe a distinct ZFC-FC split for 0 $\leq x \leq$ 0.036, but not for the higher $x$ samples which show magnetic order down to $T$ = 1.8 K (0.056 $\leq x \leq$ 0.097). This may imply that the magnetic state associated with the ordered state, has a small but finite FM component. This will be further discussed in the Appendix.

For $x$ = 0.105, no feature associated with a transition is observed down to 1.8 K. Finally, for the highest doped samples, $0.125 \leq x \leq 0.181$, we observe a $M(T)$ behavior more characteristic of a paramagnet. Thus, the antiferromagnetic long range magnetic ordering in ${\text{La}_{2}\text{(Cu}_{x}\text {Ni}_{1-x})_7}$ is suppressed below the base temperature, or, completely, due to Cu substitution for $x > 0.097$. 

In order to more quantitatively parametrize the number and location of ordering temperatures we plot the polycrystalline average $\frac{d \chi T}{dT}$ for the $x \leq 0.097$ samples shown in Fig \ref{fig:TNeel} in the Appendix. The magnetic transition temperatures inferred are plotted in the $T-x$ phase diagram in Fig \ref{fig:phase diagram_27} below.

\begin{figure*}
    \centering
    \includegraphics[width=0.95\textwidth]{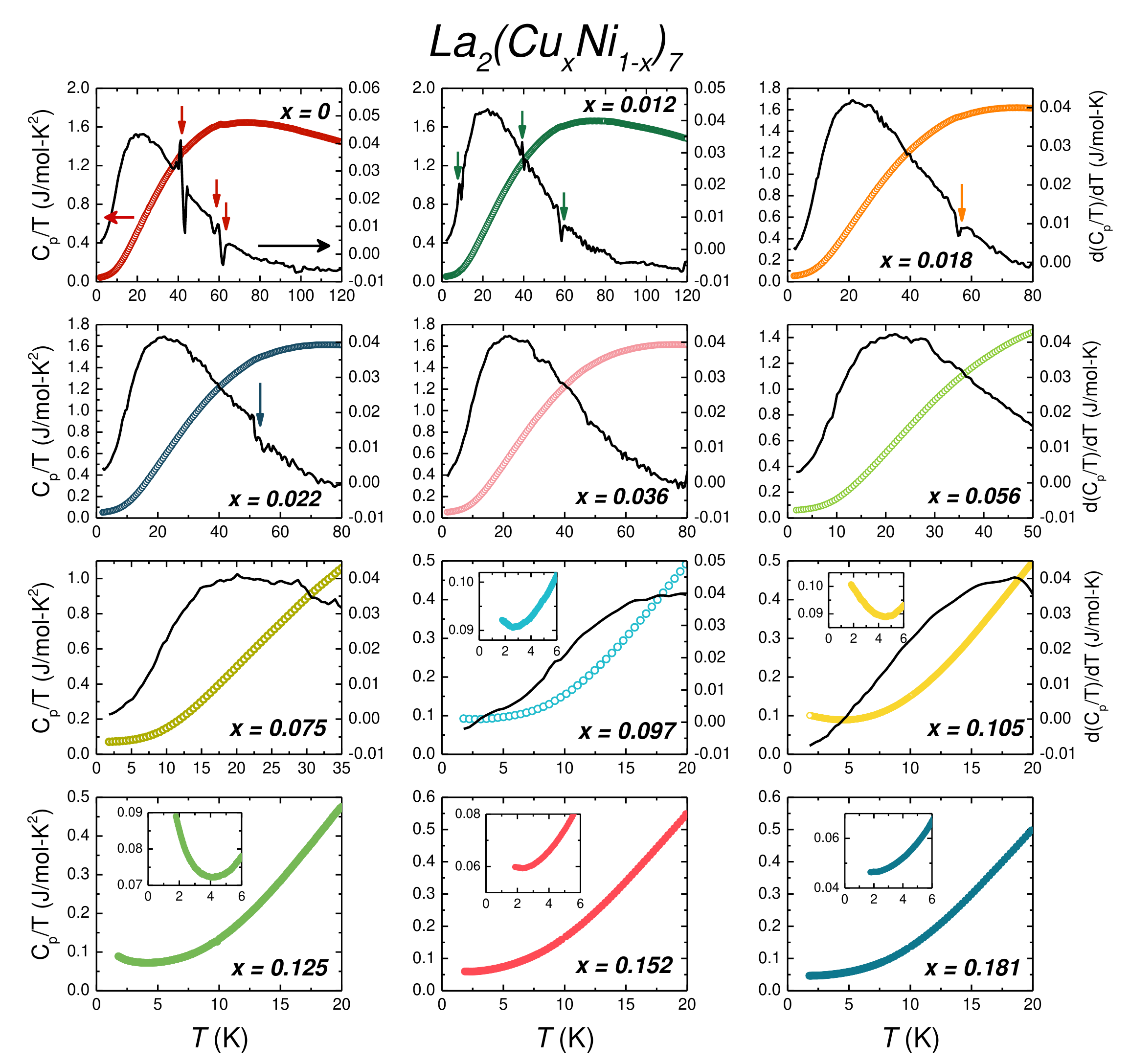}
    \caption{\footnotesize{(Color online) Temperature dependent, zero field, $C_p/T$ data of the ${\text{La}_{2}\text{(Cu}_{x}\text {Ni}_{1-x})_7}$ single crystals on the left axis (colored curves) and its corresponding derivative plotted on the right axis (black curves) in separate panels for each concentration. The derivatives are not plotted for the three highest dopings, 0.125 $\leq x \leq$ 0.181, as there are no distinct features related to any transition. The magnetic ordering feature(s) is visible in the derivative for 0 $\leq x \leq$ 0.022 and is shown using arrows. $x$ = 0.036 and 0.056 samples have broader features in their derivatives which may be related to the transition but are not clearly resolvable. For 0 $\leq x \leq$ 0.097, the data is plotted from temperatures above their magnetic transition down to 1.8 K. For 0.105 $\leq x \leq$ 0.181, the data are plotted from 20 K down to the base temperature. The data set for the parent $x = 0$ is from Ref \cite{Ribeiro2022Small-moment/math}. For $x \geq$ 0.097, the plots have insets which show the low temperature behavior of the data.}}
    \label{fig:heat capacity_27}
\end{figure*}

\subsubsection*{$\rho(T)$ in zero applied magnetic field}

Figure \ref{fig:R-T separate} shows the temperature dependent in-plane ($I || ab-$ plane) resistivity of the ${\text{La}_{2}\text{(Cu}_{x}\text {Ni}_{1-x})_7}$ single crystals in separate panels. For $0 \leq x \leq 0.097$, the data is plotted over temperature ranges that highlight the magnetic ordering transitions, with a base temperature of 1.8 K. For the 0 $\leq x \leq$ 0.097 samples, we observe metallic behavior for temperatures above the magnetic ordering and for $0.105 \leq x \leq 0.181$, metallic behavior is observed for $T \geq 15$ K . The residual resistance ratio $(RRR = \rho(300~\text{K})/ \rho(1.8~\text{K}))$ decreases abruptly from $\sim$ 19 for $x = 0$ to $\sim$ 6 for $x = 0.012$ and then slowly decreases to attain a value of 1.8 for $x = 0.105$. 

\par

The resistivity of the ${\text{La}_{2}\text{(Cu}_{x}\text {Ni}_{1-x})_7}$ single crystals changes its behavior with Cu substitution, especially the higher doped samples. For $0 \leq x \leq 0.097$, a modest, but clear loss of spin-disorder scattering is seen in the $\rho(T)$ data upon cooling, indicating the onset of magnetic ordering, with the associated feature more evident for the low dopings ($0 \leq x \leq 0.036$). This is also evident from the $\frac{d \rho}{dT}$ data that are shown along the right hand axis using black curve in Fig \ref{fig:R-T separate} and Fig \ref{fig:TNeel} in the Appendix.

\par

The resistivity for the highest doped ${\text{La}_{2}\text{(Cu}_{x}\text {Ni}_{1-x})_7}$ samples, $0.125 \leq x \leq 0.181$, have a different low temperature, behavior and exhibits a minimum at around $T \sim$ 12 K, such that $\rho(T)$ decreases and then further increases as the samples are cooled down to $T$ = 1.8 K. The $x$ = 0.105 sample is intermediate to the higher and lower $x$-samples. As temperature decreases, the $x$ = 0.105 data appears to be going through a resistive minimum, similar to higher $x$-value samples, but then, as temperature decreases, the resistivity decreases further in a linear manner. Further analysis of this sample and its transport data will be given below.  

\par

To summarize the results of the resistivity measurements on ${\text{La}_{2}\text{(Cu}_{x}\text {Ni}_{1-x})_7}$. $\textit{(i)}$ We initially see suppression of the magnetic order by observing loss of spin-disorder scattering occurring at lower temperatures as $x$ increases in the lower doped samples $(0 \leq x \leq 0.097)$. $\textit{(ii)}$ We see a resistive upturn in the highest doped samples $(0.125 \leq x \leq 0.181)$, which may be an indication of a Kondo-like behavior in this system and will be discussed in detail in a separate section, below. $\textit{(iii)}$ The $x$ = 0.105 sample lies at the border of the two regimes with competing interactions and is expected to be in a quantum critical region of the $T-x$ phase diagram (i.e. near the QCP). 

\par

\subsubsection*{Specific heat in zero applied magnetic field}

We measured temperature-dependent specific heat on the samples to get more insight into the evolution of the transition temperature(s) and the subsequent Kondo like behavior in the ${\text{La}_{2}\text{Ni}_7}$ system due to Cu substitution. Figure \ref{fig:heat capacity_27} shows the results of $C_p$ measurements on the ${\text{La}_{2}\text{(Cu}_{x}\text {Ni}_{1-x})_7}$ single crystals plotted as $C_p/T$ vs $T$. The samples were measured over different temperature ranges, depending on their respective magnetic transitions. For each sample, we started at a temperature above their respective highest magnetic transition and measured down to $T$ = 1.8 K. As reported in \cite{Ribeiro2022Small-moment/math}, the three phase transitions for $x$ = 0 are resolvable but the features are quite small. We also plot the corresponding derivatives on the right hand axis to delineate the features due to phase transitions.

\par

For the lower dopings, 0 $\leq x \leq$ 0.022, we observe small features in $C_p/T$ and the corresponding derivative at similar temperatures as those of the transitions in the $M(T)$ (Fig \ref{fig:M(T)_100 Oe}) and $\rho(T)$ (Fig \ref{fig:R-T separate}) data, with $x$ = 0 and 0.012 having multiple features (three clear ones for $x$ = 0 and either two or three features for $x$ = 0.012) and for $x$ = 0.018 and 0.022 only a single feature. For 0.036 $\leq x \leq$ 0.075, despite having magnetic transitions in their $M(T)$ data and clear features in the $\rho(T)$ data, we cannot resolve any associated feature in the specific heat data. The difficulty in resolving specific heat anomalies at the ordering temperatures is consistent with the diminishing of an already small ordered moment and the corresponding decrease in the entropy change associated with the transition. The situation becomes different for the highest dopings, $0.097 \leq x \leq 0.181$, where the low temperature data shows minima in $C_p/T$ at $T \sim$ 5 K, followed by a lower temperature upturn. The upturn increases as we increase the Cu substitution level beyond $x$ = 0.097, and is largest (at $T$ = 2 K) for $x$ = 0.105 after which it decreases for the higher dopings, becoming very small, but still resolvable at the highest dopings.  

\par

\subsubsection*{$T-x$ phase diagram}

\begin{figure}[h]
    \centering
    \includegraphics[width=\linewidth]{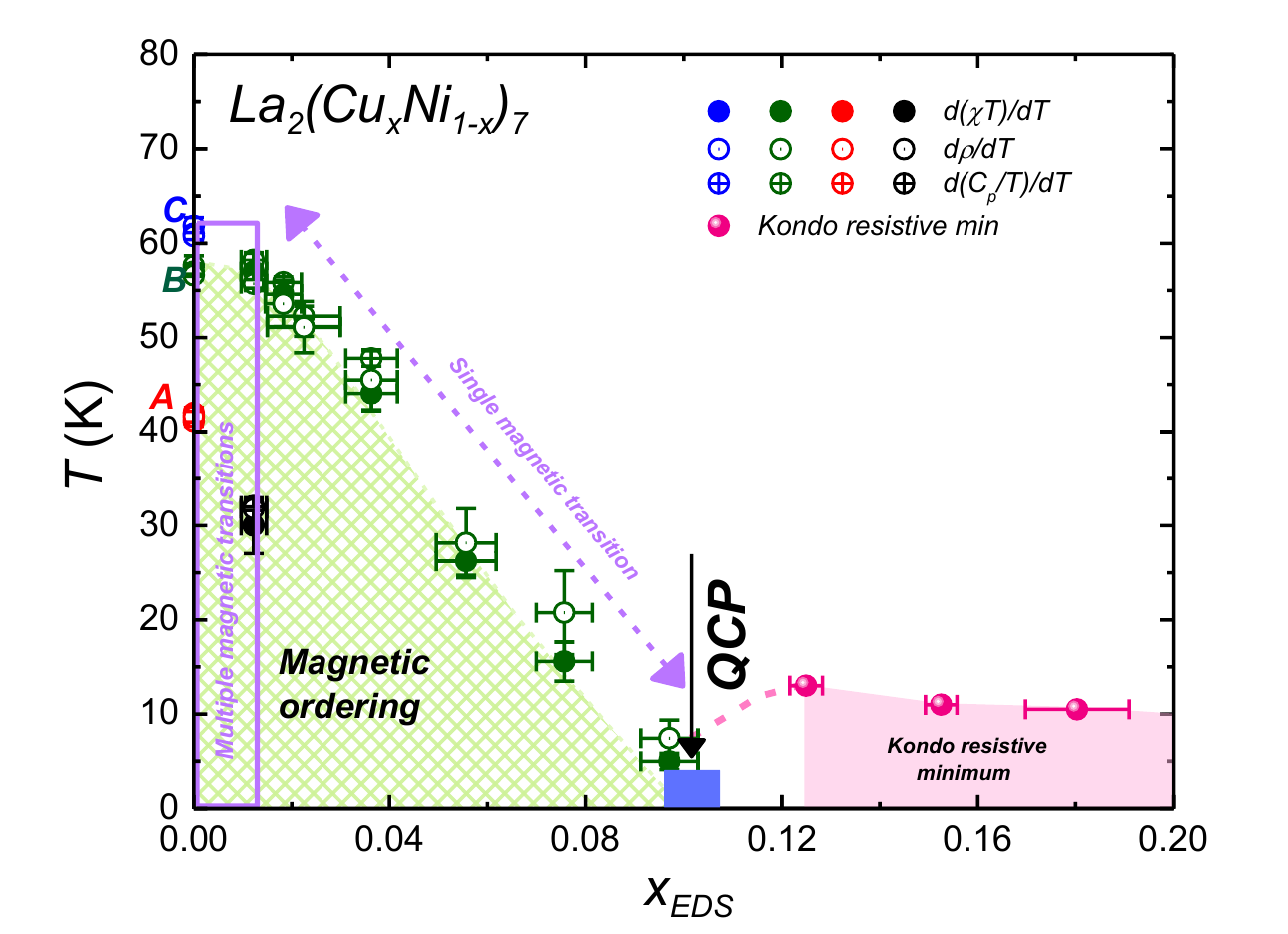}
    \caption{\footnotesize{(Color online) The $T-x$ phase diagram for Cu substituted ${\text{La}_{2}\text{(Cu}_{x}\text {Ni}_{1-x})_7}$ system constructed from $M(T)$ at 100 Oe and zero field $\rho(T)$ and $C_p/T$ data. The 0 $\leq x \leq$ 0.097 regime manifests magnetic order above 1.8 K, with the parent and the lowest doped samples ($x$ = 0 and 0.012) showing signs of multiple transitions. The three ordered regimes are labeled as $A$ (red data point), $B$ (green data point), and $C$ (blue data point) for $x$ = 0. From $M(T)$ measurements, we anticipate that the $B$ phase remains down to $x$ = 0.097 whereas phase $C$ gets suppressed with our first finite $x$-level. Whether the $A$ phase continues from $T \sim$ 42 K to $T \sim$ 35 K for $x$ = 0.012, or the 35 K (black data point) phase in this sample is different in nature remains ambiguous from our results. The regions of multiple magnetic phases and single magnetic phase are marked in the phase diagram. The three higher doped (0.125 $\leq x \leq$ 0.181) samples show a resistive upturn at low temperatures which is probably indicative of some type of Kondo behavior and is shown using the pink area. The boundary line for the Kondo area has also been extrapolated to T = 0 (with pink dashed lines) towards the low dopings. The intermediate $x$ = 0.105, is in the vicinity of the QCP for this system.}}
    \label{fig:phase diagram_27}
\end{figure}

\begin{figure*}
    \centering
    \includegraphics[width=\textwidth]{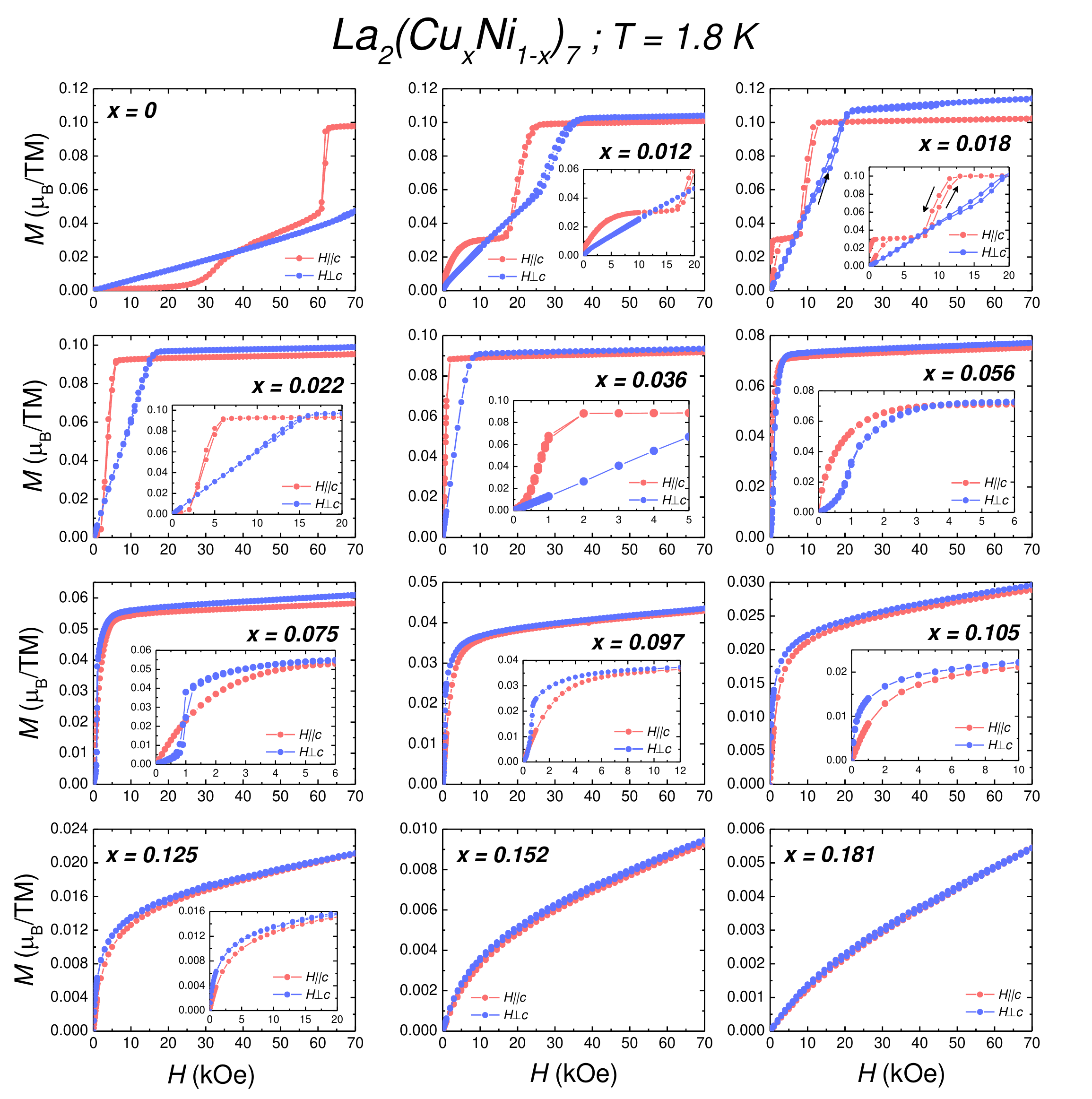}
    \caption{\footnotesize{(Color online) Anisotropic field dependent magnetization ($M(H)$) measured at $T$ = 1.8 K for ${\text{La}_{2}\text{(Cu}_{x}\text {Ni}_{1-x})_7}$ single crystals. $M(H)$ for each Cu substituted sample is plotted in separate panels for both $H||c$ and $H \perp c$. The samples were cooled down to 1.8 K in the absence of an external field (ZFC) and then measured first for increasing from zero and then followed by decreasing field to zero. Some of the panels ($x$ = 0.012, 0.018, 0.022, 0.036, 0.056, 0.075, 0.097, 0.105, and 0.125) have insets to show the low field behavior in greater detail.}}
    \label{fig:M(H)}
\end{figure*}

Figure \ref{fig:phase diagram_27} shows the $T-x$ phase diagram of the ${\text{La}_{2}\text{(Cu}_{x}\text {Ni}_{1-x})_7}$ series constructed from $M(T)$ measured at 100 Oe and zero field $\rho(T)$ and $C_p(T)$ data. The phase diagram can be broadly divided into three distinct regimes. $\textit{(i)}$ The magnetically ordered lower dopings, 0 $\leq x \leq$ 0.097, $\textit{(ii)}$ the higher substituted 0.125 $\leq x \leq$ 0.181 regime which show a low $T$ upturn in resistivity and $C_p/T$ data, suggesting Kondo-like behavior, and $\textit{(iii)}$ the intermediate $x \simeq $ 0.105, very close to where the QCP is inferred for this system. We will discuss the behavior of each of these regimes separately.

For the parent $x$ = 0, there are three AFM phases as was deduced from previous work \cite{Ribeiro2022Small-moment/math, Wilde2022Weak/math} as well as our measurement results (Fig. \ref{fig:M(T)_100 Oe}). For the lowest doped sample, $x$ = 0.012, we still observe multiple phase transitions with associated features in the low field $M(T)$, zero field $\rho(T)$ and $C_p(T)$ data. However, when we compare the $M(T)$ data for these two samples, there is a distinct difference in their behavior. This is evident in Fig. \ref{fig:MT_0 and 1} in the Appendix especially when measurements are done at a field of $H$ = 1 kOe in the easy direction of $H || c$. The $B$ and $C$ phases of the parent compound appear to merge together and we observed a much broader feature for the doped sample. This feature is qualitatively and quantitatively similar to the phase $B$. As of now, we do not know how the magnetism evolves in between these two samples and we need more measurements such as neutron diffraction to know the nature of magnetism in ${\text{La}_{2}\text{(Cu}_{x}\text {Ni}_{1-x})_7}$ with Cu substitution.

The 0.018 $\leq x \leq$ 0.097 region shows only a single magnetic transition, with the AFM feature very similar to the phase $B$. From our previous work on the parent $\text{La}_{2}\text{Ni}_{7}$ \cite{Ribeiro2022Small-moment/math}, we expect these phases to be primarily antiferromagnetic with some small ferromagnetic component. In fact, the spins are canted in this system as it was established from neutron diffraction results \cite{Wilde2022Weak/math}. As discussed earlier, the polycrystalline $\frac{d(\chi T)}{dT}$, together with $\frac{d\rho}{dT}$, and $\frac{d C_p/T}{dT}$ data are used to determine the transition temperatures \cite{Fisher1962RelationAntiferromagnet, Fisher1968ResistivePoints}. The position of the local extrema is each of these data sets determines the value of $T_N$. Figures \ref{fig:R-T separate} and \ref{fig:heat capacity_27} show the $\frac{d \rho}{dT}$ and $\frac{d C_p/T}{dT}$ plots, respectively, and Fig. \ref{fig:TNeel} in the Appendix shows the polycrystalline $\frac{d(\chi T)}{dT}$. There is good agreement between the $T_N$ values inferred from $\chi(T)$, $\rho(T)$, and $C_p(T)$ data. In the phase diagram, we anticipate that phase $B$ decreases from a value of $\sim$ 56 K for $x$ = 0 to a value of $\sim$ 5 K for $x$ = 0.097. The phase $C$ gets suppressed almost instantly with Cu doping and the fate of the lowest temperature phase, for the lowest Cu-doped doped samples remains ambiguous as of now. Thus, the magnetically ordered regime can be subdivided into 'multiple phase transitions' and 'single phase transition' regime. 
 
\par

For the three highest Cu dopings, (0.125 $\leq x \leq$ 0.181) where the resistive upturn is observed, the resistive minima is plotted as a proxy for a cross over temperature into the low temperature state. The upturn regime, which maybe due to the Kondo effect, for these substitutions is shown by the pink area in Fig \ref{fig:phase diagram_27}. $x$ = 0.105, which is in the vicinity of the QCP for this system is also marked in the phase diagram. The Kondo regime is extrapolated to $T$ = 0 for the low dopings as a possible extension of this low temperature state.

\subsection*{Higher field magnetization measurements}

The $T-x$ phase diagram was instrumental in categorizing the global features in the ${\text{La}_{2}\text{(Cu}_{x}\text {Ni}_{1-x})_7}$ system. The low dopings (0 $\leq x \leq$ 0.097) have magnetic ordering for temperatures above 1.8 K with $x$ = 0 and 0.012 showing signatures of multiple magnetic phases. The nature of these magnetic phases can be further illuminated by discussing $M(H)$ data as well as $M(T)$ measured at higher fields. 

\par

\subsubsection*{$M(H)$ at base temperature}

Figure \ref{fig:M(H)} shows the anisotropic field dependent magnetization ($M(H)$) measured at $T$ = 1.8 K for ${\text{La}_{2}\text{(Cu}_{x}\text {Ni}_{1-x})_7}$ for each Cu substituted sample on separate panels. The $M(T)$ data at 100 Oe (Fig \ref{fig:M(T)_100 Oe}) shows that the magnetic state for 0 $\leq x \leq$ 0.097 is primarily antiferromagnetic in nature. The data in Fig \ref{fig:M(H)} also indicates that the base temperature magnetic state in ${\text{La}_{2}\text{(Cu}_{x}\text {Ni}_{1-x})_7}$ remains fundamentally antiferromagnetic (AFM) for these Cu substitutions (In the appendix we will discuss what appears to be an exceptionally small FM component to some of these low field states). The low field $H||c$ data is linear through the origin, followed by a metamagnetic transition, all consistent with AFM behavior. The metamagnetic transition field $H_{C}$ changes rapidly as we increase the Cu substitution level, as shown in Fig \ref{fig:H-x} in the Appendix. For $H||c$, the transition field to the saturated paramagnetic state decreases rapidly from $\sim$ 60 kOe to $\sim$ 20 kOe just for $x$ = 0.012 and is completely suppressed by $x$ = 0.036. A lower field transition is observed near 30 kOe for the parent compound which is suppressed even more rapidly with Cu substitution and is no longer observed for $x > 0.02$. For $H \perp c$, assuming there is no in-plane anisotropy, the transition to the paramagnetic state is observed and is suppressed more gradually and only dropping below our base temperature at a much higher Cu substitution level of $x$ = 0.097. We should also point out that there is a decreasing value of the high field saturation of the $M(H)$ data that rolls over to a more Brillouin-type saturation with Cu substitution, becoming more and more linear-like with decreasing slope (susceptibility). This also indicates that with increasing Cu substitution we are suppressing the magnetism in this system. 

\par

\begin{figure}[h]
    \centering
    \includegraphics[width=\linewidth]{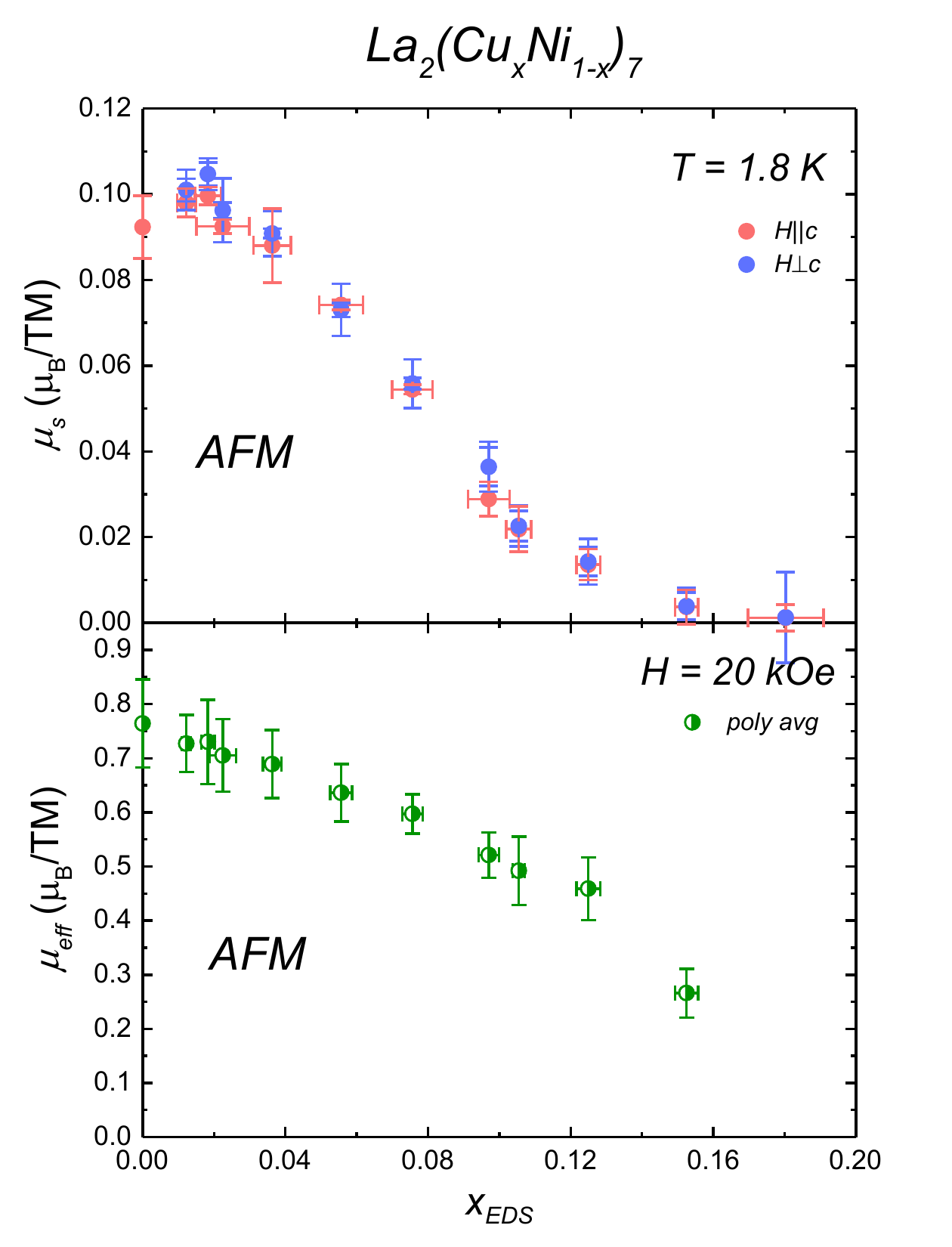}
    \caption{\footnotesize{The evolution of the extrapolated moment $\mu_s$ and the effective moment $\mu_{eff}$ with $x$ in ${\text{La}_{2}\text{(Cu}_{x}\text {Ni}_{1-x})_7}$. The value of $\mu_s$ is almost an order of magnitude smaller than the $\mu_{eff}$ confirming the itinerant nature of magnetism in this system.}}
    \label{fig:effective}
\end{figure}

\subsubsection*{Evolution of $\mu_s$ and $\mu_{eff}$ with Cu-substitution}

We can further analyze the change in the magnetic ordering with Cu substitution by studying the change in the high-field, extrapolated moment ($\mu_s$) and the effective moment $\mu_{eff}$ obtained from both $M(H)$ and $M(T)$ measurements respectively. The top panel of Fig. \ref{fig:effective} tracks the evolution of the extrapolated $\mu_s$ moment with $x$. $\mu_s$ is estimated by extrapolating the high-field (40 kOe $\leq H \leq$ 70 kOe) linear behavior of the $M(H)$ data to zero field as shown in Fig \ref{fig:extrapolated} and discussed in details in the Appendix. We see that $\mu_s$ decreases as we increase the Cu substitution in ${\text{La}_{2}\text{(Cu}_{x}\text {Ni}_{1-x})_7}$ from a value $\sim$ 0.1 $\mu_{B}/TM$ for the parent compound to essentially zero for $x$ = 0.181. 

\par

We also track the evolution of the effective moment $\mu_{eff}$ with $x$ from the polycrystalline average of the $\chi(T)$ data by doing a Curie-Weiss analysis with an added temperature independent contribution, fitting the data in the high temperature paramagnetic regime to  

\begin{equation}
    \chi=\frac{C}{T-\Theta} + \chi_{0}
    \label{eqn:CW fit}
\end{equation}

 where, $C$ is the Curie constant, $\Theta$ is the Weiss temperature and $\chi_0$ is the temperature independent contribution to the magnetic susceptibility. $\mu_{eff}$ is obtained from the fitted Curie constant $C$ using $C=\frac{N_A \mu^2_{eff}}{3 k_B}$, where $N_A$ is the Avogadro number and $k_B$ is the Boltzmann constant.

\par

Previously, Curie-Weiss analysis on the parent ${\text{La}_{2}\text {Ni}_7}$ was conducted using data measured at $H$ = 1 kOe \cite{Ribeiro2022Small-moment/math}. Here, however, the signal becomes increasingly poor at high temperatures as the Cu fraction increases (likely due to the decreasing moment). To achieve a better signal, and reliable fit, we measured the $M(T)$ of each sample at $H$ = 20 kOe (Fig. \ref{fig:CW data} in the Appendix) and used these data sets for the Curie-Weiss fits. Despite this, the data for the highest Cu substituted sample ($x$ = 0.181), does not fit well to a Curie-Weiss analysis, most likely arising from the low value of the magnetic moment. More details on this will be discussed in the Appendix. The lower panel of Fig \ref{fig:effective}  shows the evolution of the effective moment in ${\text{La}_{2}\text{(Cu}_{x}\text {Ni}_{1-x})_7}$. $\mu_{eff}$ also decreases as we increase the Cu substitution; the trend for 0 $\leq x \leq$ 0.152 is similar to that of the evolution of the extrapolated moment $\mu_s$ in the system. The magnitude of $\mu_{eff}$ is an order of magnitude higher. This indicates that the magnetism in this system is itinerant in nature. We also obtain the Weiss temperature $\Theta$ and $\chi_0$ from the same fit and Table \ref{tab:magnetic} in the Appendix shows the values of all these parameters.

\begin{figure}[htbp]
 \centering
 \includegraphics[width=1.0\linewidth]{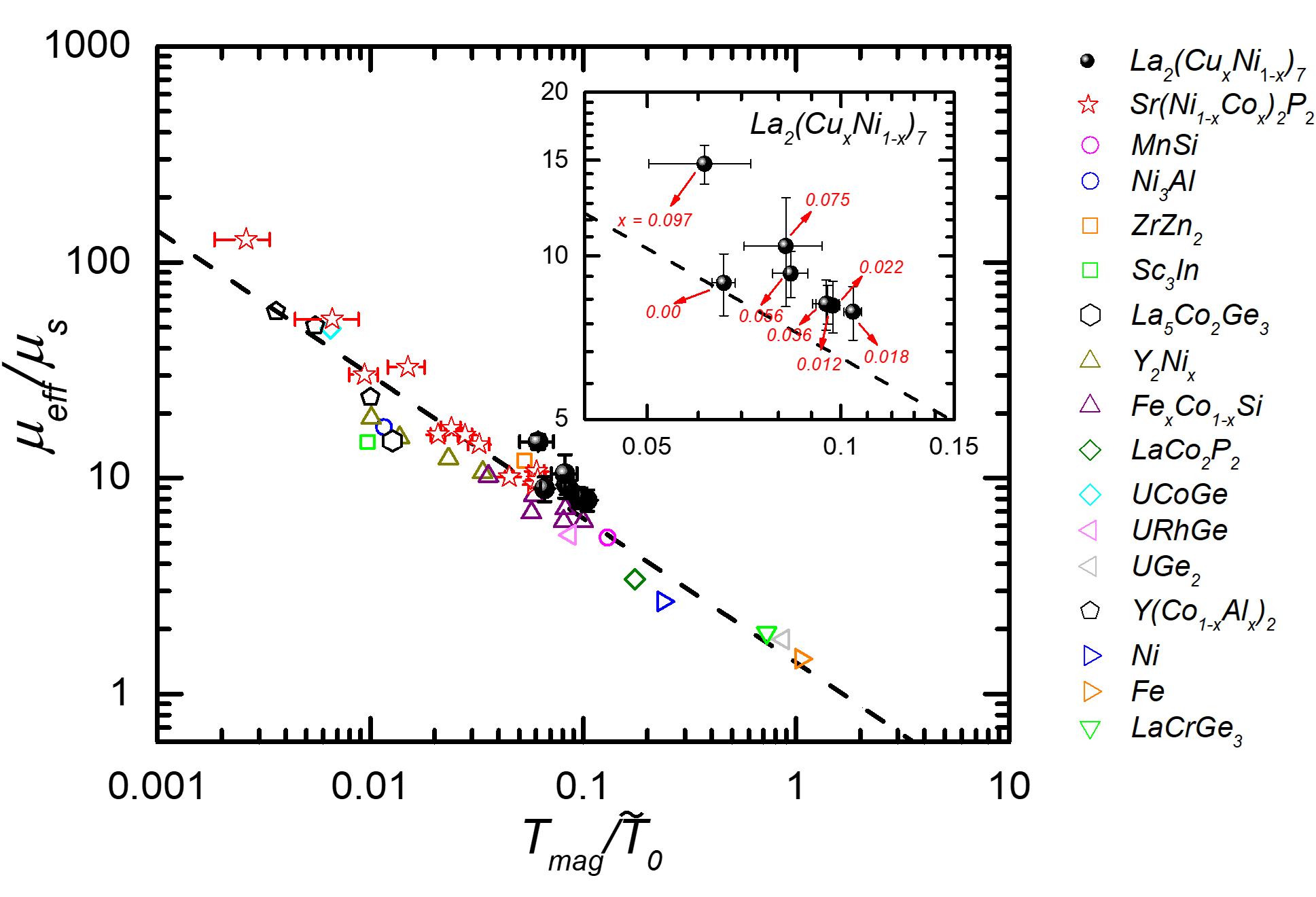}
 \caption{\footnotesize{Modified Deguchi-Takahashi plot for ${\text{La}_{2}\text{(Cu}_{x}\text {Ni}_{1-x})_7}$ with $0\leq x \leq 0.097$ and other weak itinerant magnets \cite{Schmidt2023,Takahashi,Saunders2020ExceedinglyLa5Co2Ge3,Xu2023}. The dashed line represents the expected theoretical behavior \cite{Takahashi}. Inset: enlarged scale showing only the compositions explored in this study, where the labels with red arrows are indicating the Cu fraction corresponding to each point.}}
 \label{fig:Takahashi}
\end{figure}

To further examine the itinerant character of magnetism in this system we performed a Deguchi-Takahashi analysis \cite{Schmidt2023,Takahashi,Saunders2020ExceedinglyLa5Co2Ge3,Xu2023}. The values of $\mu_{eff}$, $\mu_s$, $T_N$ and the modified spectral parameter, $\tilde{T}_0$, were combined into the modified Deguchi-Takahashi plot as shown in Fig. \ref{fig:Takahashi} (black circles). The $M(H)$ data set at $T=1.8$ K for $0\leq x\leq 0.097$ were used to estimate the values of $\tilde{T}_0$ as outlined in Ref. \cite{Schmidt2023}. This plot also includes other known itinerant ferromagnets and antiferromagnets \cite{Schmidt2023,Takahashi,Saunders2020ExceedinglyLa5Co2Ge3,Xu2023}, as well as a dashed line for the expected theoretical behavior according to Takahashi's spin-fluctuation therory for itinerant electron magnetism \cite{Takahashi}. Since this plot includes both ferromagnets and antiferromagnets, the symbol $T_{mag}$ is used to generally represent $T_C$ and $T_N$, respectively. The fact that the ${\text{La}_{2}\text{(Cu}_{x}\text {Ni}_{1-x})_7}$ compounds with $0\leq x\leq 0.097$ follow the expected trend is a good indication that, despite being antiferromagnetic at zero field, they exhibit typical characteristics of weak itinerant ferromagnets under a finite applied field \cite{takahashi1985, Sangeetha2019}. 

\par

\subsection*{Quantum Critical Point, nFL-region and beyond}

${\text{La}_{2}\text{(Cu}_{x}\text {Ni}_{1-x})_7}$ has a magnetically ordered low temperature state for $x <$ 0.097, whereas we observe a Kondo-like behavior for the higher doped samples. The suppression of both $T_N$ and the size of the ordered moment, $\mu_s$, along with the emergence of apparent Kondo-like behavior in the paramagnetic state suggests that there maybe a quantum critical point in between these two competing regimes. From all the $M(H,T)$, $\rho(T)$, and $C_p(T)$ data discussed so far, the $x$ = 0.105 sample appears to be in the vicinity of an AFM-QCP. 

\par

Given that the $x$ = 0.105 $\rho(T)$ data appears to be starting to go through a resistive minima upon cooling and then, only at lower temperatures, rolls over into a linearly decreasing $\rho(T)$, it would appear that $x$ = 0.105 may be slightly beyond the critical doping for the precise QCP. Despite this, we can still examine the low temperature behavior for a quantum critical behavior; a well-established way to do this is to use a power law analysis of the resistivity data. The $\rho(T)$ data can be analyzed by using,

\begin{equation}
    \rho(T) = \rho_{0} + AT^n
    \label{eqn:power law}
\end{equation}

 where $\rho_0$ is the residual resistivity and $A$ is a co-efficient and can be interpreted as the quasiparticle scattering cross section. Alternatively, we can use the obtained value of $\rho_0$ and plot $\ln{(\rho-\rho_0)}$ vs $\ln{T}$. A low temperature linear fit can be done to this data which corresponds to,

 \begin{equation}
     \ln{(\rho-\rho_{0})} = n\ln{T} + \ln{A}
     \label{eqn: log power_law}
 \end{equation}

The slope of the fit gives the value of $n$. In this paper, we have estimated the value of $n$ using both Eqns. \ref{eqn:power law} and \ref{eqn: log power_law}, and the values of each are within the uncertainty limit. From this point onwards in this paper, we will refer to the value of $n$ obtained from Eqn \ref{eqn: log power_law} only. The exponent, $n$, indicates whether the system is in a Fermi liquid (FL) regime $(n=2)$ where electron-electron scattering is dominant, or, in a nFL $(n\sim1)$ regime which is the behavior expected in the vicinity of a quantum critical point (QCP) \cite{stewart2001}.

\begin{figure}[h]
    \centering
    \includegraphics[width=\linewidth]{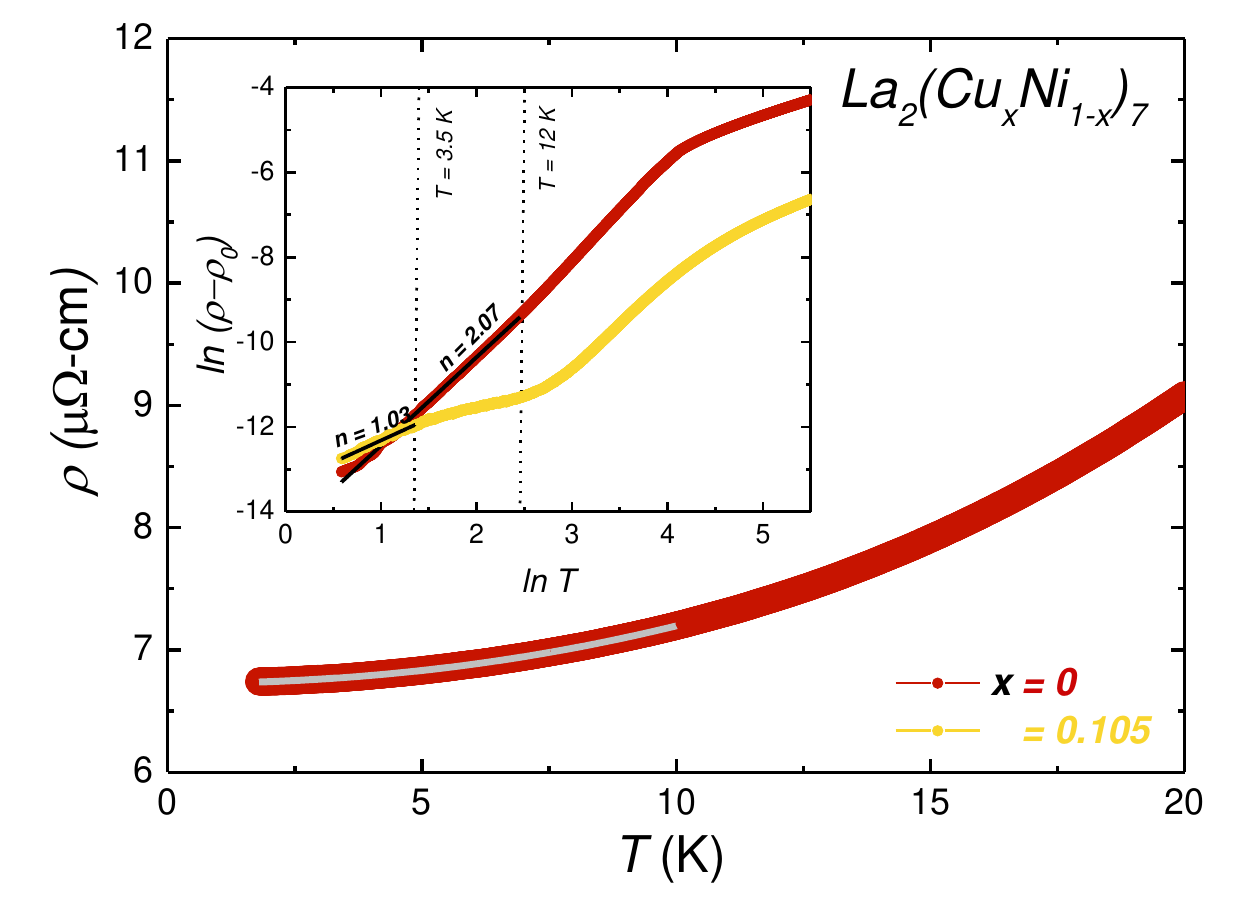}
    \caption{\footnotesize{Plot of $\rho(T)$ for the parent $x$ = 0, ${\text{La}_{2}\text{(Cu}_{x}\text {Ni}_{1-x})_7}$, with the grey line showing the fit of Eqn. \ref{eqn:power law}. We obtain $\rho_0$ from this fit, which can be further used to plot $\ln{(\rho-\rho_0)}$ vs $\ln{T}$. (Inset: $\ln{(\rho-\rho_0)}$ vs $\ln{T}$ for $x=0$ and $x=0.105$. A linear fit is done to low temperature data between 1.8 K $\leq T \leq$ 10 K for $x$=0 and 1.8 K $\leq T \leq$ 3.5 K for $x$=0.105 to obtain $n$ and $A$ (See text for details). The fit line is shown in black. The slope of this fit gives the value of $n$ = 2.07(2) for $x$ = 0 and $n$ = 1.03(6) for $x$ = 0.105.)}}
    \label{fig:ln rho 0 and 10Cu}
\end{figure}

\par

We perform such an analysis for $x < $ 0.105 samples only. The potential contribution from magnons below the ordering temperature is acknowledged, but given the very small size of the ordered moment, as well as the small resistive anomaly associated with the loss of spin disorder scattering, we assume that magnon scattering is minimal compared to the electron-electron scattering. For the higher doped $0.125 \leq x \leq 0.181$ samples, the presence of the low temperature upturn in the data inhibits us from doing such an analysis.

Figure \ref{fig:ln rho 0 and 10Cu} shows the plot of $\rho(T)$ with the fit line to Eqn. \ref{eqn:power law} for $x$ = 0 and $\ln{(\rho-\rho_{0})}$ vs $\ln{T}$ for the parent compound as well as the other extreme member: $x = 0.105$. Similar data are plotted for the other samples are shown in the Appendix in Fig \ref{fig:power law all}. Using Eqn. \ref{eqn: log power_law}, we obtain $n$ = 2.07 (2) and 1.03 (6) for $x$ = 0 and 0.105 respectively which suggests that the parent $x$ = 0 sample behaves as a Fermi liquid, whereas $n \sim$ 1 for $x$ = 0.105 sample is consistent with nFL behavior. We emphasize here that for $x$ = 0.105, the fit could only be performed over a small temperature window. However, it is clear from Fig. \ref{fig:ln rho 0 and 10Cu} that the low temperature behavior of the $x$ = 0.105 sample does not follow a Fermi liquid behavior.

\begin{figure}[h]
    \centering
    \includegraphics[width=\linewidth]{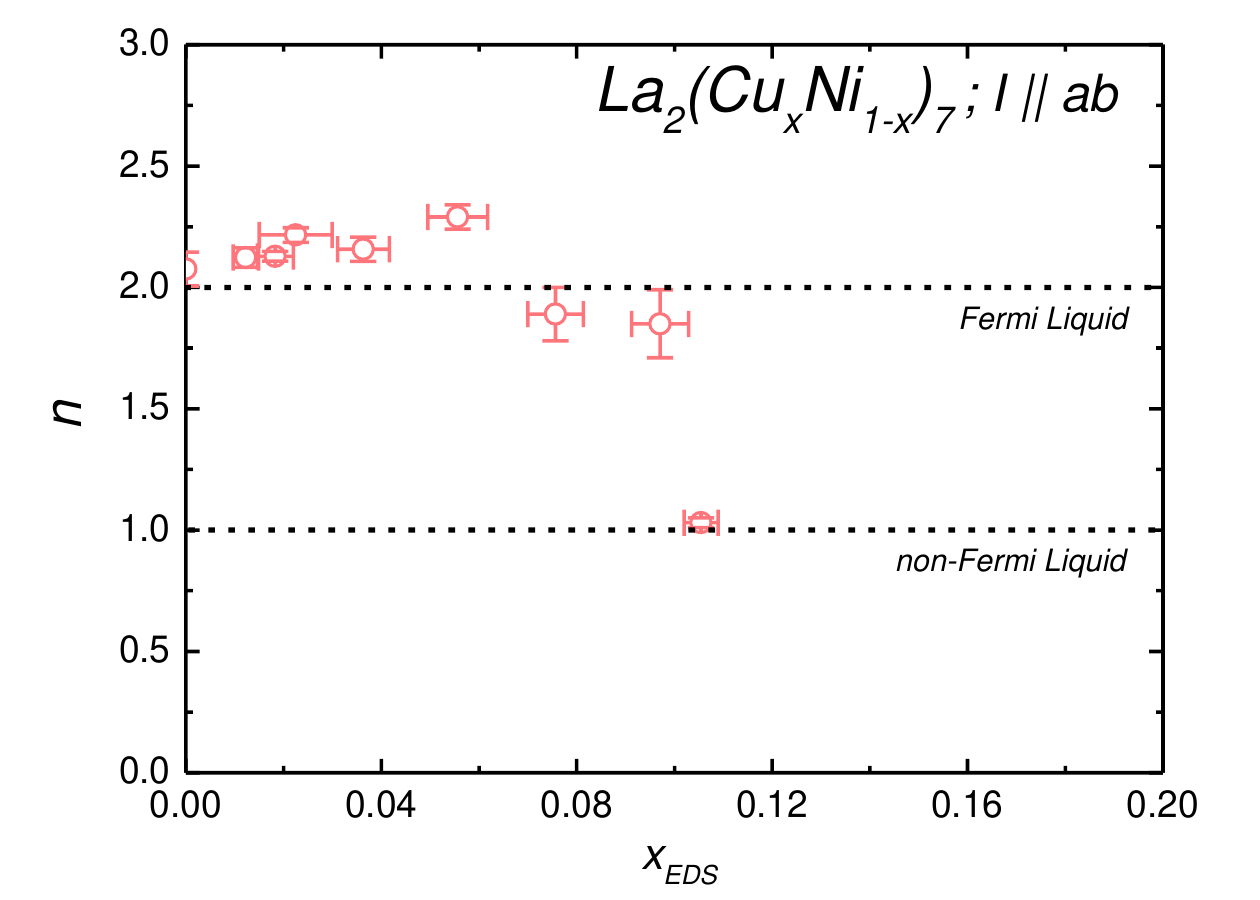}
    \caption{\footnotesize{Plot of the slope $n$ of the linear fit to the low temperature $\ln{(\rho-\rho_{0})}$ vs $\ln{T}$ data to the ${\text{La}_{2}\text{(Cu}_{x}\text {Ni}_{1-x})_7}$ samples for 0 $\leq x \leq$ 0.105 obtained from their respective zero field, in-plane $\rho(T)$ data. A value of $n$ = 2 corresponds to Fermi liquid behavior whereas $n$ = 1 corresponds to a non-Fermi liquid behavior and is shown using horizontal lines. Values of $n$ for $x>$ 0.105 are not shown due to Kondo-like upturn in their $\rho(T)$ data. }}
    \label{fig:n and A}
\end{figure}

\par 

\begin{figure}[h]
    \centering
    \includegraphics[width=\linewidth]{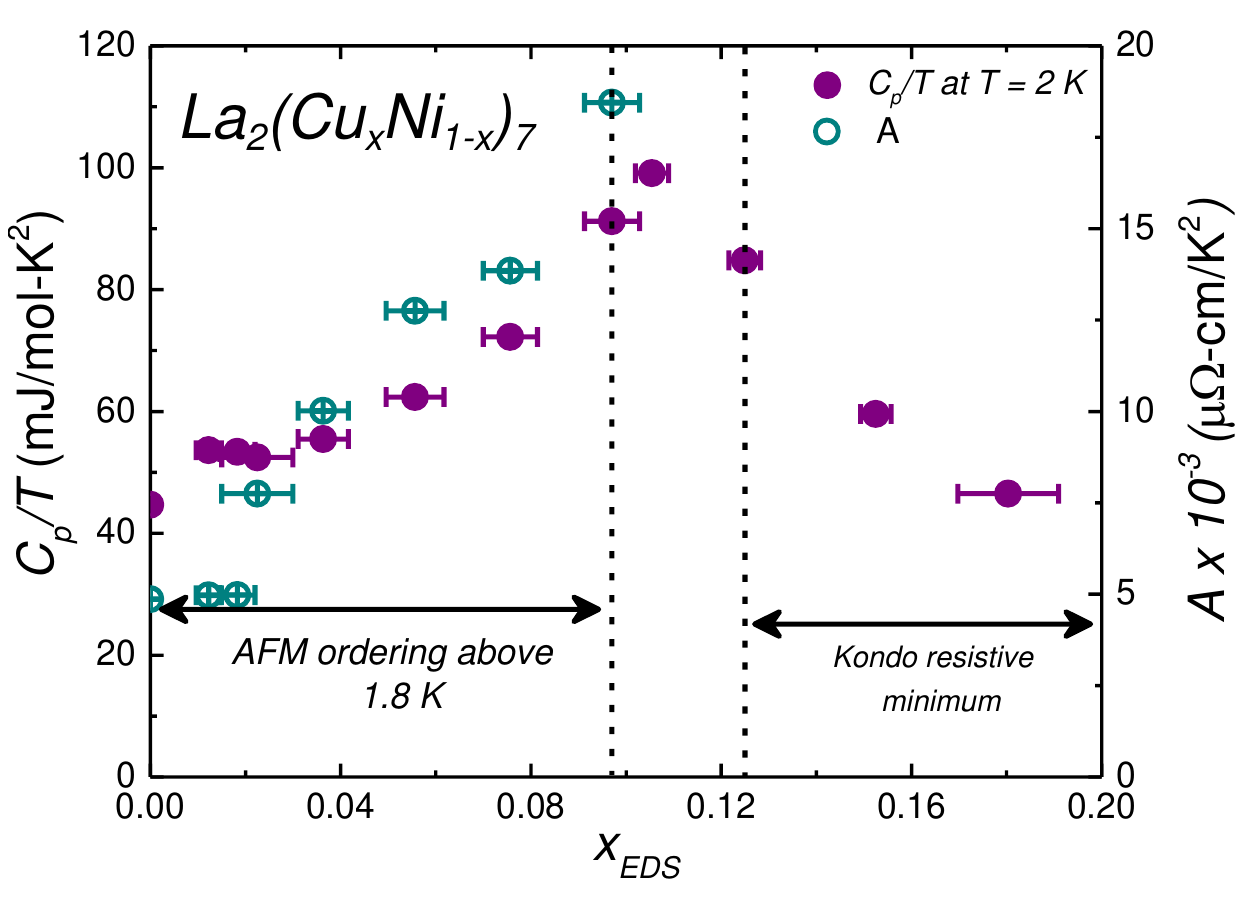}
    \caption{\footnotesize{(Color online) ${a}$ Plot of $C_p/T$ measured at $T$ = 2.0 K vs $x$ for the ${\text{La}_{2}\text{(Cu}_{x}\text {Ni}_{1-x})_7}$ single crystals plotted on the left axis (solid circles) and $A$ obtained from fitting Eqn. \ref{eqn:power law}, with a fixed $n$=2 to the low temperature $\rho(T)$ data on the right axis (hollow circles). The AFM ordered above base temperature ($T$ = 1.8 K) and the Kondo-like regime is marked.  }}
    \label{fig:gamma}
\end{figure}

\par

Figure \ref{fig:n and A} shows the evolution of the slope $n$ of the linear fit to the low temperature $\ln{(\rho-\rho_{0})}$ vs $\ln{T}$ data to the ${\text{La}_{2}\text{(Cu}_{x}\text {Ni}_{1-x})_7}$ samples for 0 $\leq x \leq$ 0.105. The value of $n$ is $\sim$ 2 for the magnetically ordered 0 $\leq x \leq$ 0.097 samples and then decreases to a value of 1.03 for $x$ = 0.105, which may suggest nFL behavior for this doping. We again emphasize the limited temperature range used for these fits restrict the confidence with which we can draw conclusions; however, given that the inferred value of $n$ changes rather sharply at the exact Cu concentration for which magnetic order is suppressed, it is reasonably to infer a crossover between Fermi liquid and non-Fermi liquid regimes may occur near $x$ = 0.105.

\par

In order to explore this apparent QCP, we can see how the prefactor $A$ of $\rho(T)~=~\rho_0~+AT^2$ behaves as we approach $x =$ 0.105 from below. Given that for $x <$ 0.105 we find $n\sim$ 2, we fit our low temperature $\rho(T)$ data to a classical $T^2$ temperature dependence. In Fig. \ref{fig:gamma} we plot $A$ as a function of $x$ and see that it diverges in a manner very similar to the $T$ = 2.0 K $C_p/T$ data. In fact, we can see in Fig \ref{fig:kadowaki}, in the Appendix, that our data are consistent with the Kadowaki-Woods scaling of $A$ with $\gamma$, another hallmark of Kondo-like systems. This divergence in both $A$ and $C_p/T$ with $x$ near the QCP is reminiscent of the behavior observed in YbRh$_2$Si$_2$, reported in \cite{YbRh2Si2_2002}, but is different from the behavior in CeIn$_3$ where the coefficient $A$ does not diverge near the QCP \cite{CeIn3_knebel}.  

\par

\begin{figure}[h]
    \centering    
    \includegraphics[width=\linewidth]{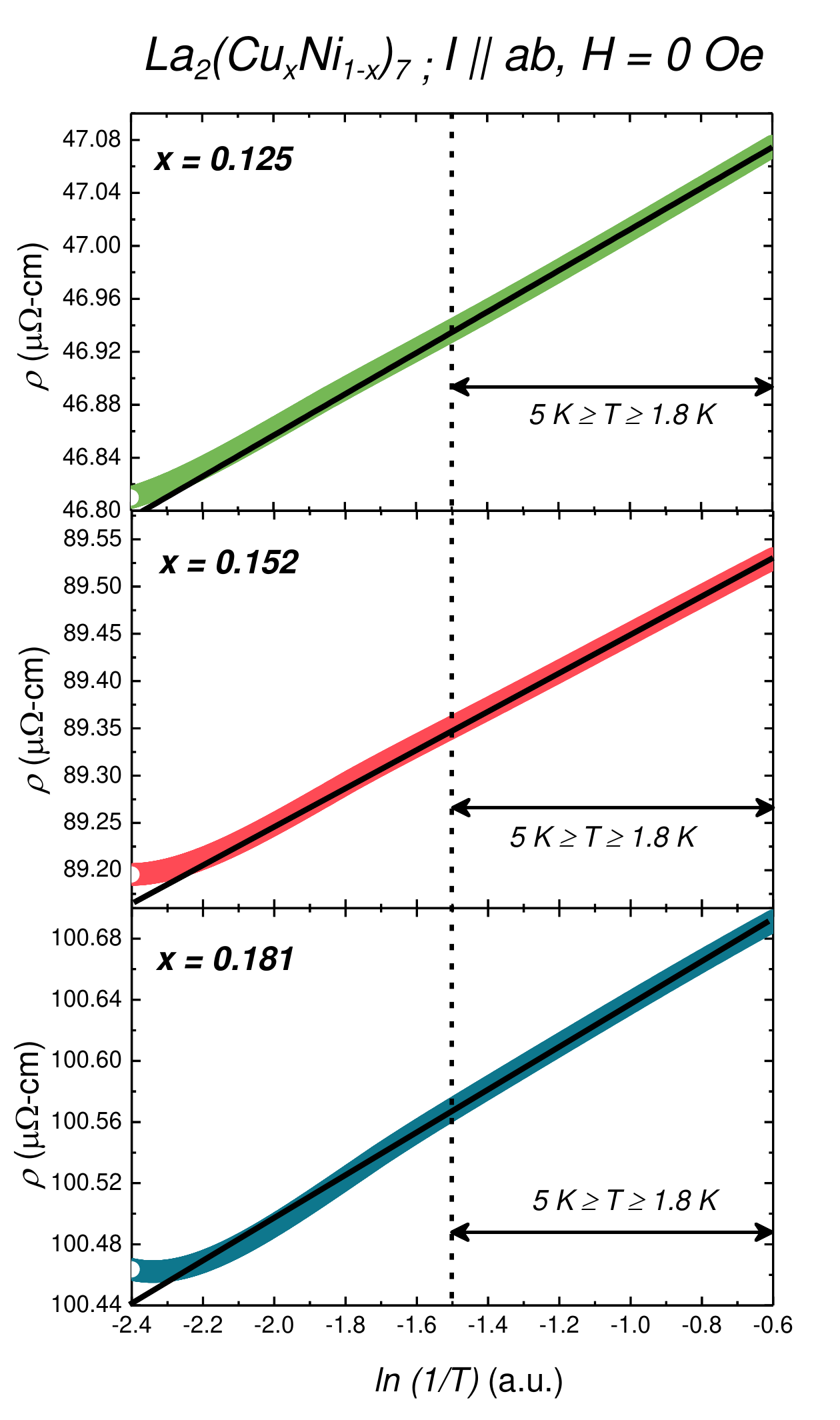}
    \caption{\footnotesize{(Color online) Plot of $\rho$ vs $\ln{1/T}$ for $x$ = 0.125, 0.152, and 0.181 ${\text{La}_{2}\text{(Cu}_{x}\text {Ni}_{1-x})_7}$ single crystals in the temperature range 1.8 K $\leq x \leq $ 11 K. A linear fit to the temperature range 1.8 K $\leq T \leq$ 5 K, shown using the black line, to each of these data shows a linear $\ln{1/T}$ behavior which is expected for single ion Kondo effect. There is a clear deviation from linearity as we increase the temperature.}}
    \label{fig:Kondo log}
\end{figure}

The $\rho(T)$ behavior for $x$ = 0.105 sample is intermediate to the lower $x$-values, with magnetic ordering and power law exponent values closer to 2, and higher $x$-values which manifest no evidence of magnetic ordering and have a Kondo-like minimum in $\rho(T)$. Returning, once more, to the data for $x$ = 0.105 shown in Fig \ref{fig:R-T separate}, in the higher temperature range it looks like the $\rho(T)$ will go through a Kondo minimum, but just as it reaches the bottom of the Kondo minimum, it instead transitions into the linear nFL behavior. This suggests that the true QCP $x$-value for this system may be slightly below the $x$ = 0.105 sample we measured.

\par

We  further investigate the possible nFL behavior with specific heat measurements. When a system shows nFL behavior in the vicinity of a QCP, an upturn in $C_p/T$ at low temperatures and enhanced Sommerfeld coefficient, $\gamma$ are often observed \cite{stewart2001}. $\gamma$ is usually obtained from the linear extrapolation of the low temperature $C_p/T$ vs $T^2$ data to $T$ = 0. But from Fig \ref{fig:heat capacity_27} and Fig \ref{fig:cp_T2} in the Appendix, we see that the low temperature upturn makes it difficult to do such extrapolation in a consistent manner.

Figure \ref{fig:gamma} shows the change of the $C_p/T$ value measured at $T$ = 2.0 K with $x$ (left axis). It should be noted that the phonon contribution to 
$C_p/T$ at 2 K is  negligible, $\sim 5$ mJ/mol K$^2$, as estimated from the high temperature (10 K $\leq T \leq$ 15 K) behavior of the $C_p/T$ data. Given this small phonon contribution, we did not subtract it from the total heat capacity at 2 K. For $x$ = 0, its value is close to the reported value of $\gamma$ at $\sim$ 40 mJ/mol K$^2$ \cite{Ribeiro2022Small-moment/math}. For 0 $\leq x \leq$ 0.075, $C_p/T$ increases to $\sim$ 70 mJ/mol K$^2$ as expected for a fragile moment (correlated electron) system as its magnetic transition temperature is reduced. For $x$ = 0.097 and 0.105, $C_p/T$ increases more rapidly and reaches its maximum value of $\sim $100 mJ/mol-K$^2$ for $x$ = 0.105, very close to  where we infer the QCP in this system. For the highest dopings 0.125 $\leq x \leq$ 0.181, $C_p/T$ decreases monotonically as the upturn in the data decreases and returns toward more linear behavior (See Fig \ref{fig:cp_T2}).  

\par

Although the change in the 2.0 K value of $C_p/T$ is smaller than would be expected in the case of QCPs associated with Ce- or Yb- based moments, there is a clear and substantial increase in $C_p/T$ that can be expressed as either, "near doubling" or "increase by roughly 50 mJ/mol-K$^2$". The observation that this peaked increase in $C_p/T$ occurs for the $x$-value close to the extrapolation of the suppressed $T_N$ line and the $x$- value for which we find nFL-like behavior in the low temperature resistivity strongly suggest that all three of these phenomenon arise from a QCP associated with a d-band based fragile magnetism.

\par

\subsection*{Possible low temperature single-ion Kondo-like region; 0.125 $\leq x \leq$ 0.181}

The three highest Cu substituted samples, 0.125 $\leq x \leq$ 0.181, show anomalous low temperature resistivity and specific heat behavior that is reminiscent of the Kondo effect. In this section we will discuss this behavior in greater detail.

\par

Such an increase in the resistivity could in principle have many potential sources. For three-dimensional systems, an increase in resistivity with decreasing temperature should follow either a $T^{-p/2}$ dependence if it is due to weak localization (p = 3/2, 2, 3, depending upon the scattering mechanism) \cite{Barua2017SignaturesVSe2, maritano2006, Lee1985DisorderedSystems}, or a $\ln{(1/T)}$ dependence if it is due to the Kondo effect \cite{Barua2017SignaturesVSe2, Wang2021Weak/math, kumar2002}. To ascertain the nature of the low T resistivity upturn in our samples, we plot the resistivity with respect to $T^{-p/2}, (p = 3/2,2,3) $ and $\ln{(1/T)}$ for $x = 0.125, 152,$ and 0.181 samples. The data for $\ln{(1/T)}$ behavior is shown in Fig \ref{fig:Kondo log} whereas the data plotted against $T^{-p/2}$ dependencies is shown in Fig \ref{fig:kondo test} in the Appendix. A linear fit is done between 1.8 K $\leq T \leq$ 5 K to each of the data sets, and a good fit is only obtained for $\ln{(1/T)}$, suggesting the Kondo effect is indeed the origin of the resistance upturn.

\begin{figure}[h]
    \centering
    \includegraphics[width=\linewidth]{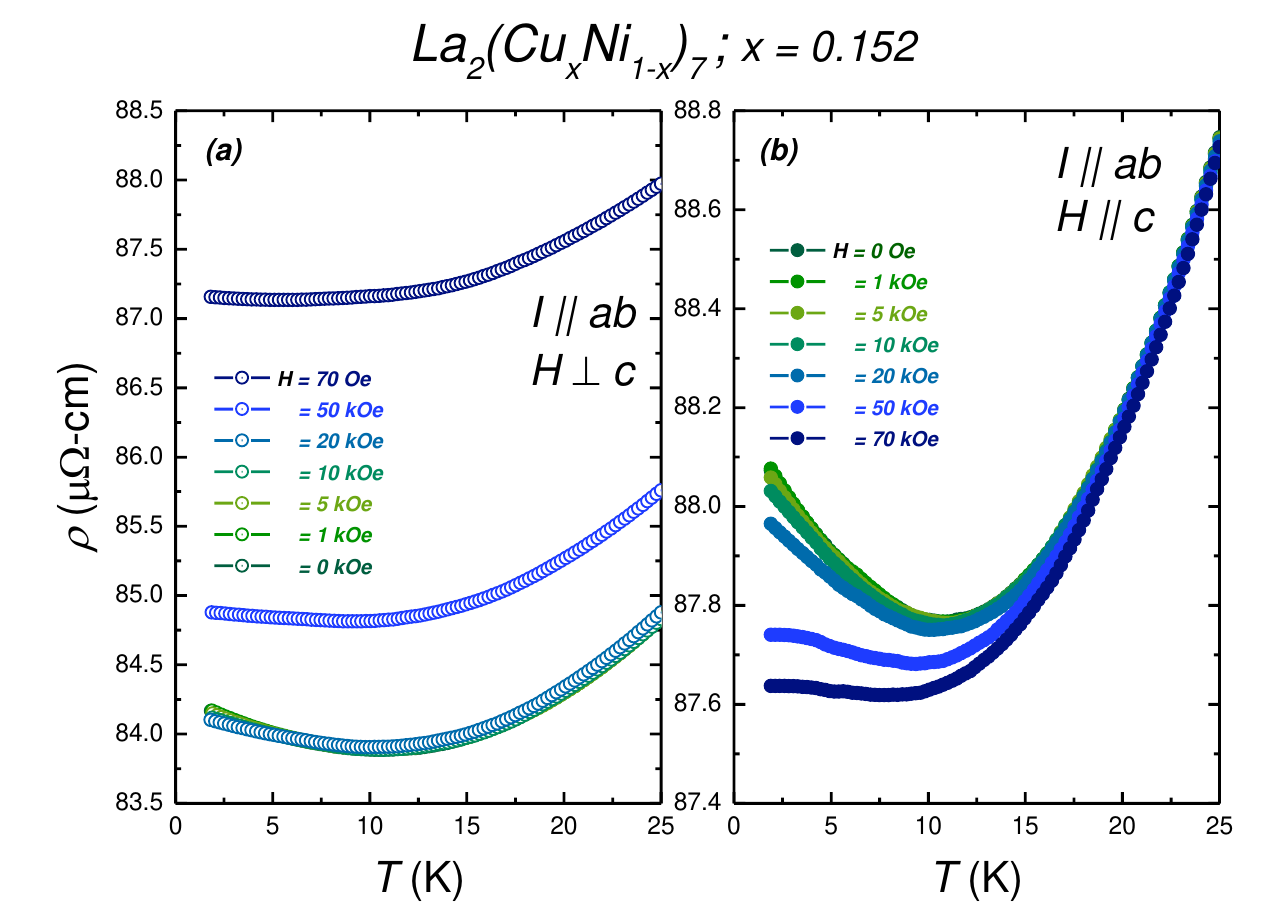}
    \caption{\footnotesize{Plot of the transverse temperature dependent resistivity $\rho(T)$ under an external applied magnetic field for $H \perp c$ in panel $(a)$ and $H \|| c$ in panel $(b)$ for $x$ = 0.152 ${\text{La}_{2}\text{(Cu}_{x}\text {Ni}_{1-x})_7}$ sample. In panel $(a)$, the data sets for 0 kOe $\leq H \leq$ 20 kOe essentially overlap and are not resolvable in this plot.}}
    \label{fig:RT_under field}
\end{figure}

\par

 We also measured $\rho(T)$ under an external applied field for both $H \perp c$ and $H \|| c$ directions for $x$ = 0.152 sample. If the resistive upturn is due to Kondo scattering, then it is expected to be suppressed gradually with increasing magnetic field as it was observed in the case of weak Kondo effect in $\text{Zr} \text{Te}_2$ \cite{Wang2021Weak/math} or $\text{La}_3 \text{Cu}_4 \text{P}_4 \text{O}_2$ \cite{Szymon2025} or YbPtBi \cite{Mun2013Magnetic-field-tunedYbPtBi}. Figure \ref{fig:RT_under field} shows the results of $\rho(T)$ under an external field, where the resistance minima is clearly suppressed by the applied field, nearly vanishing at 70 kOe. All together, given that the specific heat as well as the $\rho(T)$ are consistent with Kondo-like behavior, it appears that after the suppression of the fragile magnetism in ${\text{La}_{2}\text{(Cu}_{x}\text {Ni}_{1-x})_7}$, a 3d-based single-ion Kondo-like state emerges in this system.

\section{Discussion and Conclusions}

To summarize, we synthesized single crystals of ${\text{La}_{2}\text{(Cu}_{x}\text {Ni}_{1-x})_7}$ with $x$ varying between 0 and 0.181. Powder x-ray diffraction and EDS measurements confirmed the phase and the concentration of Cu respectively. Based on a suite of anisotropic temperature and field dependent magnetic, transport, and specific heat measurements, we constructed the $T-x$ phase diagram for this system. For 0 $\leq x \leq$ 0.097, the system remains magnetically ordered with transition temperature(s) above ($T$ = 1.8 K) with the parent $x$ = 0 and the lowest doped $x$ = 0.012 showing signs of multiple magnetic phases. For higher Cu-doping, a single magnetic phase transition is suppressed monotonically to below $T$ = 1.8 K. From our results it appears most likely that the $B$ phase extends all the way to $x$ = 0.097 sample, the $C$ phase gets suppressed almost instantaneously with Cu doping, and the evolution of the $A$ phase remains ambiguous. The AFM transition temperature of phase $B$ decreases from a value of $\sim$ 56 K for $x$ = 0, to a value of $\sim$ 5 K for $x$ = 0.097. The high-field saturated moment $\mu_s$ and the effective moment $\mu_{eff}$ also decreases with increase in $x$, with $\mu_s$ vanishing alongside $T_N$. A modified Deguchi-Takahashi analysis on the ${\text{La}_{2}\text{(Cu}_{x}\text {Ni}_{1-x})_7}$ series, points to a trend typically expected for itinerant magnetism under a finite applied field according to Takahashi's spin-fluctuation theory. However, we also discuss later the possibility a local moment picture of the magnetism. In addition, we find that there is good Kadowaki-Woods scaling of $C_p/T$ and the $T^2$-coefficient $A$, of the low temperature $\rho(T)$; another hallmark of correlated electron behavior.

\par

The three highest dopings studied, 0.125 $\leq x \leq$ 0.181, have no magnetic ordering but exhibit an upturn in resistivity with decreasing temperature below $T \approx$ 12 K, which we have interpreted as an indication of a type of Kondo behavior. The sample with an intermediate $x$ = 0.105 is in the proximity of the quantum critical regime for this system. As such, then, ${\text{La}_{2}\text{(Cu}_{x}\text {Ni}_{1-x})_7}$ is a manifestation of a fragile magnetic system that can be tuned to an AFM-QCP by chemical substitution.

\par

The nFL behavior for $x$ = 0.105 is suggested by power law study to the zero-field $\rho(T)$ and $C_p(T)$ data which point to an enhanced electron mass and possible nFL behavior as $x$ approaches $\sim$ 0.1 from both low and high dopings. Analysis of the temperature dependence of $\rho$ shows a $\ln{(1/T)}$ behavior below the resistance minimum, and measurements at different fields show that this minima is weakened with $H$ (Fig. \ref{fig:RT_under field}). Both of these observations suggest that the Kondo effect is the origin of the anomalous low temperature behavior.

On one hand, the single ion-Kondo effect was first discovered in metals (such as gold) with minute amounts of $3-d$ transition metal impurities (such as Fe) \cite{brewer1955progress, Kondo1964ResistanceAlloys}. On the other hand, Kondo physics, manifested by an anomalous upturn in resistivity data in the vicinity of a QCP is traditionally associated with rare earth, especially Ce or Yb, atoms in intermetallic compounds, where the localized $f$ -orbitals hybridize with the surrounding $s-,~ p-,$ or $d-$ orbital bands often revealing exotic phenomenon such as unconventional superconductivity, non-Fermi liquid, etc. \cite{Gegenwart2008QuantumMetals, QunatumcriticalityHVL}. Thus, the observation of Kondo-like behavior when tuning a purely $3d$ system is surprising. It is interesting to consider whether this behavior may be rooted in the kagome network formed from the Ni4 and Ni5 atoms (See Fig.\ref{fig:crystal}). Compact localized states associated with the kagome structure have been suggested to produce orbital currents that may act like localized moments and have been proposed to provide a mechanism for $d-$ band based Kondo physics \cite{Chen2024EmergentCorrelations, Ye2024}.

This said, having a d-shell based fragile magnetic system, by itself, is not unusual; the curious aspect of the ${\text{La}_{2}\text{(Cu}_{x}\text {Ni}_{1-x})_7}$ system is the change from Fermi-liquid-like resistivity for low $x$ that evolves to a nFL-like behavior near a QCP to a single-ion Kondo type of resistance for $x > 0.105$. Usually after the suppression of the AFM ordering and the quantum critical region, there is another Fermi-liquid region with the putative Kondo temperature rising and the $T^2$ coefficient (and $\gamma$) dropping. In the case of ${\text{La}_{2}\text{(Cu}_{x}\text {Ni}_{1-x})_7}$, for $x > 0.105$
 we find that the $\rho(T)$ behavior that has more resemblance to a system in the single ion Kondo impurity limit. Operationally, this simply means that any putative coherence temperature for the Kondo-lattice has dropped below our base temperature, but even with this said, x $\sim$ 0.1 would seem to be a small doping level for this to happen at. 

\par

One possible explanation relates to the crystal structure of La$_2$Ni$_7$, where two of the five Ni-sites form kagome-like layers. These two kagome layers accommodate only 9 of the 42 Ni-sites per unit cell. We propose that Cu dopants preferentially occupy these kagome layers. Statistically, while Cu dopants can enter either of the two kagome layers, we assume there is a bias toward one layer. Neutron diffraction experiments will be essential to verify whether Cu
preferentially occupies the Ni4 or Ni5 sites.

\par

Each kagome layer is distinguished by its own characteristic single-ion Kondo temperature—associated with the onset of local Kondo screening observable as a resistivity minimum—and lattice Kondo temperature—signifying the emergence of a hybridization gap, detectable, for instance, via optical conductivity. Additionally, each layer possesses distinct coherence temperatures related either to single-ion coherence or full lattice coherence, reflecting the formation of a coherent Fermi liquid state with fully screened local moments. Typically, single-ion temperatures differ from lattice temperatures in each layer due to detailed microscopic interactions. Given our assumption that Cu dopants preferentially enter one kagome layer, we anticipate that this layer transitions into a dilute single-ion Kondo regime, while the other layer remains in a Kondo lattice regime. Consequently, the layer in the dilute limit will exhibit suppressed lattice Kondo and coherence temperatures, whereas the other layer, retaining Kondo lattice physics, will
maintain comparatively higher lattice Kondo and coherence temperatures.

\par

Within this framework, the absence of a resistivity upturn at low Cu doping levels occurs because neither kagome layer has yet reached the dilute single-ion Kondo regime. However, with increased doping, two crucial effects become evident: (1) one kagome layer becomes sufficiently dilute, entering the single-ion Kondo regime and causing a low temperature resistivity upturn; (2) the other kagome layer, characterized by topological flat bands, undergoes strong quantum fluctuations dominated by Kondo lattice physics. These fluctuations push the system toward a quantum critical point (QCP), manifesting as non-Fermi-liquid (nFL) behavior. The latter scenario is consistent with recent theoretical predictions for kagome systems exhibiting Kondo lattice physics, quantum criticality, and nFL behavior driven by topological flat bands \cite{Si-Chen2023}. 

\par

Beyond the quantum critical point, both single-ion and lattice Kondo effects coexist. As mentioned above, the coherence temperatures associated with these two effects can differ. Experimental data indicate that single-ion coherence emerges below the lowest measured temperatures. It is possible the lattice coherence temperature is similarly low. Alternatively, the lattice coherence temperature may be higher, suggesting a recovery of Fermi-liquid behavior. However, the expected FL temperature dependence of resistivity could be obscured. Instead, transport properties exhibit a dominant logarithmic temperature dependence, $\ln (1/T)$, arising from the kagome layer now hosting dilute Ni ions in the single-ion Kondo regime. Such a masking of conventional FL scaling which can be understood as follows:
hybridization between the two kagome planes strongly indicates that the Fermi surface includes contributions from Ni atoms in both layers. Since the transport scattering rate receives contributions from the entire Fermi surface, the total scattering rate can be approximated by an averaged sum of these individual contributions across the Fermi surface (more precisely, each contribution is weighted by a factor $(1 - \cos{\theta})$, emphasizing larger-angle scattering processes). Consequently, the stronger single-ion Kondo scattering contribution dominates at low temperatures, overshadowing the Fermi-liquid contribution and leading to the observed logarithmic temperature dependence of resistivity. Thus the double kagome layered structure of La$_2$Ni$_7$ provides a rare instance where single-impurity and lattice-based Kondo effects coexist and interact, offering a unified explanation for the observed quantum criticality, nFL behavior, and single-impurity Kondo signatures.

A key ingredient of our framework is that flat bands from kagome layers can mimic states of localized magnetic moments, while the other dispersive kagome and non-kagome bands form the conduction electron bath. Density functional theory (DFT) calculations explicitly highlight the significant role played by flat bands located close to, but slightly above, the Fermi energy \cite{Lee2023Electronicsub7/sub}. Their proximity to the Fermi level strongly indicates their involvement in driving the observed phenomena. Theoretical studies of model kagome systems \cite{Si-Chen2023, Chen2024EmergentCorrelations} further support this scenario, demonstrating that correlation-induced renormalization can shift these flat bands closer to the Fermi level, thus
promoting local moment physics. Such a situation appears plausible for ${\text{La}_{2}\text{(Cu}_{x}\text {Ni}_{1-x})_7}$ despite the DFT flat bands
being above the Fermi energy. Similar band renormalizations were found to occur in pyrochlore lattice systems \cite{Yi2024}. 

\par

\begin{figure*}[htbp]
    \centering
    \includegraphics[width=1.0\textwidth]{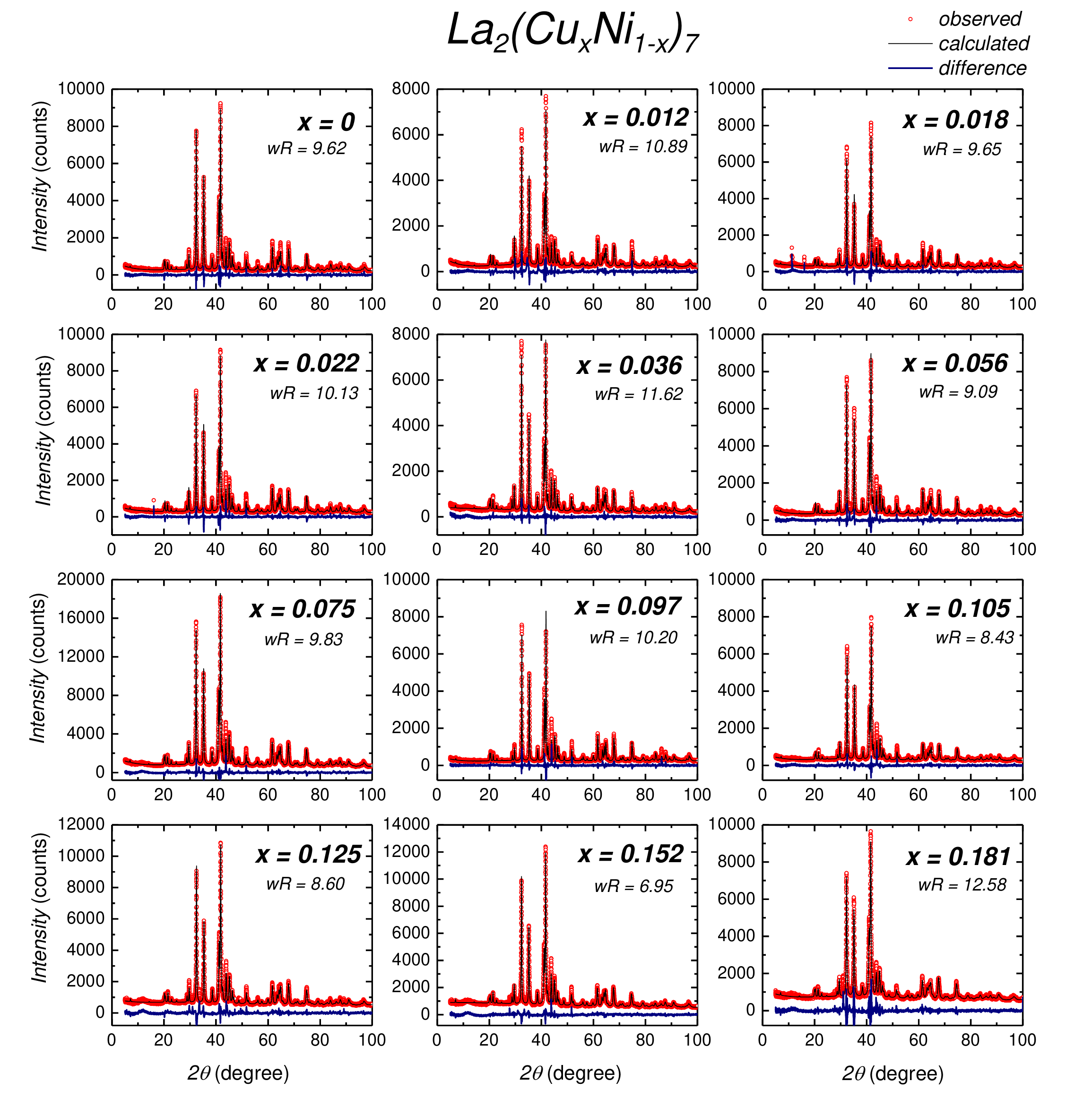}
    \caption{\footnotesize{(Color online) Plot of the room temperature powder x-ray patterns of the ${\text{La}_{2}\text{(Cu}_{x}\text {Ni}_{1-x})_7}$ samples in separate panels. The patterns are refined using GSAS-II. The difference between the observed and the calculated data is shown in blue in each panel. The wR to each data is mentioned. For $x \geq$ 0.022, there is an unindexed peak is from elemental La, the primary component of the solution the samples are grown from. The data for $x$ = 0.105 is the same as that in the main text in Fig \ref{fig:EDS}.}}
    \label{fig:xrd_all}
\end{figure*}

Importantly, these local moments do not arise from conventional atomic orbitals but rather from exponentially localized Wannier orbitals that respect all lattice and time-reversal symmetries. Exponential localization of Wannier orbitals is a necessary condition for local moment formation and is guaranteed when the flat bands are trivial.
However, when the individual bands are topological, as is typically the case with kagome flat bands in the presence of spin-orbit coupling, there is generally an obstruction to a local moment representation due to the internal Berry
curvature of the bands. Nevertheless, in cases pertinent to kagome and double kagome lattices, it is possible to “cancel” this obstruction by accommodating energetically nearby bands with the opposite Berry curvature. Resolving the topological obstructions via appropriate Wannierization facilitates a coherent Kondo lattice description, where the coupling between magnetic moments embedded in a non-kagome conduction electron environment can be effectively
tuned by doping.

\par

This study also raises a broader and crucial question regarding the viability of Kondo screening within metallic
antiferromagnets. Resolving whether the observed quantum criticality in ${\text{La}_{2}\text{(Cu}_{x}\text {Ni}_{1-x})_7}$ originates from spin-density wave/Landau order parameter-type fluctuations or local quantum criticality linked to Kondo breakdown is essential.
The emergence of the Kondo minimum precisely at the quantum critical point, rather than within the magnetically ordered phase, strongly suggests suppression of the Kondo effect in one kagome layer by the proximate magnetic order. Such an observation tends to favor local quantum critical scenarios over conventional Landau-type order parameter fluctuations. Nevertheless, alternative scenarios cannot be entirely discounted. Competing mechanisms, such as partial screening \cite{Vekhter2008} of Ni flat-band moments, or even full screening \cite{Ogata2007} under certain conditions, depend critically on the intricate interplay between Kondo physics and antiferromagnetic order. Further studies such as quantum oscillations and Hall effect measurements of Fermi volume jumps across the QCP could further narrow these possibilities.

\par

Another important aspect is the relative magnitude and interplay of key energy scales – specifically, the coherence temperature $(T_{coh})$ versus the Kondo temperature $(T_{K})$. Elucidating the conditions under which lattice coherence effects dominate over individual Kondo screening and vice-versa in the individual kagome layers can provide valuable insights into the underlying physics, a topic highlighted by previous studies \cite{Zlatic2009, Shim2019, Shim2020, SI200623, coleman2010frustration}.

\par

Within the broader global phase diagram for Kondo lattice materials, the suppression of magnetism in ${\text{La}_{2}\text{(Cu}_{x}\text {Ni}_{1-x})_7}$ could be driven either by variations in the Kondo coupling strength $(J_K)$ or through frustration and dimensionality effects (parameterized by $G$) \cite{Paschen-Si2021}. Identifying the exact positioning of ${\text{La}_{2}\text{(Cu}_{x}\text {Ni}_{1-x})_7}$ within this phase diagram would
substantially enhance the understanding of its quantum critical behavior. Furthermore, transport measurements rule out significant contributions from weak localization or other electron-electron interactions, sharpening the focus on the unique interplay among Kondo phenomena, flat-band physics, and magnetism.

Regardless of the origin of the single-ion Kondo-like resistivity behavior for $x > 0.105$, or perhaps because of it, our detailed work on the ${\text{La}_{2}\text{(Cu}_{x}\text {Ni}_{1-x})_7}$ system firmly identifies it as an interesting family for the study of d-shell based AFM-QCP. In addition, this system offers the possibility of examining the role of Ni-Kagome-layers in the moment formation, ordering and possible single ion Kondo scattering. 

\par

\section*{ACKNOWLEDGEMENTS}
\label{sec:ACKNOWLEDGEMENTS}

We would like to thank R. Flint for useful discussions. All EDS measurements was performed using instruments in the Sensitive Instrument Facility in Ames National Laboratory. This work was done at Ames National Laboratory and supported by the U.S. Department of Energy, Office of Science, Basic Energy Sciences, Materials Sciences and Engineering Division. Ames National Laboratory is operated for the U.S. Department of Energy by Iowa State University under Contract No. DE-AC02-07CH11358. 

\section*{DATA AVAILABILITY}
\label{sec:DATA AVAILABILITY}

The data that supports the findings are openly available in \cite{das_2025_17643700}.

\section*{APPENDIX}
\label{sec:Appendix}

Powder x-ray patterns of all the Cu substituted samples of ${\text{La}_{2}\text{(Cu}_{x}\text {Ni}_{1-x})_7}$ were refined (unit cell, sample displacement,etc.) using the Rietveld method and the lattice parameters ($a,~c, \text{V}$) were obtained. In the main text we have shown the change in these parameters with $x$ in Fig \ref{fig:EDS}. Here we present the values of lattice parameters ($a,~c, \text{V}$) and the associated uncertainties in Table \ref{tab:lattice_La2Ni7}. Figure \ref{fig:xrd_all} shows the observed x-ray patterns after refinement and the residual showing the goodness of the refinement. $wR$ also helps to parametrize the quality of the refinement. The intensities in the powder observed and the calculated XRD spectra pattern do not match well, which might either be due to preferential orientation in the powder sample, or due to the doped Cu not distributed randomly on the Ni sites. To address the Cu distribution in the samples, we tried to refine the pattern by assuming that Cu occupies a preferred Wyckoff position. But given the nearly identical scattering strengths of Cu and Ni, when we attempted this for the x = 0.125 sample, we did not observe significant improvement in the refinement. For $x \geq$ 0.022, there is an unindexed peak is from elemental La, the primary component of the solution the samples were grown from. The data for $x$ = 0.105 were already shown in the main text.

\par

% Table generated by Excel2LaTeX from sheet 'lattice parameters'
\begin{table}[h!]
  \centering
  \setlength{\tabcolsep}{2.8pt}
  \renewcommand{\arraystretch}{2.0}
    \begin{tabular}{r|rrrr}
    \hline
    \hline
    \multicolumn{1}{c|}{$x$} & \multicolumn{1}{c}{$a (\textrm{~\AA})$} & \multicolumn{1}{c}{$c (\textrm{~\AA})$} & \multicolumn{1}{c}{$V (\textrm{~\AA}^{3}$)} & \multicolumn{1}{c}{$\textit{wR}$}\\
    \hline
    %\midrule
    0 & 5.0585(11) & 24.667(3) & 545.36(19) & 9.62 \\
    0.012(2) & 5.0588(29) & 24.667(9) & 546.82(26) & 10.89 \\
    0.018(4) & 5.0624(15) & 24.687(4) & 547.93(22) & 9.65 \\
    0.022(9) & 5.0602(9) & 24.679(3) & 547.27(16) & 10.13 \\
    0.036(5) & 5.0654(8) & 24.681(3) & 548.38(17) & 11.62 \\
    0.056(6) & 5.0655(9) & 24.707(3) & 549.03(56) & 9.09 \\
    0.075(6) & 5.0682(18) & 24.713(6) & 549.76(31) & 9.83 \\
    0.097(6) & 5.0719(7) & 24.719(7) & 550.48(13) & 10.20 \\
    0.105(3) & 5.0729(27) & 24.727(4) & 551.09(21) & 8.43 \\
    0.125(7) & 5.0729(15) & 24.732(1) & 551.11(25) & 8.60 \\
    0.152(6) & 5.0811(12) & 24.748(4) & 553.35(14) & 6.95 \\
    0.181(11) & 5.0899(16) & 24.780(3) & 555.68(24) & 12.58 \\
    %\bottomrule
    \hline
    \hline
    \end{tabular}%
    \caption{\footnotesize{The values of the lattice parameters $a$, $c$, and the volume of the unit cell V of the different $x$ in ${\text{La}_{2}\text{(Cu}_{x}\text {Ni}_{1-x})_7}$ with the values of $x$ obtained from EDS measurements in the leftmost column. As ${\text{La}_{2}\text{(Cu}_{x}\text {Ni}_{1-x})_7}$ manifests a hexagonal unit cell, the lattice parameters $a$ and $b$ are equal. The associated uncertainties are indicated in the respective parentheses. The wR parameter is the goodness of the fit obtained.}}
  \label{tab:lattice_La2Ni7}%
\end{table}%

\begin{figure*}[htbp]
    \centering
    \includegraphics[width=1.0\textwidth]{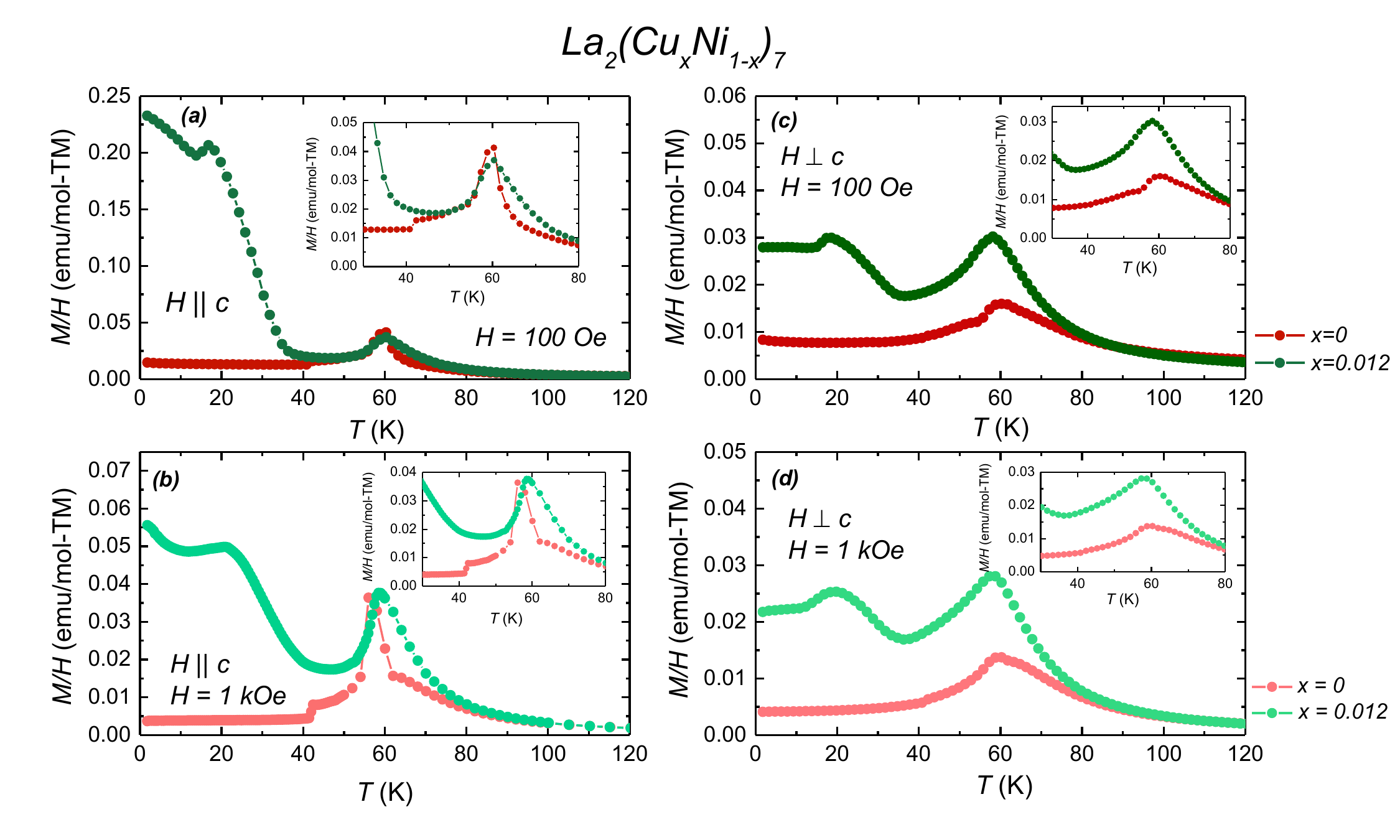}
    \caption{\footnotesize{(Color online) Comparison of the anisotropic $M(T)$ data for $x$ = 0 and 0.012. $(a)$ $M(T)$ data at $H$ = 100 Oe for $H||c$ direction. $(b)$ $M(T)$ data at $H$ = 1 kOe for $H||c$ direction. $(c)$ $M(T)$ data at $H$ = 100 Oe for $H \perp c$ direction. $(d)$ $M(T)$ data at $H$ = 1 kOe for $H \perp c$ direction. The insets to each panel show the data in the temperature region where we observe the transition features.}}
    \label{fig:MT_0 and 1}
\end{figure*}

\begin{figure*}[htbp]
    \centering
    \includegraphics[width=\textwidth]{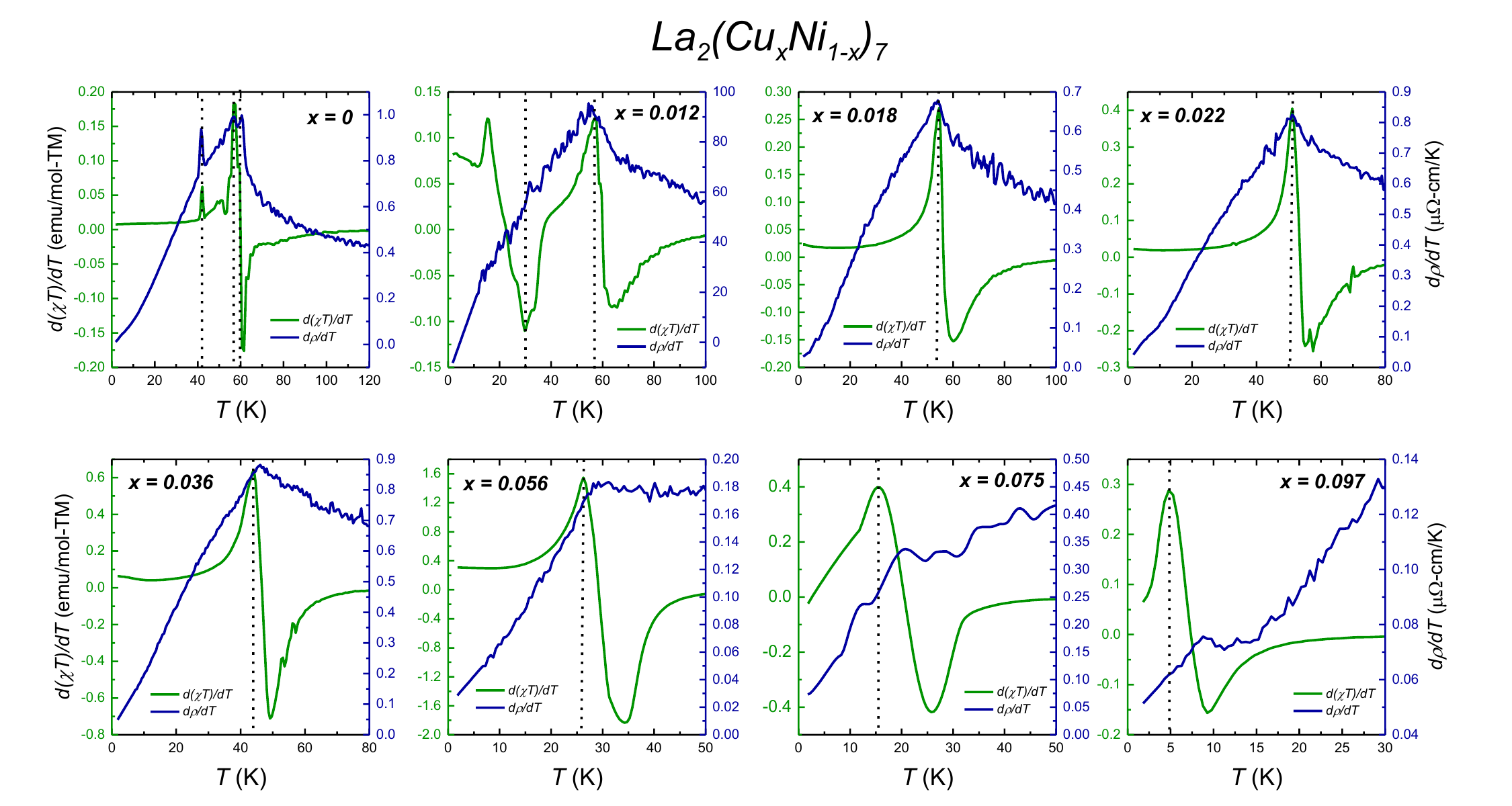}
    \caption{\footnotesize{(Color online) Plot of polycrystalline averaged $\frac{d(\chi T)}{dT}$ at an external field of $H$=100 Oe and in-plane, zero field $\frac{d \rho}{dT}$, for the ${\text{La}_{2}\text{(Cu}_{x}\text {Ni}_{1-x})_7}$ single crystals which show magnetic transitions (Fig. \ref{fig:M(T)_100 Oe}) and (Fig. \ref{fig:R-T separate}). The AFM magnetic transition temperatures, $T_N$ for these Cu doped samples are determined using both polycrystalline $\frac{d(\chi T)}{dT}$ and $\frac{d \rho}{dT}$ (See text for details). The vertical line(s) in each panel is the $T_N$ determined from the $M(T)$ data and agree with that obtained from the $\rho(T)$ data especially for the lower $x$-values. The parent compound ($x$=0) and lowest doped ($x$=0.012) show signs of multiple transitions whereas the other doped systems show only one.}}
    \label{fig:TNeel}
\end{figure*}

For the parent $x$ = 0, there are three AFM phases as was deduced from previous work \cite{Ribeiro2022Small-moment/math, Wilde2022Weak/math} as well as our measurement results (Fig. \ref{fig:M(T)_100 Oe}). For the lowest doped sample, $x$ = 0.012, we still observe multiple phase associated features in the low field $M(T)$, zero field $\rho(T)$ and $C_p(T)$ data. However, when we compare the $M(T)$ data for these two samples, there is a distinct difference in their behavior. This is especially evident when measurements are done at a field of $H$ = 1 kOe in the easy direction of $H || c$ (Fig. \ref{fig:MT_0 and 1}). The $B$ and $C$ phases of the parent compound appear to merge together and we observed a much broader feature for the doped sample. This feature is qualitatively and quantitatively similar to the phase $B$. In addition, the behavior of phase $A$ in $x$ = 0 and the lowest temperature feature in $x$ = 0.012 is different, with the feature in the doped sample having a larger low temperature value. As of now, we cannot determine how the magnetism evolves in between these two samples. Further measurements, such as neutron diffraction, will be needed to better understand the evolution of magnetic ordering with low Cu substitution in this system.

\begin{figure}[h]
    \centering
    \includegraphics[width=\linewidth]{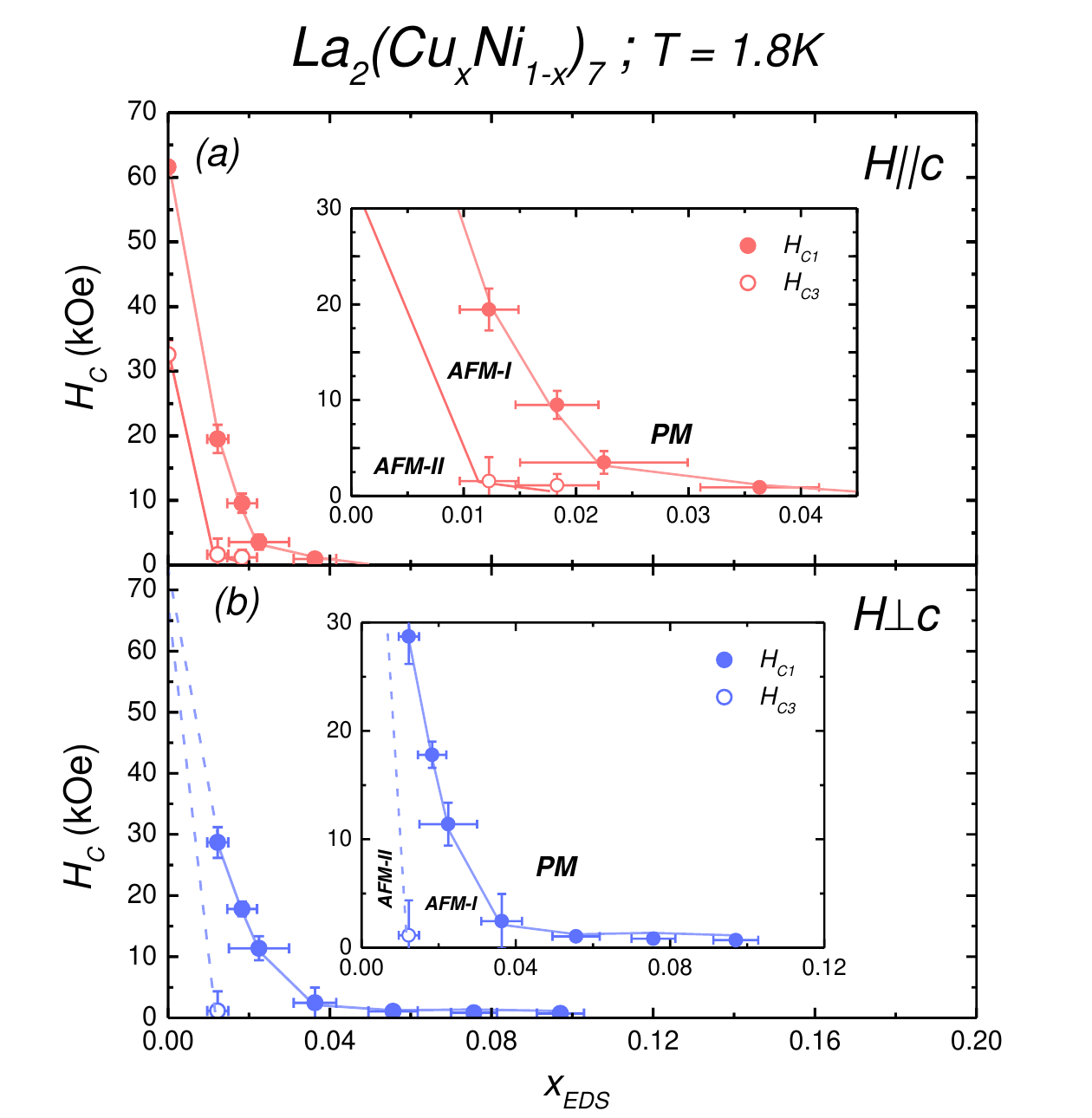}
    \caption{The metamagnetic transition field $H_C$ vs $x$, obtained from the $M(H)$ data measured at T=1.8 K for both (a) $H||c$ and (b) $H \perp c$ for ${\text{La}_{2}\text{(Cu}_{x}\text {Ni}_{1-x})_7}$ single crystals. The data has been measured for increasing magnetic field. The insets to both the panels show the low $x$ behavior up to which $H_C$ is suppressed. For both $H||c$ and $H \perp c$ direction, we observe two magnetically ordered regimes which are labeled as $(AFM-I)$ and $(AFM-II)$ areas. The lines are drawn as guide to the eye. $H_{C1}$ and $H_{C3}$ for $x=0$ for $H \perp c$ direction is $>$ 70 kOe as shown in \cite{Ribeiro2022Small-moment/math} and is not marked in this plot. We extrapolate the guide lines (dashed lines) for this direction to H$>$ 70 kOe for $x=0$. }
    \label{fig:H-x}
\end{figure}

\par

\begin{figure}[h!]
    \centering
    \includegraphics[width=1.02\linewidth]{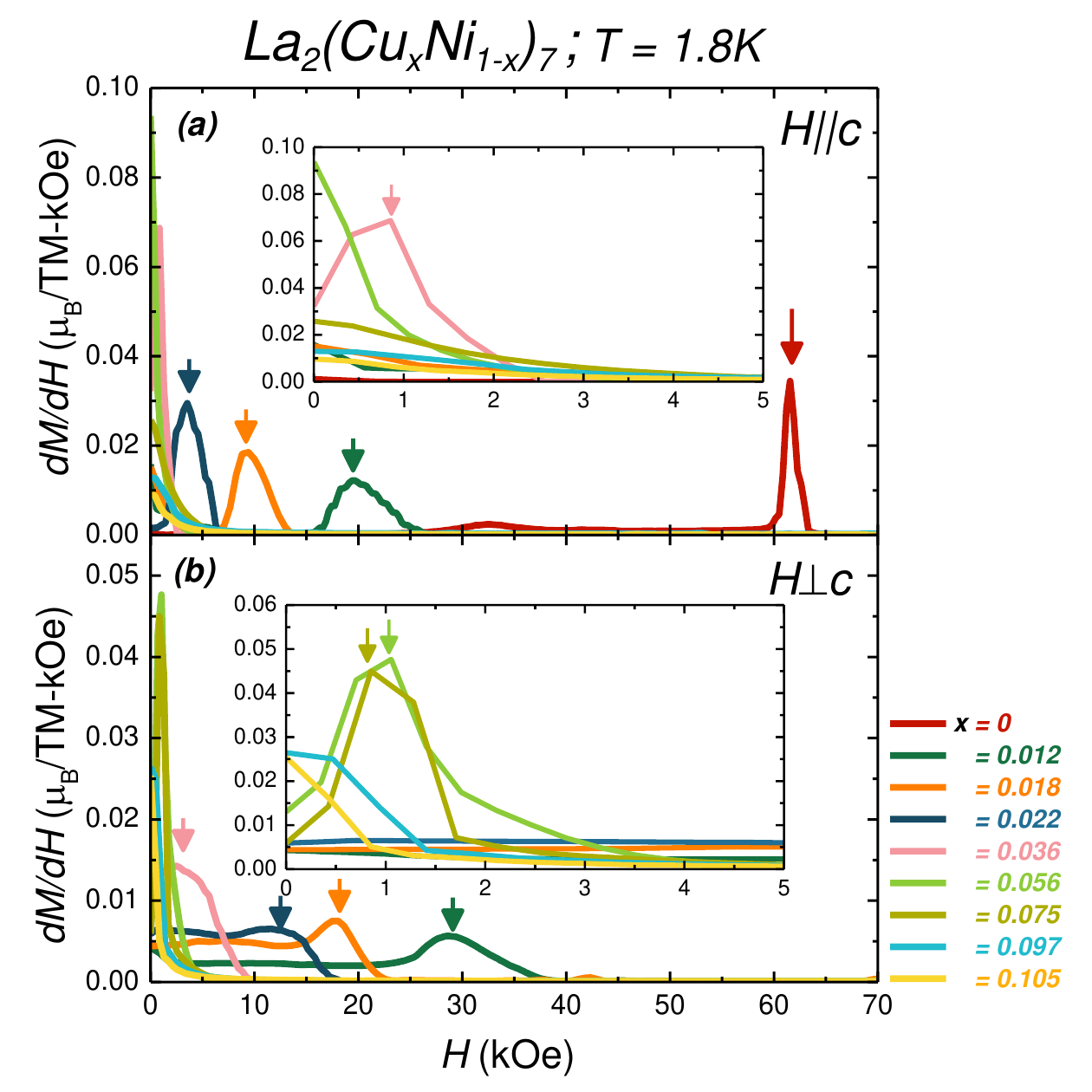}
    \caption{\footnotesize{(Color online) Derivative of the anisotropic $M(H)$ with respect to $H$ measured at $T$ = 1.8 K for the magnetically ordered single crystals of ${\text{La}_{2}\text{(Cu}_{x}\text {Ni}_{1-x})_7}$, for both $H || c$ in panel $(a)$, and $H \perp c$ in panel $(b)$ respectively. For $H \perp c$ direction. The inset to both the panels shows the low field behavior for clarity, especially for the higher doped samples. }}
    \label{fig:dM/dH}
\end{figure}

\par

The magnetic transition temperature $T_N$ of the lower doped ${\text{La}_{2}\text{(Cu}_{x}\text {Ni}_{1-x})_7}$ samples (0 $\leq x \leq$ 0.097) is determined from the  $\frac{d(\chi T)}{dT}$ at 100 Oe, zero field $\frac{d\rho}{dT}$ and zero field $C_p(T)$ data. Fig \ref{fig:TNeel} shows the polycrystalline averaged $\frac{d(\chi T)}{dT}$ obtained by measuring anisotropic $M(T)$ at $H$ = 100 Oe. Not only the signatures of multiple transitions for $x$ = 0 and 0.012, but also the single transition for the other substitutions are clearly observed in the data. The value of $T_N$ obtained from $M(T)$ is similar to that obtained from the $\rho(T)$ measurements; we have included the $\frac{d\rho}{dT}$ (originally shown in Fig. \ref{fig:R-T separate} in the main text) data for easier comparison.

Field dependent $M(H)$ measured at the base temperature ($T$ = 1.8 K) (Fig \ref{fig:M(H)}) has signatures of metamagnetic transitions for 0 $\leq x \leq$ 0.097 samples. The presence of metamagnetic transitions in the $M(H)$ data confirms an antiferromagnetic magnetic order in the system at this temperature. We track the change of the metamagnetic transition field $H_C$ with $x$ in Fig. \ref{fig:H-x}. We will refer to the higher field transition as $H_{C1}$ and the lower field one as $H_{C3}$. $H_C$ is obtained from the peak in the $\frac{dM}{dH}$ plot which is shown in Fig \ref{fig:dM/dH}.

\begin{figure}[h]
    \centering
    \includegraphics[width=\linewidth]{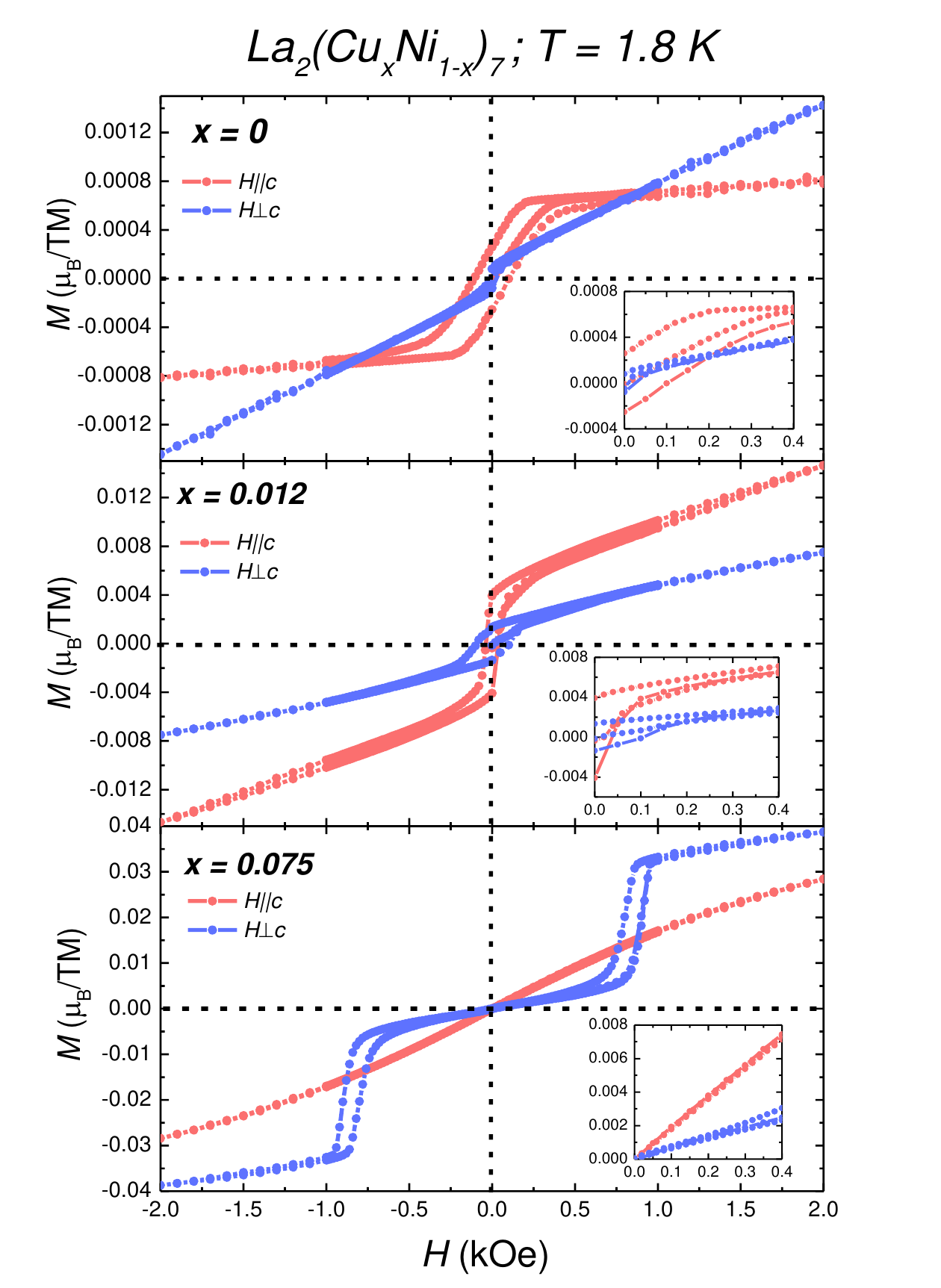}
    \caption{\footnotesize{5 segment $M(H)$ measured at $T$ = 1.8 K for the parent ($x$ = 0) as well as $x$ = 0.012 and 0.075 ${\text{La}_{2}\text{(Cu}_{x}\text {Ni}_{1-x})_7}$ for both $H || c$ and $H \perp c$ directions. (Inset to each panel shows the zoomed in data to show the low field feature.}}
    \label{fig:5 segment}
\end{figure}

\begin{figure}
    \centering
    \includegraphics[width=\linewidth]{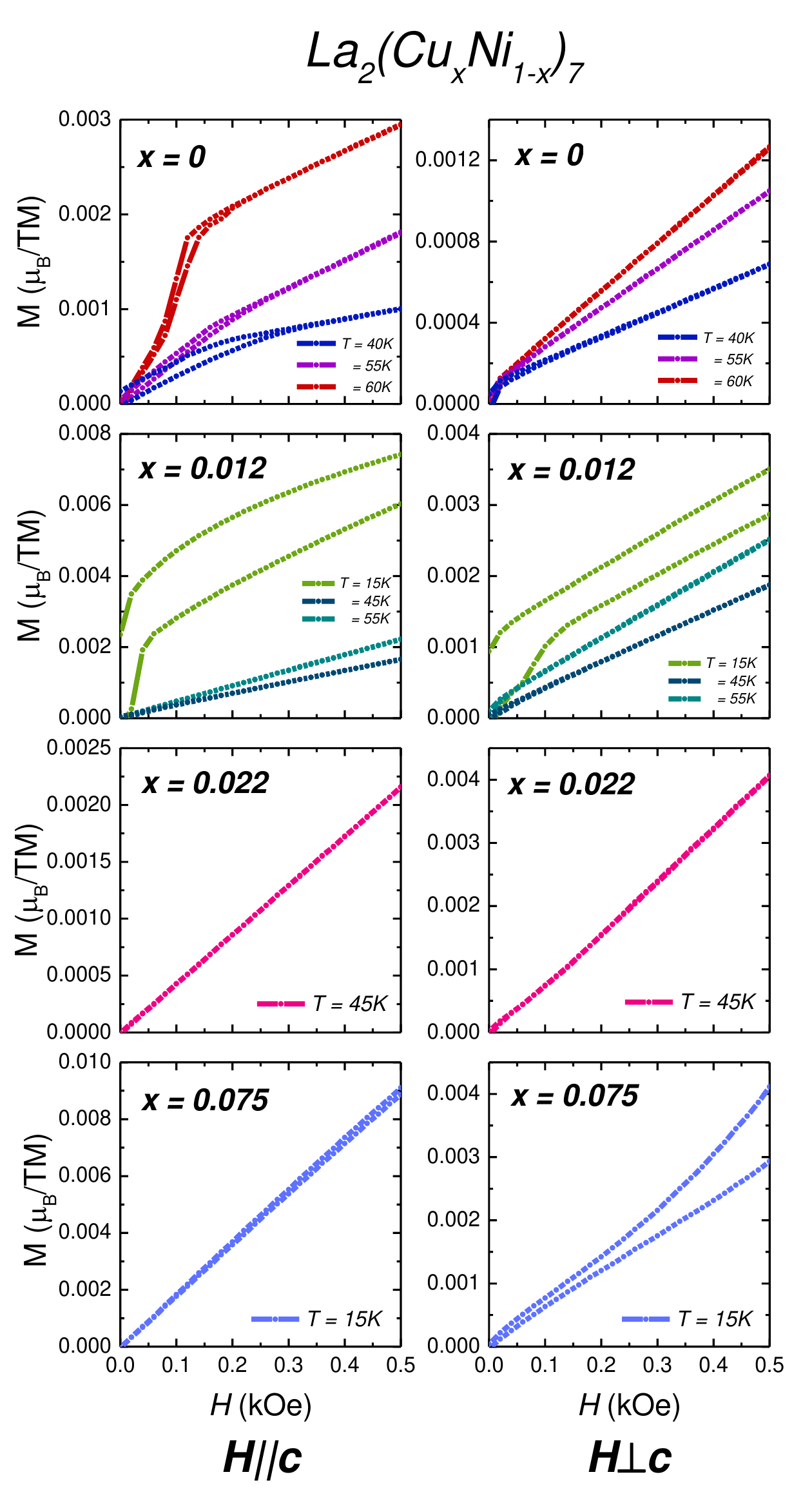}
    \caption{\footnotesize{$M(H)$ measured at temperatures slightly below the transition temperatures for $x$ = 0, 0.012, 0.022, and 0.075 for both $H || c$ and $H \perp c$ directions. $M(H)$ is measured for both increasing and decreasing field directions to test for possible FM component to the AFM transition. }}
    \label{fig:FM_test}
\end{figure}

\begin{figure}
    \centering
    \includegraphics[width=\linewidth]{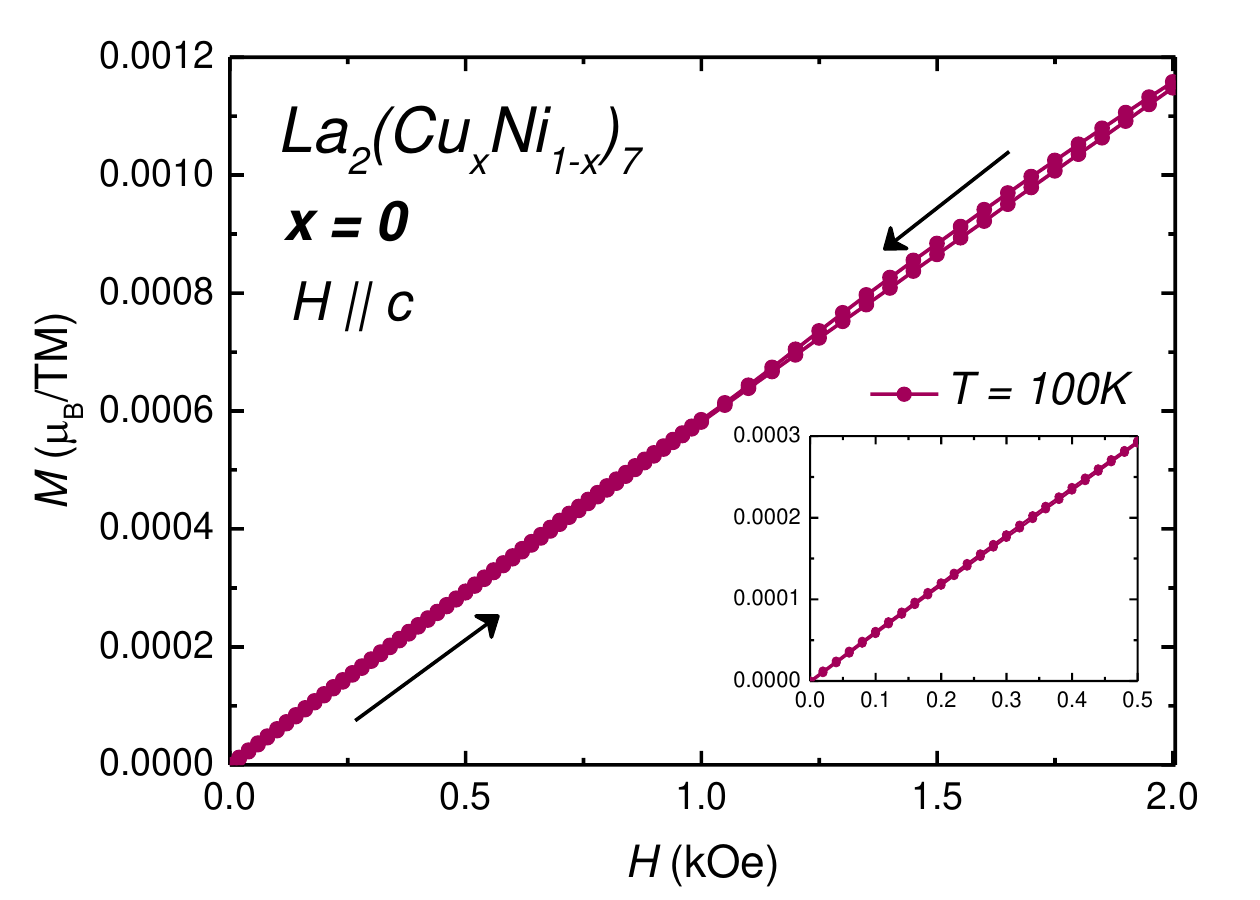}
    \caption{\footnotesize{$M(H)$ for the parent $x$ = 0 measured at $T$ = 100 K for $H || c$ direction. The data was measured for both increasing and decreasing fields. (Inset: zoomed in data to show the low field behavior for clarity.) }}
    \label{fig:MH_parent_100K}
\end{figure}

\begin{figure}[h]
    \centering
    \includegraphics[width=\linewidth]{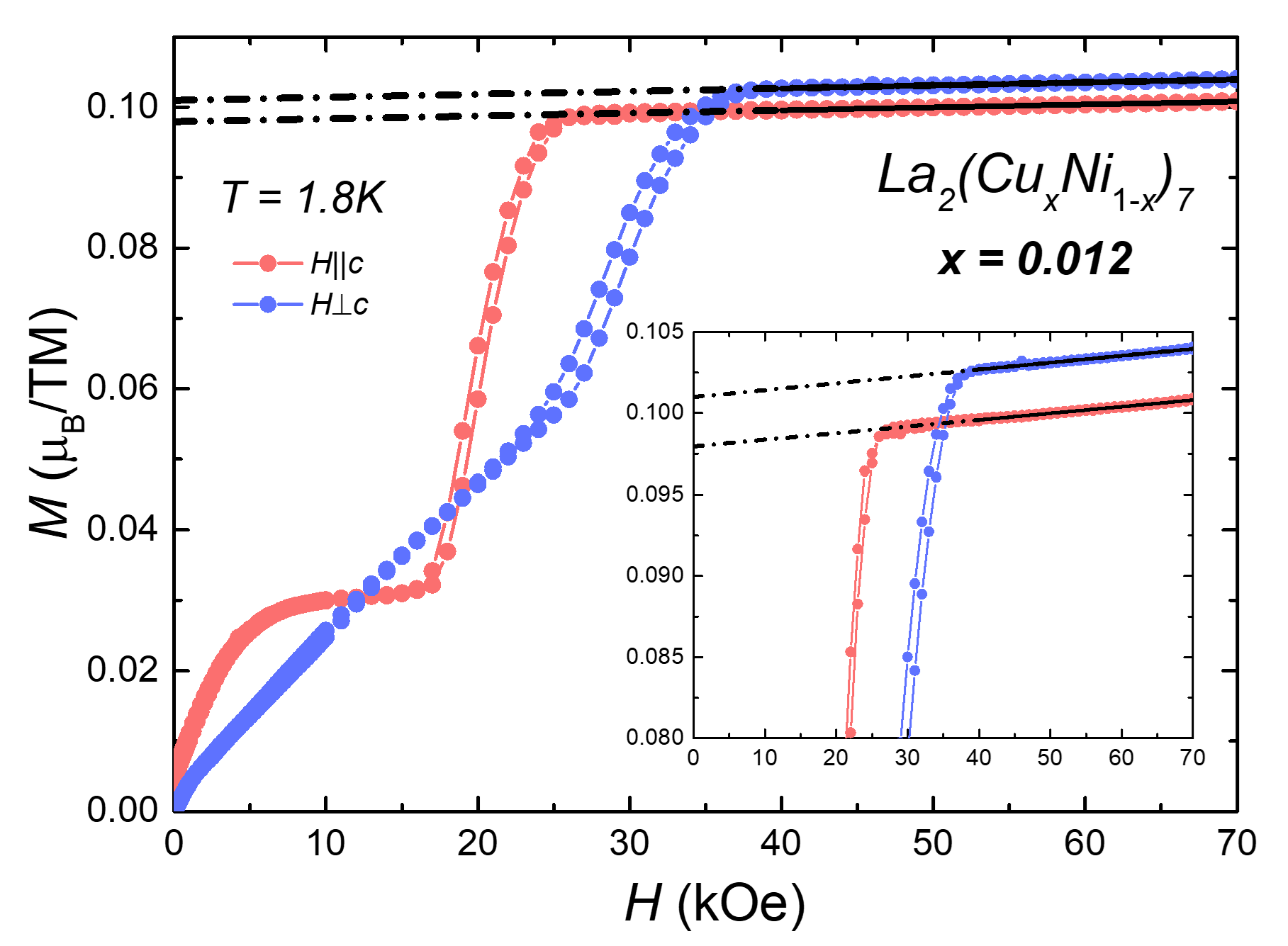}
    \caption{\footnotesize{The extrapolated moment $\mu_s$, obtained by extrapolating the high-field behavior to zero field for $x$ = 0.012 for both $H || c$ and $H \perp c$. The black line intercepting the $y-$ axis is the value of the extrapolated moment. $\mu_s$ for all other $x$ in ${\text{La}_{2}\text{(Cu}_{x}\text {Ni}_{1-x})_7}$ are also obtained in the same manner. The inset shows the zoomed in y-axis to highlight the higher field behavior.}}
    \label{fig:extrapolated}
\end{figure}

\begin{table}[htbp]
  \centering
  \setlength{\tabcolsep}{1.0pt}
  \renewcommand{\arraystretch}{1.5}
    \begin{tabular}{r|rrrrr}
    \hline
    \hline
    \multicolumn{1}{c|}{$x$} & \multicolumn{2}{c|}{$\mu_S$} & \multicolumn{1}{c}{$\mu_{eff}$} & \multicolumn{1}{c}{$\Theta_{CW}$} & \multicolumn{1}{c}{$\chi_0 (\times 10^{-5})$}\\
    \multicolumn{1}{c|}{} & \multicolumn{2}{c|}{$(\mu_B/TM)$} & \multicolumn{1}{c}{$(\mu_B/TM)$} & \multicolumn{1}{c}{$(K)$} & \multicolumn{1}{c}{emu/mol-TM}\\
    \hline
    \multicolumn{1}{c|}{} & \multicolumn{2}{c|}{$T$=1.8 K} & \multicolumn{1}{c}{} & \multicolumn{1}{c}{$H$=20 kOe} & \multicolumn{1}{c}{}\\
    \hline     
    \multicolumn{1}{c|}{} & \multicolumn{1}{c}{$H || c$} & \multicolumn{1}{c|}{$H \perp c$} & \multicolumn{1}{c}{} & \multicolumn{1}{c}{poly avg} & \multicolumn{1}{c}{}\\
    \hline    
    %\midrule
    \multicolumn{1}{c|}{0} & \multicolumn{1}{c}{~0.092(7)} & \multicolumn{1}{c|}{-} & \multicolumn{1}{c}{0.84(9)} & \multicolumn{1}{c}{-71.5(3)} & \multicolumn{1}{c}{-4.1(8)}\\
    \multicolumn{1}{c|}{0.012} & \multicolumn{1}{c}{~0.097(3)} & \multicolumn{1}{c|}{~0.100(4)} & \multicolumn{1}{c}{0.80(6)} & \multicolumn{1}{c}{-75.1(2)} & \multicolumn{1}{c}{-3.4(3)}\\
    \multicolumn{1}{c|}{0.018} & \multicolumn{1}{c}{~0.099(2)} & \multicolumn{1}{c|}{~0.105(3)} & \multicolumn{1}{c}{0.78(9)} & \multicolumn{1}{c}{-75.4(4)} & \multicolumn{1}{c}{-2.1(6)}\\
    \multicolumn{1}{c|}{0.022} & \multicolumn{1}{c}{~0.092(2)} & \multicolumn{1}{c|}{~0.096(2)} & \multicolumn{1}{c}{0.75(8)} & \multicolumn{1}{c}{-74.4(3)} & \multicolumn{1}{c}{-1.1(5)}\\
    \multicolumn{1}{c|}{0.036} & \multicolumn{1}{c}{~0.088(8)} & \multicolumn{1}{c|}{~0.091(1)} & \multicolumn{1}{c}{0.72(7)} & \multicolumn{1}{c}{-71.6(3)} & \multicolumn{1}{c}{4.2(4)}\\
    \multicolumn{1}{c|}{0.056} & \multicolumn{1}{c}{~0.074(1)} & \multicolumn{1}{c|}{~0.073(2)} & \multicolumn{1}{c}{0.66(6)} & \multicolumn{1}{c}{-61.7(4)} & \multicolumn{1}{c}{3.9(2)}\\
    \multicolumn{1}{c|}{0.075} & \multicolumn{1}{c}{~0.054(1)} & \multicolumn{1}{c|}{~0.056(1)} & \multicolumn{1}{c}{0.57(12)} & \multicolumn{1}{c}{-45.9(15)} & \multicolumn{1}{c}{6.3(9)}\\
    \multicolumn{1}{c|}{0.097} & \multicolumn{1}{c}{~0.029(4)} & \multicolumn{1}{c|}{~0.036(4)} & \multicolumn{1}{c}{0.53(4)} & \multicolumn{1}{c}{-32.1(4)} & \multicolumn{1}{c}{8.0(7)}\\
    \multicolumn{1}{c|}{0.105} & \multicolumn{1}{c}{~0.022(5)} & \multicolumn{1}{c|}{~0.023(5)} & \multicolumn{1}{c}{0.49(5)} & \multicolumn{1}{c}{-17.7(6)} & \multicolumn{1}{c}{6.9(8)}\\
    \multicolumn{1}{c|}{0.125} & \multicolumn{1}{c}{~0.014(4)} & \multicolumn{1}{c|}{~0.014(5)} & \multicolumn{1}{c}{0.47(4)} & \multicolumn{1}{c}{-11.9(14)} & \multicolumn{1}{c}{8.7(8)}\\
    \multicolumn{1}{c|}{0.152} & \multicolumn{1}{c}{~0.004(4)} & \multicolumn{1}{c|}{~0.004(4)} & \multicolumn{1}{c}{0.29(4)} & \multicolumn{1}{c}{-13.7(17)} & \multicolumn{1}{c}{13.9(7)}\\
    \multicolumn{1}{c|}{0.181} & \multicolumn{1}{c}{0.001(3)} & \multicolumn{1}{c|}{0.001(4)} & \multicolumn{1}{c}{-} & \multicolumn{1}{c}{-} & \multicolumn{1}{c}{-}\\
    %\bottomrule
    \hline
    \hline
    \end{tabular}%
    \caption[\footnotesize]{The values of the extrapolated moment $\mu_S$, effective moment $\mu_{eff}$, Weiss temperature $\Theta_{CW}$ and $\chi_0$ obtained from the magnetization measurements on the ${\text{La}_{2}\text{(Cu}_{x}\text {Ni}_{1-x})_7}$ single crystals. $\mu_S$ is obtained from $M(H)$ measurements at 1.8 K and has distinct values along $H||c$ and $H \perp c$ whereas the other parameters ($\mu_{eff}$, $\Theta_{CW}$ and $\chi_0$) are obtained from the Curie-Weiss fit to the polycrystalline average of the $M(T)$ data taken at an applied field of 20 kOe. The uncertainties associated with each parameter is shown in the respective parentheses.} 
  \label{tab:magnetic}%
\end{table}%

\begin{figure*}[htbp]
    \centering
    \includegraphics[width=\textwidth]{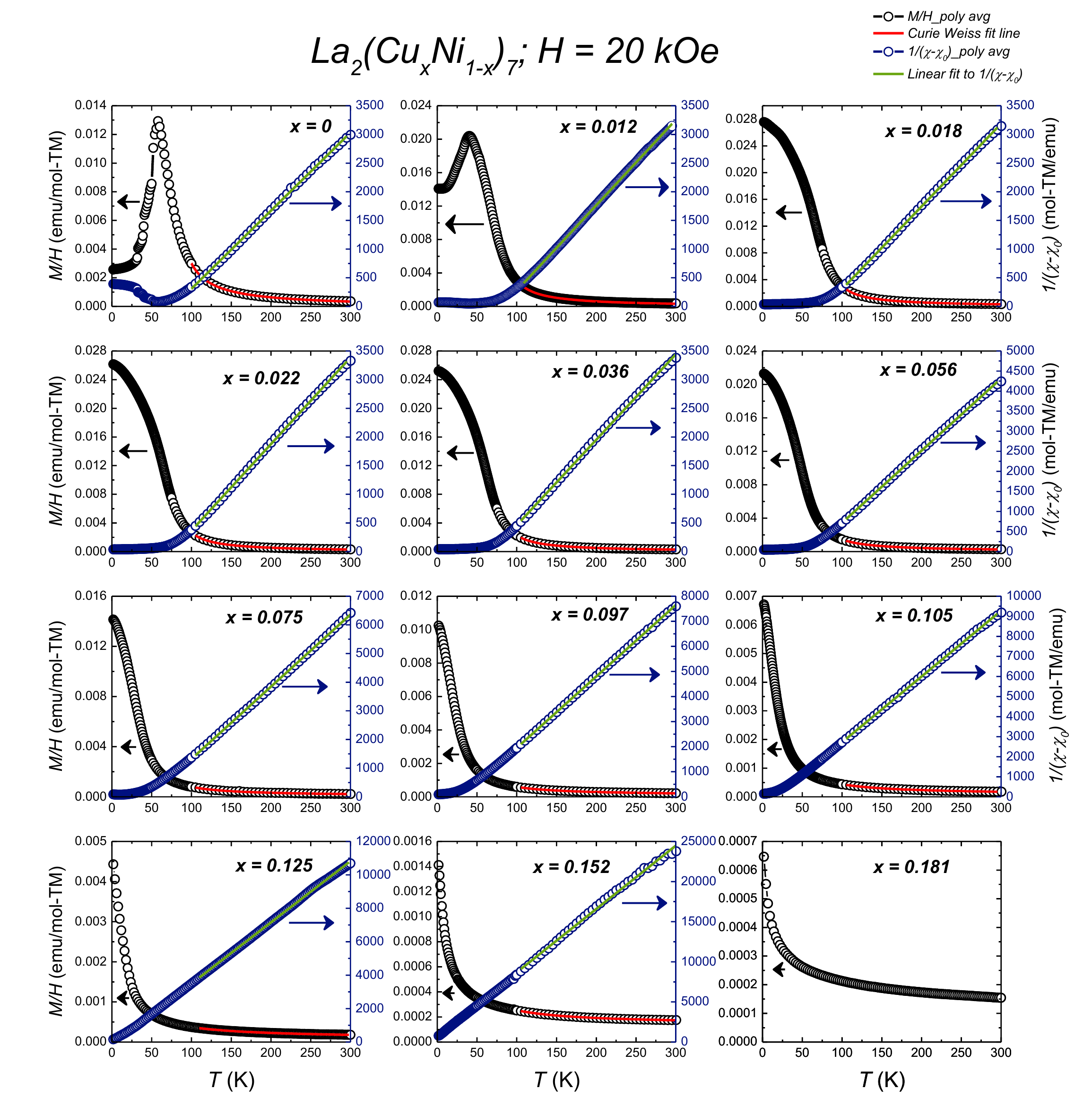}
    \caption{\footnotesize{Temperature dependent magnetization $(M(T))$ measured at $H$ = 20 kOe presented as the polycystalline average along the left axis for all ${\text{La}_{2}\text{(Cu}_{x}\text {Ni}_{1-x})_7}$ single crystals and $(\chi - \chi_0)^{-1}$ along the right axis for 0 $\leq x \leq$ 0.152 samples shown in separate panels. The procedure to obtain the poly avg $(M(T))$ data is described in the main text. A Curie-Weiss like fit, as described by Eqn \ref{eqn:CW fit} is used in the high temperature regime (100 K $\leq T \leq $ 300 K) of the $M(T)$ data for 0 $\leq x \leq$ 0.152, ${\text{La}_{2}\text{(Cu}_{x}\text {Ni}_{1-x})_7}$ samples and the fit line is shown in red. The obtained value of $\chi_0$ from the fit is used to plot $(\chi - \chi_0)^{-1}$ and is expected to have linear behavior. A linear fit in the same temperature range (100 K $\leq T \leq $ 300 K) is done to the $(\chi - \chi_0)^{-1}$ plots and is shown using a green line.}}
    \label{fig:CW data}
\end{figure*} 

From the anisotropic $H_C-x$ phase diagram, we observe that $H_C$ is suppressed by $x \sim$ 0.04 for $H || c$ whereas at a higher doping level of $x \sim$ 0.10 for $H \perp c$. However for both directions we observe a lower field AFM region suppressed by $x \sim$ 0.012. The two magnetically ordered regimes are shown in Fig \ref{fig:H-x}. For both directions, the higher [lower] field region will be referred as $AFM-I$ [$AFM-II$] respectively. %The behavior of $H_C-x$ phase diagram for ${\text{La}_{2}\text{(Cu}_{x}\text {Ni}_{1-x})_7}$ is qualitatively similar to that of the magnetically ordered regime of the $T-x$ phase diagram (Fig \ref{fig:phase diagram}). 
The low dopings (0 $\leq x \leq$ 0.012) exhibit more than one ordered regime ($A,B$ and $C$ in the $T-x$ and $AFM-I, AFM-II $) in the $H_C-x$) and the higher Cu doped samples exhibit only a single AFM regime. Neutron diffraction studies will be important to determine the exact character of the magnetic ordering as a function of $x$ and $T$. 

\par

Given that we find finite ZFC-FC splitting in the 100 Oe $M(T)$ data as shown in Fig.\ref{fig:M(T)_100 Oe} for lower $x$ concentrations, at low temperatures, we measured a 5-quadrant $M(H)$ at the base temperature of 1.8 K for $x$ = 0, 0.012, and 0.075 for both $H || c$ and $H \perp c$ directions which is shown in Fig \ref{fig:5 segment}. The data for $x$ = 0 and 0.012 have a more significant hysteresis in their base temperature $M(H)$ isotherms for both directions whereas for $x$ = 0.075 no observable hysteresis is seen at $H$ = 0. It should be noted that as the low field extrapolation of $H$ goes to zero, saturated moment associated with the apparent FM component is 0.0006 $\mu_B /TM$ for the parent $x$=0. This is two orders of magnitude smaller than the weak FM component inferred for the higher temperature $C$ phase ($\sim 0.05 \mu_B /TM$) \cite{Ribeiro2022Small-moment/math}. As such then, this seems to be a remarkably small component. We speculate that this may well be associated with dislocations of defects in the non-trivial, long wavelength antiferromagnetic order. 

\par

\begin{figure*}[htbp]
    \centering
    \includegraphics[width=\textwidth]{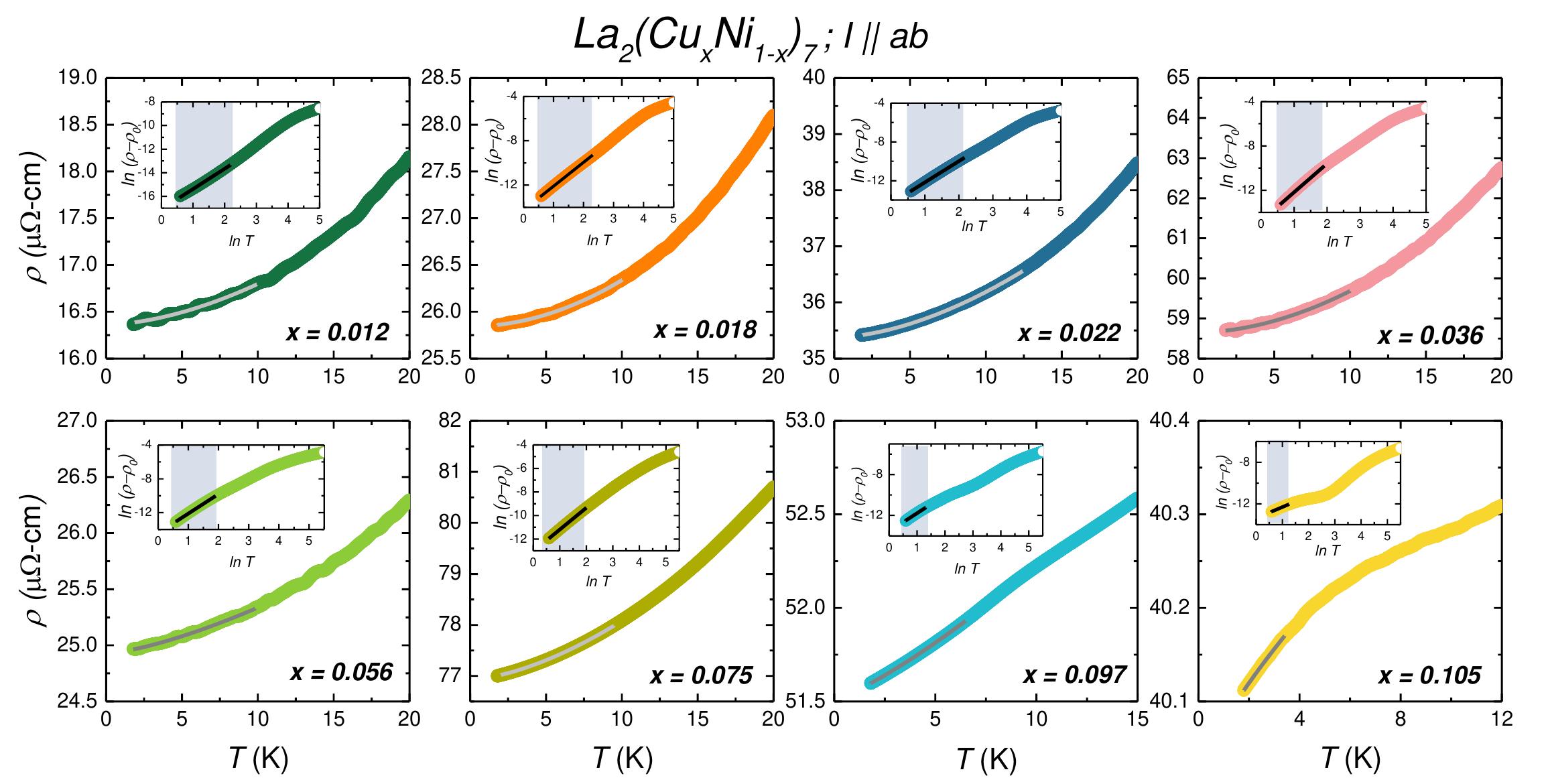}
    \caption{\footnotesize{Low temperature $\rho(T)$ data for 0.012 $\leq x \leq$ 0.105 ${\text{La}_{2}\text{(Cu}_{x}\text {Ni}_{1-x})_7}$ samples, shown using colored curves and the fit to Eqn. \ref{eqn:power law} shown using a grey line. We obtained $\rho_0$ from these fits and used to plot $\ln{(\rho - \rho_{0})}$ vs $\ln{T}$ for each of these substitutions as shown in the respective insets. A linear fit corresponding to Eqn. \ref{eqn: log power_law}, shown in black is done to the $\ln{(\rho - \rho_{0})}$ vs $\ln{T}$ plots. The slope and the y-intercept of these linear fits give us values of $n$ and $A$. The temperature range over which we do the linear fit is shown using the grey area in the inset.}}
    \label{fig:power law all}
\end{figure*}

Figure \ref{fig:FM_test} addresses the question whether there is a FM component to the AFM magnetic ordering that is found just below the highest (or only) transition temperature for representative Cu substituted samples. We measure $M(H)$ at temperatures slightly below the AFM transition temperatures for $x$ = 0, 0.012, 0.022, and 0.075 for both increasing and decreasing field directions. After demagnetization, the sample was cooled and then data was collected for increasing and then decreasing the field to 0. For $x$ = 0, and 0.012, although the most significant split in the data occurs at the lowest $T$ phase, a much smaller but observable split is seen for the higher $T$ states also for both $H || c$ and $H \perp c$ directions. For the two higher dopings $x$ = 0.022, and 0.075, the situation is slightly different. We do not observe any split in the data for $x$ = 0.022 taken at $T$ = 45 K but a small split is seen for $x$ = 0.075 measured at $T$ = 15 K only for the $H \perp c$ direction.

\par

$M(H)$ for the parent ${\text{La}_{2}\text {Ni}_7}$ was also measured at the paramagnetic regime at a temperature of $T$ = 100 K and to confirm the absence of any magnetic impurity with a $T_C$ higher than 100 K that could lead to the small FM signal we find. Figure \ref{fig:MH_parent_100K} shows the $M(H)$ for $x$ = 0 measured at $T$ = 100 K for both increasing and decreasing fields for $H || c$ direction only. The data has no finite moment and resembles that of a system in non-FM state suggesting that the FM component is intrinsic in nature. Thus, from the measurements shown in Figs \ref{fig:5 segment}-\ref{fig:MH_parent_100K} we get an indication that all three magnetic phases ($A$, $B$, and $C$) for $x$=0, the multiple phases for $x=$ 0.012, and phase $B$ for the other Cu-doped samples may have a small FM component associated with the primary AFM phase. We may speculate this arises due to AFM domain wall and stacking faults in the long wavelength of the AFM ordered phase. 

\par

The extrapolated moment, $\mu_s$ can be estimated as a proxy for the saturated moment in an itinerant magnetic system. We obtain $\mu_s$ by doing a linear fit to the high field (40 kOe $\leq H \leq$ 70 kOe) $M(H)$ data, except for the parent $x$ = 0 where we used 60 kOe $\leq H \leq$ 70 kOe as the interval. The $H$ = 0 intercept of this fit gives the value of $\mu_s$. Figure \ref{fig:extrapolated} shows the procedure of obtaining $\mu_s$ for $x$ = 0.012 as an example. $\mu_s$ for all the other ordered samples (0 $\leq x \leq$ 0.097) is obtained in a similar manner and are plotted in Fig \ref{fig:effective} of the main text.

\begin{figure}[h!]
    \centering
    \includegraphics[width=\linewidth]{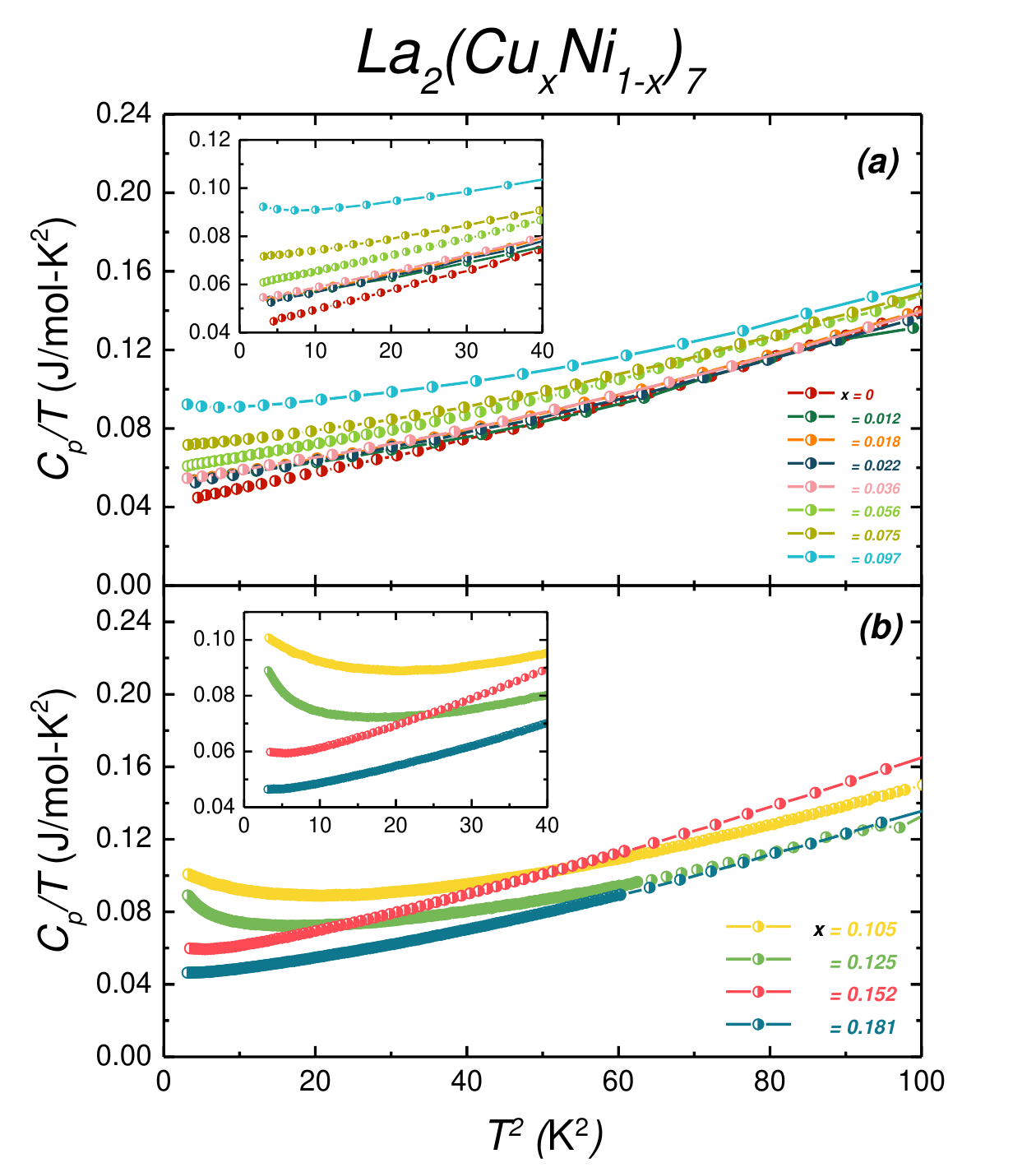}
    \caption{\footnotesize{(Color online) Plot of $C_p/T$ vs $T^2$ for the ${\text{La}_{2}\text{(Cu}_{x}\text {Ni}_{1-x})_7}$ samples divided into two separate panels based on their low T behavior. Panel $(a)$ shows the low doped 0 $\leq x \leq$ 0.097 which are almost linear down to the lowest measured temperature. The inset shows the zoomed in data for clarity. Panel $(b)$ shows the $C_p$ vs $T^2$ data for 0.105 $\leq x \leq$ 0.181 which is no longer linear and has a low T upturn. The insets shows the low T data for $T^2 \leq$ 40 K for clarity. }}
    \label{fig:cp_T2}
\end{figure}

We have shown the change of the effective moment with $x$ for ${\text{La}_{2}\text{(Cu}_{x}\text {Ni}_{1-x})_7}$ samples in Fig \ref{fig:effective}. $\mu_{eff}$ is obtained from doing a Curie-Weiss fit to the polycrystalline $M(T)/H$ data which has been already been explained in details in the main text. The fit was done for Cu substituted samples, 0 $\leq x \leq$ 0.152, except the highest doped sample ($x$ = 0.181), as it does not fit well to a Curie-Weiss analysis, most likely due to low value of magnetic moment for $x$ = 0.181, even when measured at an external field of 20 kOe.  Figure \ref{fig:CW data} shows the polycrystalline $M(T)$ data measured at $H$ = 20 kOe for all the Cu substituted samples along with the Curie-Weiss fit line described by Eqn. \ref{eqn:CW fit} for the 0 $\leq x \leq$ 0.152 Cu substituted  ${\text{La}_{2}\text{(Cu}_{x}\text {Ni}_{1-x})_7}$ samples. CW fit is done over the temperature range 100 K $\leq T \leq$ 300 K and the fit is shown using a red line. To visualize the quality of the fit we plotted $1/(\chi-\chi_0)$ vs $T$ after obtaining $\chi_0$ from Eqn. \ref{eqn:CW fit} for the 0 $\leq x \leq$ 0.152 samples. $1/(\chi-\chi_0)$ vs $T$ should be linear in the same temperature range as that of the CW fit (100 K $\leq T \leq$ 300 K) as shown using a green line. The values of all these parameters are shown in Table \ref{tab:magnetic}. The values for the parent $x$ = 0 in Ref \cite{Ribeiro2022Small-moment/math} were obtained from doing a CW fit to the data at $H$ = 1 kOe.

\begin{figure}[h]
    \centering
    \includegraphics[width=\linewidth]{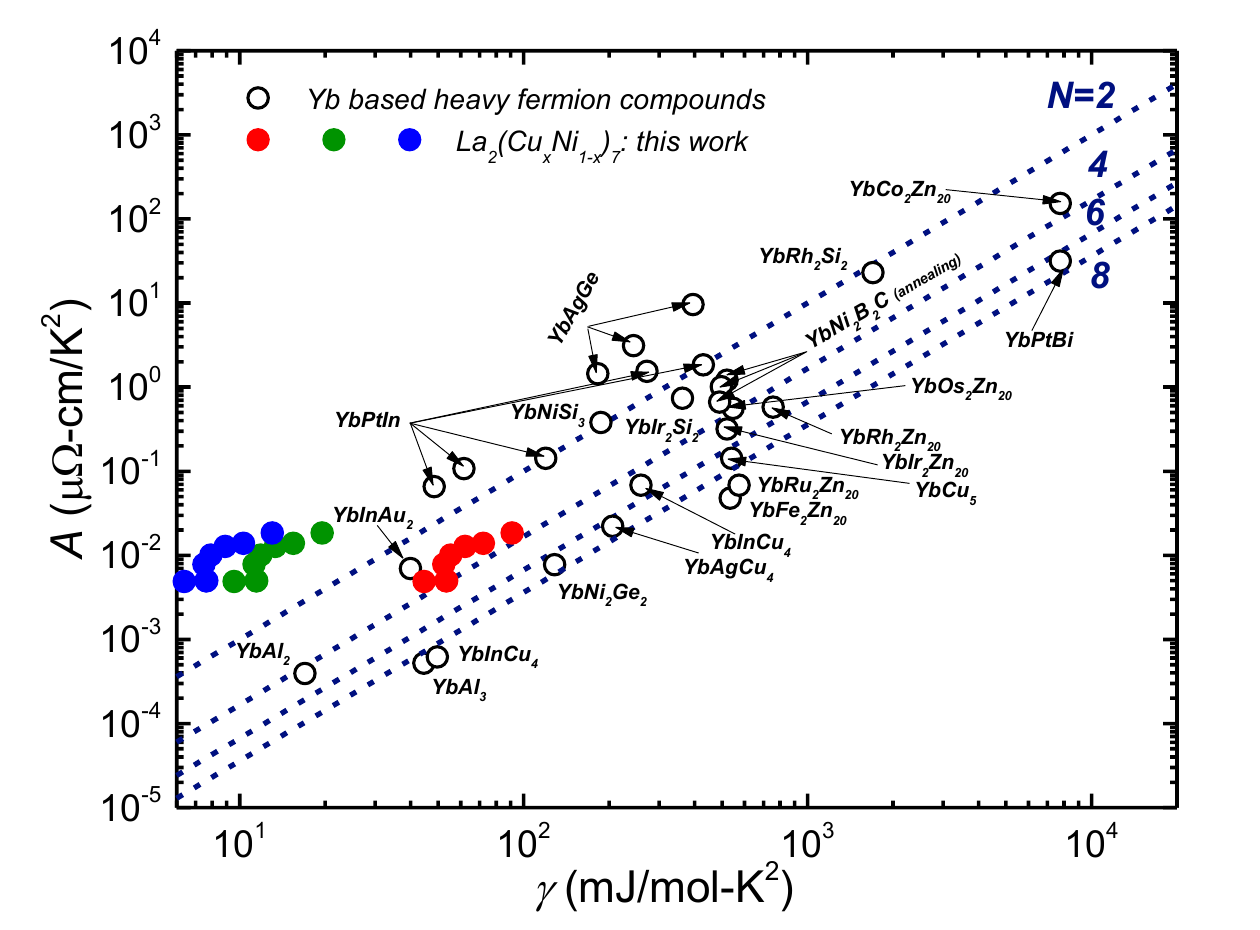}
    \caption{\footnotesize{(Color online) Generalized Kadowaki-Woods plot for Yb-based systems \cite{generalized_KW} with our data on ${\text{La}_{2}\text{(Cu}_{x}\text {Ni}_{1-x})_7}$ plotted on it three times. Once for simple per-mole-f.u.(red circles), a second time where we have $\gamma$ in terms of per-mole TM (transition metal) (i.e. divide $\gamma$ by 7) (blue circles), and finally $\gamma$ in terms of per-mole Ni-on Kagome plane (i.e. multiply $\gamma$ by 9/42) (green circles). The data for the Yb-based systems have been taken from \cite{YbT2Zn20_MT}. The $N$ values shown in the plot refer to the amount of degeneracy (for example, associated with Yb-CEF splitting) that is hybridized into correlated electron state by the Kondo-effect.}}
    \label{fig:kadowaki}
\end{figure}

In the main text we explained the power law analysis of the resistivity data. A fit to Eqn. \ref{eqn:power law} was shown only for the parent compound ($x$ = 0). In Fig. \ref{fig:power law all} we show such a fit for 0.012 $\leq x \leq$ 0.105 ${\text{La}_{2}\text{(Cu}_{x}\text {Ni}_{1-x})_7}$ samples. We obtain $\rho_0$ from these fits, and use those values to plot $\ln{(\rho - \rho_{0})}$ vs $\ln{T}$. A linear fit corresponding to Eqn. \ref{eqn: log power_law} is done on the $\ln{(\rho - \rho_{0})}$ vs $\ln{T}$ to obtain $n$ and $A$. Figure \ref{fig:n and A} shows the evolution of $n$ with $x$. Additionally, we can also obtain these parameters using Eqn. \ref{eqn:power law} to the low temperature $\rho(T)$ data. In both cases, the obtained values are within their uncertainty limits.

\par

\begin{figure*}[htbp]
    \centering
    \includegraphics[width=\textwidth]{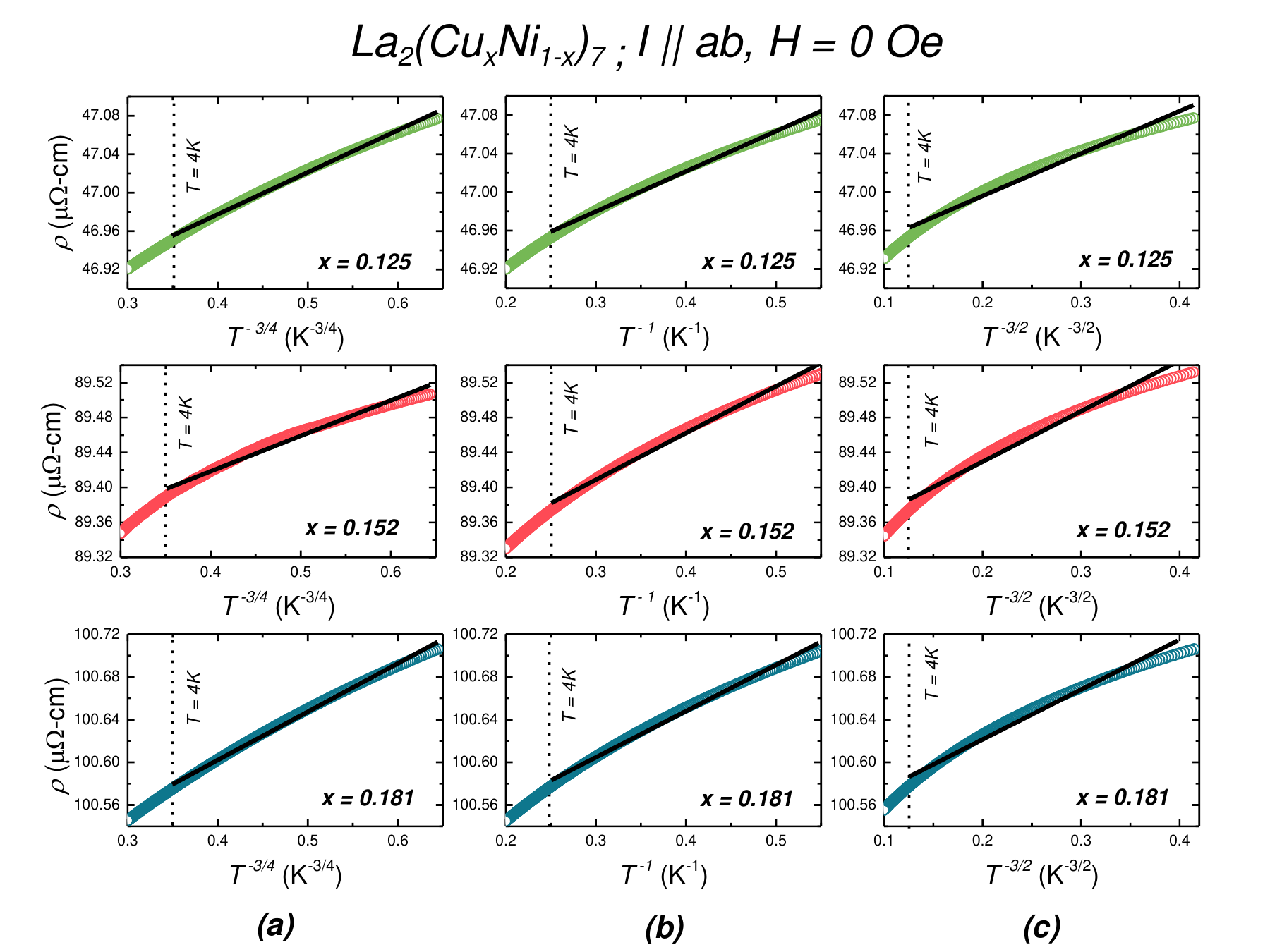}
    \caption{\footnotesize{(Color online) Plot of resistivity vs $T^{-3/4}$ in Col$(a)$, $T^{-1}$ in Col$(b)$, and $T^{-3/2}$ in Col$(c)$, for $x = 0.125, 0.152,$ and $0.181$ single crystals. These particular Cu substitutions show upturn in resistivity with decreasing T (See Fig $\ref{fig:R-T separate}$ in the main text). In each of these panels we mark the corresponding $T$ = 4 K value for reference. To verify the temperature dependence and subsequently the nature of the low T upturn, we do a linear fit to each of the panels in the temperature range 1.8 K $\leq x \leq $ 4 K. The linear fit line is shown using the black line. The low T upturn in resistivity is linear in $\ln{1/T}$ as seen in Fig \ref{fig:Kondo log} and not in any of the following T dependencies and is thus due to Kondo effect in the ${\text{La}_{2}\text{(Cu}_{x}\text {Ni}_{1-x})_7}$ crystals for $x = 0.125, 0.152,$ and $0.181$.}}
    \label{fig:kondo test}
\end{figure*}

Figure \ref{fig:cp_T2} shows the $C_P/T$ vs $T^2$ for the ${\text{La}_{2}\text{(Cu}_{x}\text {Ni}_{1-x})_7}$ samples. The data are divided into two panels. For 0 $\leq x \leq$ 0.097, $C_p/T$ vs $T^2$, starts for low $x$ with almost linear temperature dependence and starts picking up an increasingly strong, low temperature upturn as $x$ approaches 0.1. For 0.105 $\leq x \leq$ 0.181, the Kondo-like upturn is largest for $x$ = 0.105 and decreased with increasing $x$. For $x \geq$ 0.152, the $C_P/T$ vs $T^2$ is still not linear. 

\par

In order to evaluate the apparent scaling of $C_p/T$ at $T$ = 2 K with $A$ shown in Fig. \ref{fig:gamma} (in the main text), we show in Fig. \ref{fig:kadowaki} a generalized Kadowaki-Woods (KW) plot for a wide variety of Yb-based heavy fermions \cite{YbT2Zn20_MT} and our data on the ${\text{La}_{2}\text{(Cu}_{x}\text {Ni}_{1-x})_7}$ system. It is very important to note that whereas the units of $A$ do not depend on our choice of what atoms or interactions are important, the units of $C_p/T$ do. For this plot all of the Yb compounds have a single Yb ion per mole formula unit, therefore "per mole f.u." is the same as "per mole-Yb". Also, given that the correlated electron behavior is credibly coming from the Yb ions, this is a reasonable plotting choice. In the case of our system, we plot our data in three ways: first in a simple per-mole-f.u. notation, second in a per-mole-transition metal (Ni/Cu) and third in a per-mole-Ni-on-Kagome-lattice-site. The second plotting divides $C_p/T$ by 7 and the third plotting multiplies the $C_p/T$ by 9/42 given that only 9 of the 42 Ni atoms per unit cell are on Kagome planes. If we presume that all the Ni ions are responsible for the correlated electron behavior, multiplication by 1/7 is appropriate. If we presume that only Ni ions on a Kagome site are responsible, then multiplication by 9/42 is better. Each data manifold demonstrates that there is good KW scaling. That said, until it is determined which Ni-sites are responsible for enhanced $\gamma$ values, we cannot really say which KW-manifold (if any) that agree best with.

In the main text it was already discussed that an upturn in resistivity at low temperatures can be due to Kondo effect and we have observed that the higher doped ${\text{La}_{2}\text{(Cu}_{x}\text {Ni}_{1-x})_7}$ samples exhibit Kondo behavior based on a low temperature $-\ln(T)$ resistivity dependence in Fig \ref{fig:Kondo log}. We also tried to test whether the low-temperature $\rho(T)$ upturn could be explained by weak localization by plotting the resistivity of $x$ = 0.125, 0.152 and 0.181 samples with respect to $T^{-{p/2}}, (p = 3/2, 2,3)$. Figure \ref{fig:kondo test} shows these plots. A linear fit was also done to each of these temperature dependencies to each of these Cu dopings to verify the correlation. It can be observed that the low temperature resistivity does not follow a $T^{-{p/2}}$ behavior suggesting the Kondo behavior as a more reasonable explanation for our data in the ${\text{La}_{2}\text{(Cu}_{x}\text {Ni}_{1-x})_7}$ system.

%\newpage
%\clearpage

\nocite{apsrev41Control}
\bibliographystyle{apsrev4-2}
\bibliography{References}

\end{document}